\documentclass[journal]{IEEEtran}

\usepackage{graphicx}
\usepackage{amsmath}
\usepackage{amssymb}
\usepackage{cite}
\usepackage[caption=false,font=footnotesize]{subfig}

\begin{document}

\title{Comparative Performance of Graphene-Enabled Transmitarray Antenna and Reflectors for Wide-Angle Terahertz Beam Steering}
\author{
Somayeh~Komeylian,~\IEEEmembership{Member,~IEEE},
Truong~Nguyen,~\IEEEmembership{Fellow,~IEEE},
and Christopher~Paolini,~\IEEEmembership{Member,~IEEE}
\thanks{
S. Komeylian is with the Department of Electrical and Computer
Engineering, University of California San Diego, La Jolla, CA 92093
USA, and the Department of Electrical and Computer Engineering,
San Diego State University, San Diego, CA 92182 USA
(e-mail: skomeylian@ucsd.edu).
}
\thanks{
C. Paolini is with the Department of Electrical and Computer
Engineering, San Diego State University, San Diego, CA 92182 USA
(e-mail: paolini@engineering.sdsu.edu).}
\thanks{T. Nguyen is with the Department of Electrical and Computer Engineering,  University Of California, San Diego, La Jolla, CA 92093-0403 USA (e-mail: tqn001@ucsd.edu).}
}
\markboth{}%
{Komeylian \MakeLowercase{\textit{et al.}}}
\maketitle
\begin{abstract}
This work compares a hemispherical graphene-based transmitarray antenna with its planar reflector counterpart for wide-angle beam steering in the THz regime. 
The theoretical framework of the planar reflector is formulated and numerically evaluated, yielding a elevation beam-steering range of $\pm60^\circ$. 
In contrast, the transmitarray extends the elevation steering range to $\pm78^\circ$ while maintaining full $360^\circ$ azimuthal coverage. 
The planar reflector exhibits a larger HPBW variation of $26.48^\circ$, compared with $13.1^\circ$ for the transmitarray, resulting in a broader reflected-beam distribution and reduced directional power density, directivity, and gain. 
Meanwhile, the transmitarray maintains a more stable and controllable beamwidth response with greater directional power concentration over a wide steering range.

The performance advantages of the transmitarray are demonstrated through comparisons with experimental results reported in the literature for planar reflectors and antennas.
We further provide a comprehensive assessment of the performance advantages of the transmitarray over planar configurations across the remaining metrics.
\end{abstract}
\begin{IEEEkeywords}
Graphene, hemispherical transmitarray, terahertz antennas,
metasurfaces, beam steering, reconfigurable surface.
\end{IEEEkeywords}

\section{Introduction}
\IEEEPARstart{T}{he} terahertz (THz) frequency band has emerged as a promising candidate for future sixth-generation (6G) wireless communication systems because of its potential to support extremely high data rates, ultra-wide transmission bandwidths \cite{hirata2015ultrafast,he2017stochastic}, low-latency communication\cite{komeylian2023implementation}, high-resolution sensing and imaging\cite{chung2024terahertz,ait2023synthetic}, highly directional links, and dense spatial multiplexing\cite{xu2014reconfigurable}. The abundant spectrum available at THz frequencies can enable multi-gigabit-per-second to terabit-per-second wireless links, while the short wavelengths permit the realization of compact antenna arrays with a large number of elements, enabling highly precise beamforming and beam steering\cite{komeylian2024ffnn,huang2025high,karimi2024silicon,komeylian2020performance,komeylian2023overcoming,komeylian2022implementation,komeylian2023high}. 

Furthermore, the short wavelengths at THz frequencies enable highly directive radiation with narrow beamwidths, which can enhance spatial reuse by allowing multiple links to operate within the same spectral and spatial resources. 
The resulting spatial confinement reduces inter-beam interference and enhances interference management. Additionally, the directional nature of THz links boosts physical-layer security by limiting unintended signal leakage.
To overcome severe path loss, THz systems utilize highly directional beams and ultra-massive multiple-input multiple-output (UM-MIMO) architectures to maximize the received signal-to-noise ratio (SNR). Although this high-gain beamforming strategy enhances the link budget, the resulting narrow beamwidth severely restricts the spatial coverage required for tracking multiple dynamic targets. 
While conventional phased arrays enable beam steering, their implementation at terahertz frequencies is severely constrained by the prohibitive hardware complexity and high power consumption associated with large-scale phase-shifting networks\cite{song2022terahertz,deng2015320}.

Graphene-enabled reconfigurable metasurfaces overcome these limitations, offering a promising alternative for achieving dynamic beam control without relying on conventional large-scale phase-shifting networks\cite{liu2021multichannel,jiang2026breaking}. 
The high electron mobility of graphene enables rapid electrostatic modulation of its complex surface conductivity, facilitating dynamic, reconfigurable, and programmable control over the terahertz wavefront response.
Moreover, its linear energy-momentum dispersion relation yields an inherently broadband optoelectronic response, rendering graphene uniquely suited for wideband THz operations.
Electrically tuning the chemical potential of individual graphene-based array elements dynamically modulates their surface conductivity and transmission phase. This allows precise synthesis of the required spatial phase distribution, enabling electronically reconfigurable, wideband beam steering.
Owing to its atomically thin profile, graphene exhibits a limited interaction length with normally incident electromagnetic waves, resulting in relatively weak light-matter interaction.
Consequently, an isolated graphene layer may provide limited modulation depth in both the amplitude and phase of the transmitted THz wave, thereby constraining its effectiveness for dynamic wavefront control.
In the proposed hemispherical transmitarray antenna in Fig.~\ref{fig:hemispherical_transmitarray_2}, this limitation is mitigated by embedding independently biased graphene sectors within a multilayer electromagnetic unit cell comprising gold radiating patches and hBN dielectric layers. 
The graphene conductivity is electrically controlled through its chemical potential, thereby modifying the effective surface impedance and, consequently, the complex transmission coefficient of each unit cell. 

The metallic patches and surrounding dielectric media boost local electromagnetic fields near the graphene layer. 
This ensures that even modest changes in surface conductivity significantly change the transmission phase.
Furthermore, independently biasing the isolated graphene sectors provides element-level phase control. 
This localized response is subsequently synthesized across the conformal hemispherical aperture to achieve dynamic beam steering.
The hemispherical configuration further provides nearly normal incidence with respect to the local surface of each element, improving the uniformity of the element responses across the aperture. 
These design strategies mitigate the intrinsically weak light--matter interaction of atomically thin graphene, enabling effective phase control across the conformal aperture and, consequently, realizing wide-angle three-dimensional beam steering\cite{komeylian2025high,komeylian2026graphenebasedhemisphericaltransmitarrayantenna,komeylian2026activehemisphericalmetasurfacetransmitarray}.

As reported in \cite{komeylian2025high,komeylian2026graphenebasedhemisphericaltransmitarrayantenna,komeylian2026activehemisphericalmetasurfacetransmitarray}, the antenna efficiency ranges from 71\% to 80\%, with a degradation of 5\% across the entire scanning range due to practical fabrication imperfections. 
These performance metrics render the proposed transmitarray antenna highly suitable for real-time moving-target-tracking applications.

To mitigate severe propagation attenuation inherent to THz channels, Li \textit{et al.}~\cite{li2026wide} proposed a reflective beam-steering metasurface leveraging cross-polarization conversion and phase-dispersion engineering. 
The reflector provides a quasi-linear phase response with a broad phase-tuning range while maintaining high reflection efficiency over a wide frequency band, thereby enabling stable wide-angle beam steering across an angular range of $\pm45^\circ$ over a continuous frequency tuning range.
Furthermore, experimental validation successfully demonstrated its viability for 16-state quadrature-amplitude modulation (16-QAM) in a non-line-of-sight (NLOS) THz communication link.

In contrast, our proposed reflector in Fig.~\ref{fig:flat_reflector_configuration} incorporates independently tunable annular graphene sectors to achieve a $\pm60^\circ$ beam-steering range, and provide full $360^\circ$ azimuthal coverage while maintaining a simplified geometrical configuration.

Recent experimental validations have further verified the efficacy of graphene-based architectures for dynamic, ultra-wideband THz beamforming\cite{jiang2026breaking}.
The graphene-driven transmissive programmable metasurface, incorporating a dual-resonance configuration and a graphene–insulator–graphene (GIG) heterostructure, demonstrated independent gain and beam-steering control, achieving a scanning range of $\pm45^\circ$ across 187–250 GHz with a fractional bandwidth of 28.8\% \cite{jiang2026breaking}.
Despite these advances, the demonstrated architecture remains based on a planar transmissive metasurface with a binary phase-coding scheme. 

In contrast, our proposed hemispherical transmitarray antenna in Fig.~\ref{fig:hemispherical_transmitarray_2} operates at the same center frequency of 250~GHz and employs independently tunable graphene sectors to achieve an electrically reconfigurable three-dimensional beam-steering capability over a substantially wider angular range of $\pm78^\circ$, while maintaining a relative bandwidth of 40\% with $S_{11}<-10$ dB over the entire 200--300~GHz frequency for representative broadside and extreme off-axis directions.
This substantial enhancement in beam-steering capability, compared not only with its geometrically reflector counterpart, Fig.~\ref{fig:flat_reflector_configuration}, but also with other reflectors\cite{li2026wide} and transmitarray antennas\cite{jiang2026breaking}, is primarily attributed to its hemispherical configuration and the constituent fractal concentric circular (FCC) array elements. 
Moreover, the centrally located feedhorn in the proposed transmitarray establishes a normal incidence condition at the individual array elements, resulting in more uniform electromagnetic excitation across the conformal aperture. 
This spatial uniformity facilitates consistent element responses and precise, agile phase tuning, thereby enabling wide-angle two-dimensional beam steering \cite{komeylian2025high,komeylian2026graphenebasedhemisphericaltransmitarrayantenna,komeylian2026activehemisphericalmetasurfacetransmitarray}.

The theoretical framework and design considerations for the proposed hemispherical transmitarray antenna were presented in\cite{komeylian2025high,komeylian2026graphenebasedhemisphericaltransmitarrayantenna}, while its corresponding measurement results were discussed in \cite{komeylian2026activehemisphericalmetasurfacetransmitarray}. 
Complementing these previous investigations, this work comprehensively compares the proposed hemispherical transmitarray antenna with its planar reflector counterpart and with experimental measurement results reported for representative antennas and reflectors.
Although reflectors and transmitarray antennas belong to different antenna classes and employ fundamentally different electromagnetic mechanisms, the objective of this work is not to establish a general performance comparison between the two architectures. 
This comparison specifically evaluates the beam-steering performance of the proposed hemispherical transmitarray in Fig.~\ref{fig:hemispherical_transmitarray_2} relative to its geometrically corresponding reflector counterpart in Fig.~\ref{fig:flat_reflector_configuration}.

To streamline future fabrication, we aim to adopt the concept of the utilization of localized planar elements conformally arranged over a hemispherical surface, as illustrated in \cite{catalani2009ku} for our proposed hemispherical transmitarray antenna. 
This topological approximation substantially reduces fabrication and assembly complexity while preserving the electromagnetic and structural characteristics inherent to the conformal hemispherical geometry.

Another challenge to be addressed in future fabrication efforts is the routing and control complexity associated with managing 121 independently tunable graphene patches. 
To mitigate this complexity, a hierarchical biasing architecture is utilized.
Rather than assigning a dedicated bias line to each hemispherical sector, the elements are addressed through grouped or multiplexed DC control lines, thereby reducing the overall routing complexity.
Graphene patches requiring identical chemical potentials can share a common bias network, thereby reducing the number of physical interconnections while preserving the desired spatial phase distribution for beam steering.
The biasing circuitry is distributed along the peripheral rings or integrated onto a dedicated control substrate, with electrical connectivity to the graphene layers established through short local vias.
Furthermore, the bias lines will be appropriately designed and optimized to exhibit high impedance at RF frequencies, thereby minimizing their electromagnetic loading and reducing their perturbation of the antenna performance at the 250 GHz operating frequency.
This biasing architecture substantially reduces the fabrication and control complexity while preserving the wide-angle beam-steering capability and reconfigurability of the proposed hemispherical transmitarray antenna.
\begin{figure}[htbp]
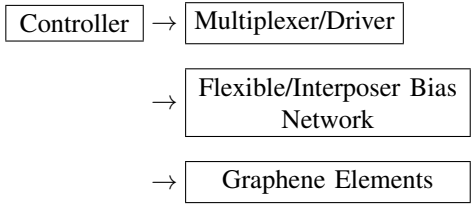

    \centering
    \[\begin{aligned}
        \fbox{\parbox{0.18\linewidth}{\centering Controller}}
        &\rightarrow
        \fbox{\parbox{0.30\linewidth}{\centering Multiplexer/Driver}}
        \\[6pt]
        &\rightarrow
        \fbox{\parbox{0.40\linewidth}{\centering Flexible/Interposer Bias Network}}
        \\[6pt]
        &\rightarrow
        \fbox{\parbox{0.40\linewidth}{\centering Graphene Elements}}
\end{aligned}
\]
\caption{The conceptual architecture of our proposed hierarchical biasing and control scheme for the graphene-based hemispherical transmitarray antenna consistent with Fig.~\ref{fig:hemispherical_transmitarray_2}}
\label{fig:biasing_architecture}
\end{figure}
The remainder of this paper is organized as follows.
Section II establishes the theoretical framework and foundational formulations for the planar reflector counterpart of the proposed hemispherical transmitarray antenna. 
Here, the term of \textit{its planar reflector counterpart} denotes a planar reflector configuration, Fig.~\ref{fig:flat_reflector_configuration}, established as a comparable reference for the proposed hemispherical transmitarray antenna, Fig.~\ref{fig:hemispherical_transmitarray_2}, while maintaining the same operating frequency, graphene tunability range, and reconfigurable control concept. 
The primary geometrical distinction is that the graphene elements are arranged on a planar surface in the reflector rather than conformally distributed over a hemispherical surface in the proposed transmitarray. 
Since the electromagnetic operating principle is changed from transmission to reflection, the material composition and multilayer configuration are correspondingly modified to accommodate the reflector architecture.
Section III is devoted to the formulations of the array factor for the planar reflector counterpart and the development of the theoretical foundations for phase synthesis and dynamic beam steering.
Section IV investigates the broadband operation and beam-Steering stability of the proposed hemispherical transmitarray antenna. The evaluation spans two distinct sub-bands: 200–300 GHz with an operating frequency of 250 GHz and 240–360 GHz with an operating frequency of 350 GHz, while preserving the identical design dimensions as those shown in Fig.~\ref{fig:hemispherical_transmitarray_2}.
Section V evaluates the significant enhancement of the half-power beamwidth (HPBW) performance achieved by the hemispherical transmitarray compared to its planar reflector counterpart.  
Section VI highlights that the incorporation of fractal concentric circle elements substantially enhances the bandwidth performance of the hemispherical transmitarray antenna.
Section VII provides a further comprehensive comparison between the hemispherical transmitarray and its planar reflector counterpart, highlighting the advantages of intrinsic alignment, a significant reduction in the edge diffraction and parasitic radiation, and also ohmic losses.
Finally, Section VIII concludes the work and summarizes the contributions of this work.
\section{Theoretical Background and Analytical Formulations for the Planar Reflector}
\begin{figure}[htbp]
    \centering
    \includegraphics[
        width=0.95\columnwidth,
        height=0.66\columnwidth
    ]{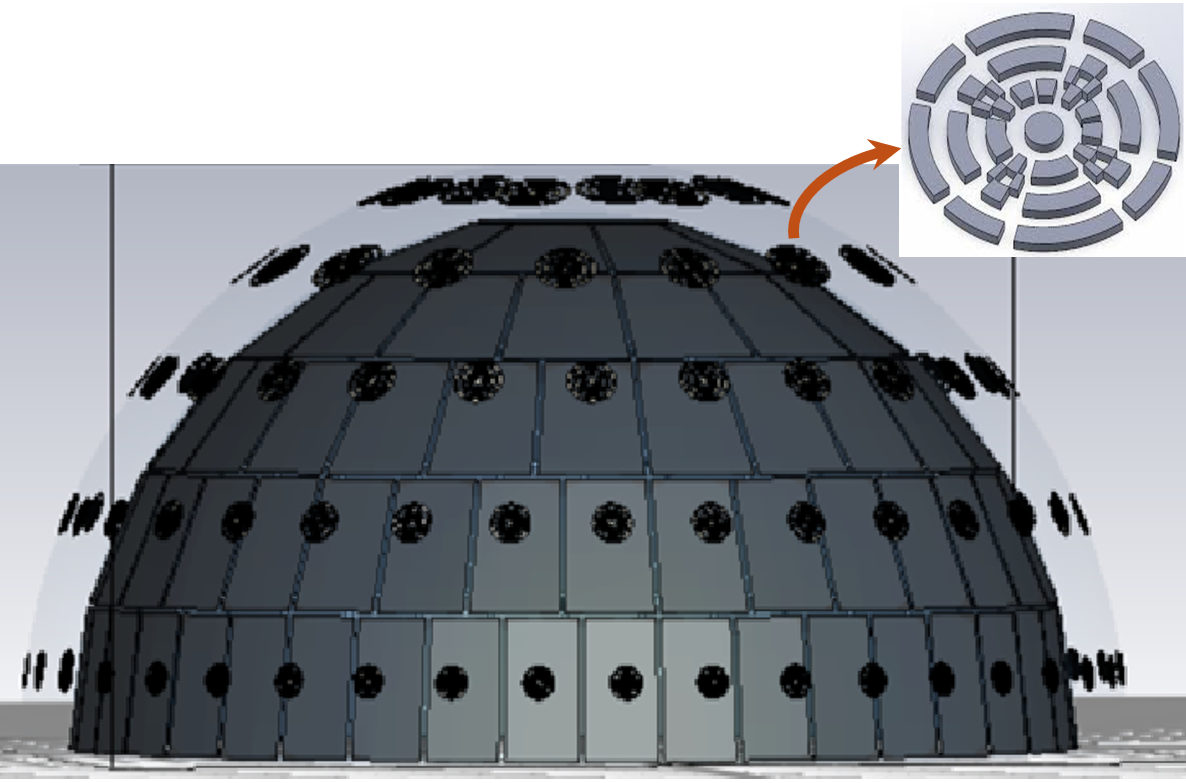}
    \caption{
    Configuration of the proposed hemispherical transmitarray antenna. The structure comprises 121 independently controlled graphene sectors conformally distributed over the hemispherical surface. Each locally planar cell is approximated by a trapezoidal geometry. The chemical potential of each graphene sector is independently tuned over the range of $0.1$--$1.0$~eV, while a carrier relaxation time of $\tau=0.1$~ps and an operating temperature of $T=300$~K are assumed for all graphene sectors. The dielectric layers consist of SiO$_2$ with $\varepsilon_{r,\mathrm{SiO_2}}=3.9$ and $t_{\mathrm{SiO_2}}=10~\mu\mathrm{m}$, and hBN with $\varepsilon_{r,\mathrm{hBN}}=4.4$ and $t_{\mathrm{hBN}_1}=t_{\mathrm{hBN}_2}=0.25~\mu\mathrm{m}$. The graphene sectors are physically and electrically isolated from one another, enabling independent electrostatic biasing and reconfigurable control of the local electromagnetic response across the conformal aperture\cite{komeylian2026graphenebasedhemisphericaltransmitarrayantenna,komeylian2025high, komeylian2026activehemisphericalmetasurfacetransmitarray}.}
    \label{fig:hemispherical_transmitarray_2}
\end{figure}

\begin{figure}[htbp]
    \centering
    \includegraphics[
        width=0.9\columnwidth,
        height=0.85\columnwidth
    ]{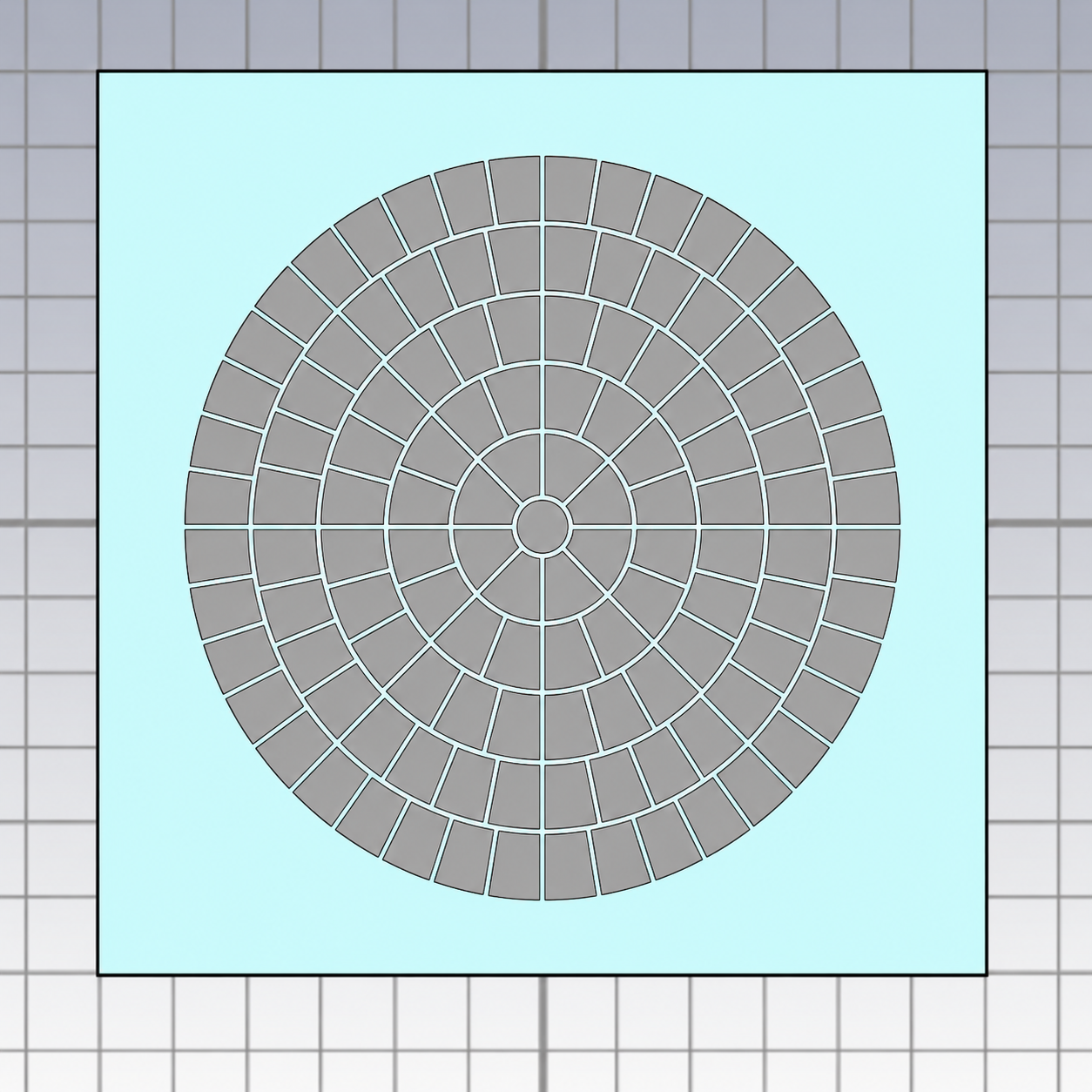}
    \caption{The geometrical reflector counterpart of the proposed hemispherical transmitarray antenna operating at 250~GHz. The reflector comprises five active concentric rings, with the number of elements in the successive rings given by $N_n=[8,16,24,32,40]$, where $n=1,2,\ldots,5$. 
    Our proposed planar reflector comprises 121 graphene sectors, arranged in an apex and the five concentric rows. The graphene chemical potential of each annular sector is independently tunable over the range of 0.1--1.0~eV, with a carrier relaxation time of $\tau=0.1$~ps and an operating temperature of $T=300$~K. The multilayer structure consists of a 1-mm-thick perfect electric conductor (PEC) ground plane, followed by a 1-$\mu$m-thick SiO$_2$ layer with $\varepsilon_r=3.85$ and a 0.1-$\mu$m-thick hexagonal boron nitride (hBN) layer with $\varepsilon_r=4.5$. A circular feedhorn aperture with an aperture diameter of 5$mm$ is positioned at a distance of $\approx 41.7\lambda_0$ from the reflector plane.}
\label{fig:flat_reflector_configuration}
\end{figure}

The discussion of this section has focused on representing the theoretical background and formulations of the planar reflector in Fig.~\ref{fig:flat_reflector_configuration}, which serves as the geometrically equivalent counterpart to the proposed hemispherical transmitarray antenna in Fig.~\ref{fig:hemispherical_transmitarray_2}.
The proposed reflector consists of 121 independently controlled graphene sectors distributed over the planar surface. An array of annular graphene sectors constitutes the planar reflector.
The graphene chemical potential of each annular sector can be independently controlled over the range of $0.1$--$1.0$~eV. All graphene sectors are characterized by a carrier relaxation time of $\tau=0.1$~ps at an operating temperature of $T=300$~K. 
The sectors are physically and electrically isolated from one another to enable independent electrostatic biasing and stable reconfigurable operation.  
\subsection{Graphene conductivity}
In the THz frequency regime, the intraband response of graphene is accurately described by the semiclassical Drude model.
Assuming the $e^{j\omega t}$ time-harmonic convention, the graphene surface conductivity of $\sigma_g\left(\omega,\mu_c^{(\mathrm{J})},\tau\right)
$ is expressed as, 
\begin{equation}
\begin{aligned}
\sigma_g\left(\omega,\mu_c^{(\mathrm{J})},\tau\right)
=\\{}&
\frac{e^2 k_B T}{\pi\hbar^2}
\left[
\frac{\mu_c^{(\mathrm{J})}}{k_B T}
+
2\ln\left(
e^{-\frac{\mu_c^{(\mathrm{J})}}{k_B T}}+1
\right)
\right]
\frac{1}{\tau^{-1}+j\omega}
\end{aligned}
\label{eq:graphene_drude_conductivity}
\end{equation}
where $\mu_c$, $\tau$, $e$, and $T$ denote the chemical potential (Fermi energy), the momentum relaxation time, the elementary charge, and the absolute temperature, respectively. 
At room temperature and for sufficiently high chemical potentials satisfying $\mu_c \gg k_B T$, interband thermal excitation across the Dirac point is negligible. Consequently, the general expression in Eq.\ref{eq:graphene_drude_conductivity} reduces to the conventional intraband Drude formulation,
\begin{equation}
\sigma_g\left(\omega,\mu_c^{(\mathrm{J})},\tau\right)
\approx
\frac{e^2\mu_c^{(J)}}{\pi\hbar^2}
\frac{1}{\tau^{-1}+j\omega}
\label{eq:graphene_drude_simplified}
\end{equation}
The chemical potential in Eq.\ref{eq:graphene_drude_simplified} is expressed in joules where $\mu_c^{(\mathrm{J})}$ is related to its value in electron volts, $\mu_c^{(\mathrm{eV})}$, through $\mu_c^{(\mathrm{J})}=e\,\mu_c^{(\mathrm{eV})}$. 
The term of $\mu_c^{(\mathrm{eV})}$ denotes the numerical value of the chemical potential expressed in electronvolts (eV). Accordingly, Eq.\ref{eq:graphene_drude_simplified} can be rewritten in terms of the chemical potential in electron volts as,
\begin{equation}
\sigma_g\left(\omega,\mu_c^{(\mathrm{eV})},\tau\right)
=
\frac{e^3\mu_c^{(\mathrm{eV})}}{\pi\hbar^2}
\frac{1}{\tau^{-1}+j\omega}
\label{eq:graphene_drude_eV}
\end{equation}
In the zero-frequency limit, the Drude conductivity reduces to its
dc value, which is given by,
\begin{equation}
\sigma_g\left(\omega=0,\mu_c^{(\mathrm{eV})},\tau\right)
=
\sigma_{dc}
=
\frac{e^3\mu_c^{(\mathrm{eV})}}{\pi\hbar^2}
\frac{1}{\tau^{-1}}
=
\frac{e^3\mu_c^{(\mathrm{eV})}\tau}{\pi\hbar^2}
\end{equation}
The DC conductivity in the zero-frequency limit establishes the foundation for deriving the graphene sheet resistance and kinetic inductance. 

\textbf{Graphene Sheet Resistance}:
The corresponding graphene sheet resistance is obtained as the reciprocal of the DC conductivity,
\begin{equation}
R_s
=
\frac{1}{\sigma_{\mathrm{dc}}}
=
\frac{\pi\hbar^2}
{e^3\mu_c^{(\mathrm{eV})}\tau}
\label{eq:sheet_resistance}
\end{equation}
\textbf{Graphene Kinetic Inductance}:
The corresponding graphene kinetic inductance per square is expressed as, \begin{equation} L_k = \frac{\tau}{\sigma_{\mathrm{dc}}} = \frac{\pi\hbar^2} {e^3\mu_c^{(\mathrm{eV})}} \label{eq:kinetic_inductance} \end{equation}
\subsection{Complete Unit-Cell Geometry in the Cylindrical Coordinate System}
\begin{figure}[htbp]
\centering
\includegraphics[
    width=1\columnwidth,
    height=1\columnwidth
]{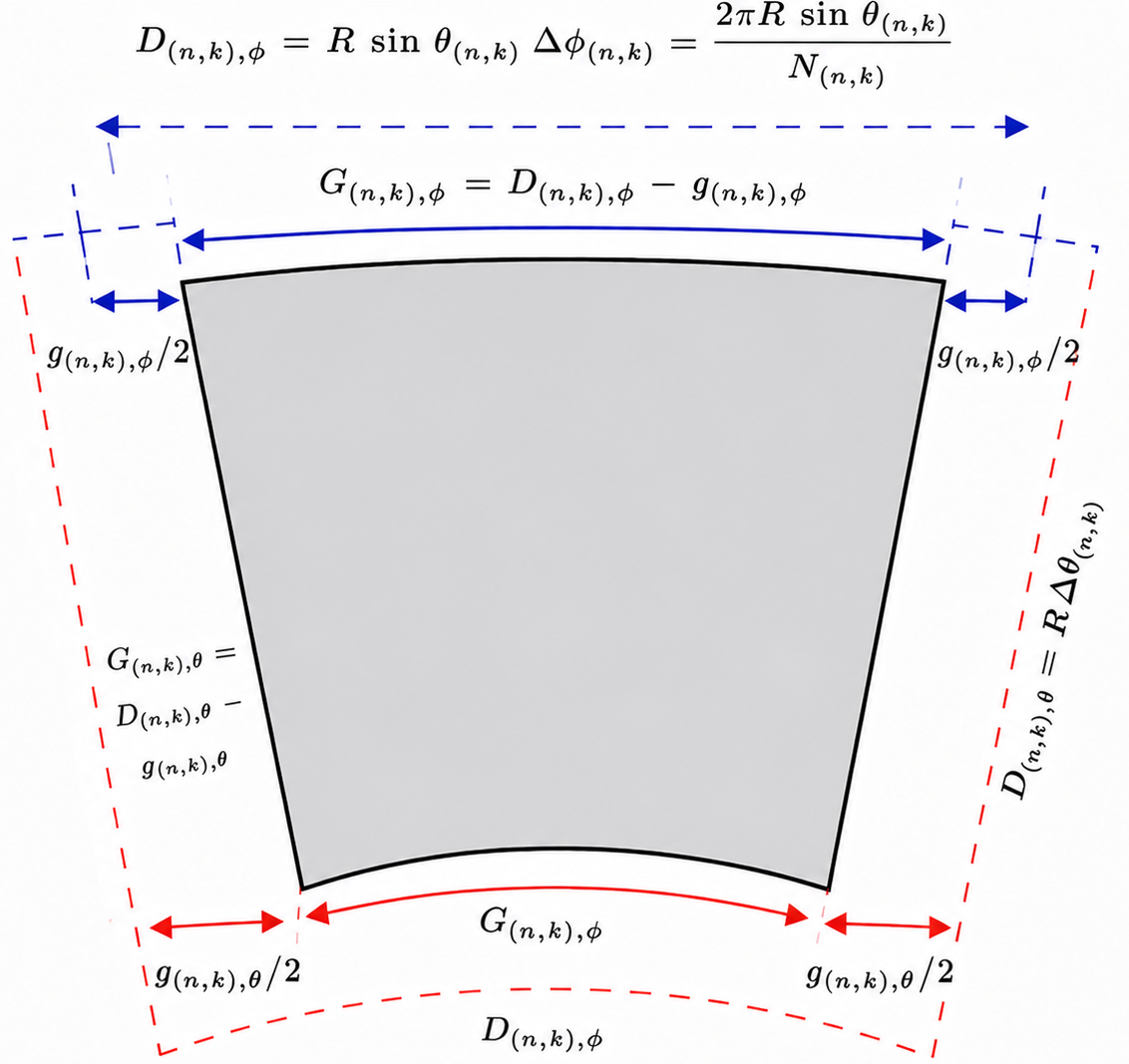}
\caption{Geometry of a representative annular graphene sector used in the planar reflector consistent with Fig.~\ref{fig:flat_reflector_configuration}, at 250 GHz.}
\label{fig:flat_reflector_sectors}
\end{figure}
The proposed planar circular reflector in Fig.~\ref{fig:flat_reflector_configuration} is defined in a cylindrical coordinate system of $(\rho, \phi, z)$, where $\rho$ denotes the radial distance from the reflector center, $\phi$ is the azimuthal angle, and the $z$-axis is normal to the reflector surface. 
The reflector is positioned in the $z = 0$ plane.
For the $(n,k)$-th graphene sector, the radial position of the $n$-th ring is defined as,
\begin{equation}
\Delta\rho=
\frac{\rho_{\max}-\rho_{\mathrm{a}}}{N_r}
\label{eq:radial_spacing}
\end{equation}
where $N_r$ denotes the total number of radial rings. 
The term of $\rho_{a}$ denotes the radius of the central disk, which is related to the radial spacing between consecutive rings by Eq.\ref{eq:radial_spacing}. Also, $\rho_{\max}$ refers to the outer radius of the planar circular reflector.
Excluding the graphene apex, the radial boundaries of the annular sectors within the $n$-th ring are defined as follows, 
\begin{equation}
\rho_{n,-}
=
a+(n-1)\Delta\rho
\label{eq:cyl_rho_minus_1}
\end{equation}
\begin{equation}
\rho_{n,+}
=
(n+1)\Delta\rho
\label{eq:cyl_rho_plus_1}
\end{equation}
The azimuthal position of the $(n,k)$-th graphene sector is given by,
\begin{equation}
\phi_{(n,k)}
=
\phi_{0,n}
+
\frac{2\pi(k-1)}{N_n},
\qquad
k=1,2,\ldots,N_n 
\label{eq:cyl_phi_nk}
\end{equation}
where $\phi_{0,n}$ is the initial azimuthal offset of the $n$-th ring. 
The number of graphene sectors in the $n$-th ring is governed by the following linear ring-population model,
\begin{equation}
N_n
=
a\left[1+\beta(n-1)\right]
\label{eq:cyl_Nn}
\end{equation}
where $a$ refers to the initial number of unit cells in the first ring and $\beta$ controls the rate at which the number of sectors increases toward the outer rings. Setting $a = 8$ and $\beta = 1$ in Eq.\ref{eq:cyl_Nn} yields the element distribution of $N_n = 8n$. This results in $N_n = [8, 16, 24, 32, 40]$ for the five active rings of $n = 1, \ldots, 5$, excluding the apex sector.
The azimuthal angular separation between adjacent sectors on the $n$-th ring is given by,
\begin{equation}
\Delta\phi_n
=
\frac{2\pi}{N_n}
\label{eq:cyl_delta_phi}
\end{equation}
The local position of the center of the $(n,k)$-th sector is expressed as,
\begin{equation}
\mathbf{r}_{(n,k)}
=
\begin{bmatrix}
\rho_n\cos\phi_{(n,k)}\\
\rho_n\sin\phi_{(n,k)}\\
0
\end{bmatrix}
\label{eq:cyl_position_vector}
\end{equation}
It is worth noting that the geometry defined by Eqs.~\ref{eq:radial_spacing}--\ref{eq:cyl_rho_plus_1} excludes the central circular graphene cell. Consequently, this apex disk is formulated separately and characterized by its own area, capacitance, impedance, reflection phase, and biasing variable.
Indeed, unlike the annular sectors, the apex sector is modeled independently and centered at the origin of the cylindrical coordinate system, i.e., at,
\begin{equation}
\rho_{0}=0,\qquad \phi_{0}=0
\label{eq:apex_center}
\end{equation}
The radial boundary of the apex sector is defined as,
\begin{equation}
0 \leq \rho \leq \rho_a 
\label{eq:apex_radial_boundary}
\end{equation}
The azimuthal boundary of the apex sector is given by, 
\begin{equation}
0 \leq \phi < 2\pi 
\label{eq:apex_azimuthal_boundary}
\end{equation}
Accordingly, the complete region of the apex sector in Fig.~\ref{fig:flat_reflector_configuration} is defined as, 
\begin{equation}
\mathcal{A}_{\mathrm{apex}}
=
\left\{
(\rho,\phi)\,\middle|\,
0 \leq \rho \leq \rho_a,\;
0 \leq \phi < 2\pi
\right\}
\label{eq:apex_region}
\end{equation}
\subsection{Local Dimensions of Each Graphene Sector}
Since the reflector lies in a planar geometry, the local radial boundary width of the $(n,k)$-th unit cell is governed by $\Delta\rho$, Figs.~\ref{fig:flat_reflector_configuration},~\ref{fig:flat_reflector_sectors}, and ~\ref{fig:flat_reflector_cross_section_1}. The dimension of the corresponding azimuthal cell is a function of the local arc length at radius $\rho_{n}$, which yields the following azimuthal periodicity,
\begin{equation}
D_{(n,k),\phi}
=
\rho_n\Delta\phi_n
=
\frac{2\pi\rho_n}{N_n}
\label{eq:D_phi}
\end{equation}
As a result, in contrast to the hemispherical transmitarray antenna, the azimuthal periodicity of the planar reflector is not scaled by the term of $\sin\theta_n$. 
With the graphene sectors physically isolated from one another, finite gaps are introduced along both the local radial and azimuthal directions, 
\begin{equation}
g_{(n,k),\rho}>0,
\qquad
g_{(n,k),\phi}>0
\label{eq:cyl_gaps}
\end{equation}
The radial dimension of the $(n,k)$-th graphene sector is therefore given by,
\begin{equation}
G_{(n,k),\rho}
=
D_{(n,k),\rho}-g_{(n,k),\rho}
=
\Delta\rho-g_{(n,k),\rho}
\label{eq:G_rho}
\end{equation}
Accordingly, the corresponding physical azimuthal dimension is expressed by,
\begin{equation}
G_{(n,k),\phi}
=
D_{(n,k),\phi}-g_{(n,k),\phi}
=
\frac{2\pi\rho_n}{N_n}
-
g_{(n,k),\phi}
\label{eq:G_phi}
\end{equation}
For a symmetric ring configuration in which all sectors within the same ring have identical dimensions, the corresponding gap parameters can be expressed as,
\begin{equation}
g_{(n,k),\rho}=g_{n,\rho}
\label{eq:ring_gap_rho}
\end{equation}
\begin{equation}
g_{(n,k),\phi}=g_{n,\phi}
\label{eq:ring_gap_phi}
\end{equation}
In this case, the modified gap dimensions of each electrically isolated graphene sector determine the corresponding equivalent RLC circuit parameters according to,
\begin{equation}
G_{n,\rho}
=
\Delta\rho-g_{n,\rho}
\label{eq:G_n_rho}
\end{equation}
\begin{equation}
G_{n,\phi}
=
\frac{2\pi\rho_n}{N_n}
-
g_{n,\phi}
\label{eq:G_n_phi}
\end{equation}
\subsection{Area of the Graphene Sector}
For a sufficiently small unit cell, the physical area of the $(n,k)$-th graphene sector can be approximated by,
\begin{equation}
A_{(n,k)}
\approx
G_{(n,k),\rho}G_{(n,k),\phi}
\label{eq:cyl_Ank}
\end{equation}
Geometrically, the exact surface area of an annular sector bounded by the radial boundaries is given by,
\begin{equation}
\Delta S_{(n,k)}
=
\frac{1}{2}
\left(
{\rho}_{n,+}^{2}
-
{\rho}_{n,-}^{2}
\right)
\Delta\phi_n
\label{eq:cyl_exact_area}
\end{equation}
where $\rho_{n,+}$ and $\rho_{n,-}$ are defined in Eqs.\ref{eq:cyl_rho_minus_1} and \ref{eq:cyl_rho_plus_1}, respectively, for the special case of $a=0$.
Substituting Eqs.\ref{eq:cyl_rho_minus_1} and \ref{eq:cyl_rho_plus_1} when $a=0$ into Eq.\ref{eq:cyl_exact_area} yields the exact surface area of the annular sector,
\begin{equation}
\Delta S_{(n,k)}
=
\frac{1}{2}
\left[
n^2-(n-1)^2
\right]
\Delta\rho^2
\Delta\phi_n
\label{eq:cyl_area_expanded}
\end{equation}
Using,
\begin{equation}
n^2-(n-1)^2
=
2n-1
\end{equation}
Equivalently, the area of an annular sector can be rewritten as, 
\begin{equation}
\Delta S_{(n,k)}
=
\frac{2n-1}{2}
\Delta\rho^2
\Delta\phi_n
\label{eq:cyl_area_n_1}
\end{equation}
Substituting the azimuthal spacing of $\Delta\phi_n = 2\pi/N_n$ into Eq.\ref{eq:cyl_area_n_1} yields,
\begin{equation}
\Delta S_{(n,k)}
=
\frac{(2n-1)\pi\Delta\rho^2}{N_n}
\label{eq:cyl_area_final}
\end{equation}
Hence, for the equivalent circuit model, the effective physical area of the $(n,k)$-th graphene sector is approximated by,
\begin{equation}
A_{(n,k)}
\approx
G_{(n,k),\rho}G_{(n,k),\phi}
\end{equation}
which accounts for the finite separation between adjacent graphene sectors rather than the entire area of the annular sector.
\subsection{Resistance of each Graphene Sector}
Although graphene is modeled as an isotropic two-dimensional conductor characterized by a scalar surface conductivity, the finite and nonuniform geometry of the electrically isolated annular sectors results in direction-dependent effective impedances. Accordingly, the unit-cell response is characterized by a local anisotropic surface-impedance tensor in the $(\rho,\phi)$ coordinate system, where the direction-dependent response results solely from the finite geometrical dimensions of the annular graphene sector or the geometrical aspect ratio of the annular graphene sector.
Thus, the corresponding local surface-impedance tensor, which accounts for the direction-dependent resistive and reactive responses of the $(n,k)$-th annular graphene sector, is expressed in the local $(\rho,\phi)$ coordinate system as,
\begin{equation}
\overline{\overline{Z}}_{\mathrm{eff},(n,k)}
=
\begin{bmatrix}
R_{(n,k),\rho} + jX_{(n,k),\rho} & 0 \\
0 & R_{(n,k),\phi} + jX_{(n,k),\phi}
\label{eq:effective_impedance_tensor}
\end{bmatrix}
\end{equation}
where $X_{(n,k),\rho}$ and $X_{(n,k),\phi}$ account for the corresponding inductive and capacitive contributions.
Within the equivalent series RLC model, the radial and azimuthal impedance components are expressed as,
\begin{equation}
Z_{(n,k),\rho}
=
R_{(n,k),\rho}
+j\omega L_{(n,k),\rho}
+\frac{1}{j\omega C_{(n,k),\rho}}
\label{eq:radial_RLC_impedance}
\end{equation}
\begin{equation}
Z_{(n,k),\phi}
=
R_{(n,k),\phi}
+j\omega L_{(n,k),\phi}
+\frac{1}{j\omega C_{(n,k),\phi}}
\label{eq:azimuthal_RLC_impedance}
\end{equation}
Accordingly, the unit-cell response is modeled by a local anisotropic surface-impedance tensor in the $(\rho,\phi)$ coordinate system, where the directional dependence arises strictly from the finite geometry of the annular graphene sector. 
Here, the geometrical aspect ratio is defined by the ratio of the radial width to the azimuthal arc length.  
If $G_{(n,k),\rho}=G_{(n,k),\phi}$, the sector has an aspect ratio of unity. 
If $G_{(n,k),\rho}>G_{(n,k),\phi}$, the sector is relatively elongated in the radial direction, and thereby the associated geometrical aspect ratio of the annular graphene sector is expressed by,
\begin{equation}
\mathrm{AR}_{(n,k),\rho}
=
\frac{G_{(n,k),\rho}}
{G_{(n,k),\phi}}
\label{eq:aspect_ratio_1}
\end{equation}
Conversely, if $G_{(n,k),\rho}<G_{(n,k),\phi}$, the sector is relatively elongated in the azimuthal direction, and thereby the associated geometrical aspect ratio of the annular graphene sector is expressed by,
\begin{equation}
\mathrm{AR}_{(n,k),\phi}
=
\frac{G_{(n,k),\phi}}
{G_{(n,k),\rho}}
\label{eq:aspect_ratio_2}
\end{equation}

Assuming the dominant current flows in the radial direction of $\rho$, the effective propagation length corresponds to the radial dimension of $L_{\rho} = G_{(n,k),\rho}$, while the effective width for current flow is approximated by the azimuthal dimension of $W_{\phi} = G_{(n,k),\phi}$. 
This approximation is based on the dominant current distribution over graphene annular sectors. 
Since the resonant current distribution may exhibit coupled radial and azimuthal components near resonance, this directional assumption can be verified via surface current density distributions or modal analysis in CST Studio Suite.
Hence, the resistance of the $(n,k)$-th graphene sector is given by,
\begin{equation}
R_{(n,k),\rho}
=
R_s
\frac{G_{(n,k),\rho}}
{G_{(n,k),\phi}}
\label{eq:cyl_R_nk}
\end{equation}
By substituting Eq.\ref{eq:ring_gap_rho} and Eq.\ref{eq:ring_gap_phi} into Eq.\ref{eq:cyl_R_nk}, and using the graphene sheet resistance, the corresponding effective radial resistance of the $(n,k)$-th graphene sector is expressed as,
\begin{equation}
R_{(n,k),\rho}
=
\frac{\pi\hbar^2}
{e^3\mu_{c,(n,k)}^{(\mathrm{eV})}\tau}
\frac{
\Delta\rho-g_{(n,k),\rho}
}{
\displaystyle
\frac{2\pi\rho_n}{N_n}
-g_{(n,k),\phi}
}
\label{eq:cyl_R_nk_expanded}
\end{equation}
If the dominant surface current is directed along the azimuthal direction, the corresponding effective azimuthal resistance of the $(n,k)$-th graphene sector is given by,
\begin{equation}
R_{(n,k),\phi}
=
R_s
\frac{G_{(n,k),\phi}}
{G_{(n,k),\rho}}
\label{eq:cyl_R_phi}
\end{equation}
Substituting Eq.\ref{eq:ring_gap_rho} and Eq.\ref{eq:ring_gap_phi} into Eq.\ref{eq:cyl_R_phi} and incorporating the graphene sheet resistance, the effective azimuthal resistance of the $(n,k)$-th graphene sector is obtained as,
\begin{equation}
R_{(n,k),\phi}
=
\frac{\pi\hbar^2}
{e^3\mu_{c,(n,k)}^{(\mathrm{eV})}\tau}
\frac{
\displaystyle
\frac{2\pi\rho_n}{N_n}
-g_{(n,k),\phi}
}{
\Delta\rho-g_{(n,k),\rho}
}
\label{eq:cyl_R_phi_expanded}
\end{equation}
For a scalar RLC representation, the effective impedance is determined according to the surface current with the dominant direction for the corresponding annular graphene sector.
\subsection{Kinetic and Geometrical Inductance}
The inductive response of each electrically isolated graphene sector consists of both kinetic and geometric contributions.
The kinetic inductance originates from the finite inertia of charge carriers in graphene and depends primarily on the chemical potential; conversely, the geometric inductance arises from the magnetic field generated by the current path.
Hence, the total inductance is modeled as the sum of the graphene kinetic
inductance and the geometrical inductance:
\begin{equation}
L_{(n,k)}
=
L_{(n,k)}^{\mathrm{kin}}
+
L_{(n,k)}^{\mathrm{geo}}
\label{eq:cyl_Ltotal}
\end{equation}
By similar arguments to derive the resistance, for the radial current orientation, the kinetic inductance of the $(n,k)$-th graphene sector is expressed as,
\begin{equation}
L_{(n,k),\rho}^{\mathrm{kin}}
=
L_k
\frac{G_{(n,k),\rho}}
{G_{(n,k),\phi}}
\label{eq:cyl_Lkin_nk}
\end{equation}Therefore,
\begin{equation}
L_{(n,k),\rho}^{\mathrm{kin}}
=
\frac{\pi\hbar^2}
{e^3\mu_{c,(n,k)}^{\mathrm{eV}}}
\frac{
\Delta\rho-g_{(n,k),\rho}
}{
\displaystyle
\frac{2\pi\rho_n}{N_n}
-g_{(n,k),\phi}
}
\label{eq:cyl_Lkin_expanded}
\end{equation}
It is worth mentioning that $\mathbf{Z}_s$ in Eqs.~\ref{eq:effective_impedance_tensor}--\ref{eq:azimuthal_RLC_impedance} denotes the graphene sheet-impedance tensor, whose components are expressed in $\Omega/\mathrm{L}$. The equivalent lumped impedances of the individual graphene sectors are subsequently obtained by accounting for their respective aspect ratios. Accordingly, the lumped impedances in Eqs.~\ref{eq:cyl_R_nk}--\ref{eq:cyl_R_phi_expanded} are expressed in $\Omega$, 
whereas $Z_s$ represents the intrinsic sheet impedance in 
$\Omega/\mathrm{L}$.
\subsection{Capacitance of an Electrically Isolated Graphene Sector}
\begin{figure}[htbp]
    \centering
    \includegraphics[
        width=0.95\columnwidth,
        height=0.3\columnwidth
    ]{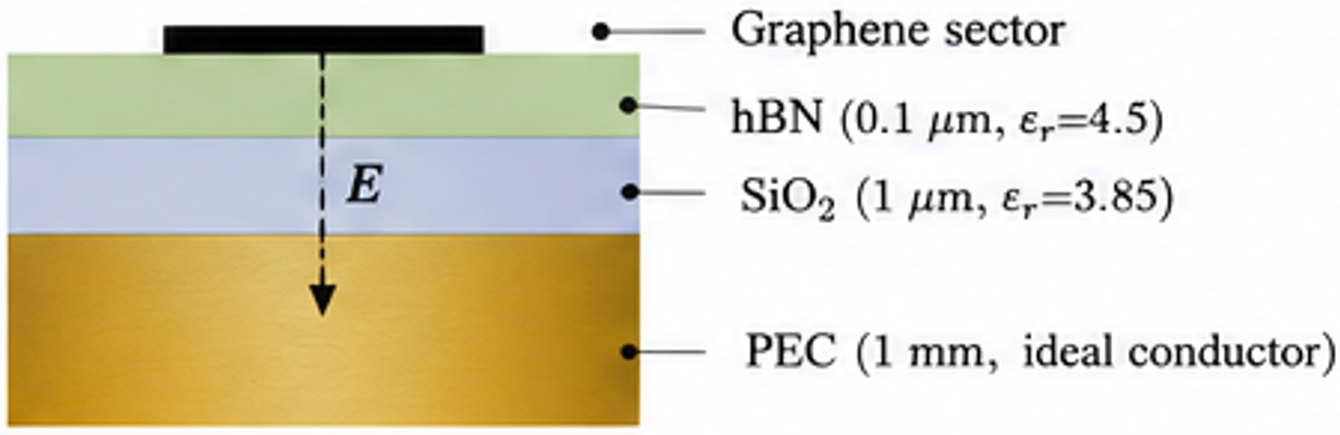}
    \caption{Cross-sectional view of the planar reflector. The flat reflector consists of graphene sectors placed on an hBN layer, followed by a SiO$_2$ dielectric layer and an ideal PEC ground plane, resulting in the layer sequence of $\mathrm{Graphene}\rightarrow\mathrm{hBN}\rightarrow\mathrm{SiO_2}\rightarrow\mathrm{PEC}$, Fig.~\ref{fig:flat_reflector_configuration}.}
    \label{fig:flat_reflector_cross_section_1}
\end{figure}

Since adjacent graphene sectors in Fig.~\ref{fig:flat_reflector_configuration} are physically isolated, no direct conductive current flows across the intervening gaps, the electromagnetic coupling between neighboring sectors occurs through the electric field and the associated displacement current.
Thus, the effective capacitance of the $(n,k)$-th electrically isolated graphene sector in Fig.~\ref{fig:flat_reflector_cross_section_1} is determined by combining several capacitive contributions as follows,
\begin{equation}
\begin{aligned}
C_{(n,k)}
={}&
\underbrace{
\frac{
C_{(n,k)}^{\mathrm{geo}} C_{(n,k)}^{Q}
}{
C_{(n,k)}^{\mathrm{geo}} + C_{(n,k)}^{Q}
}
}_{\text{graphene-to-PEC capacitance including quantum capacitance}}
\\
&+
\underbrace{
C_{(n,k)}^{\mathrm{gap}}
}_{\text{inter-sector capacitive coupling}}
+
\underbrace{
C_{(n,k)}^{\mathrm{edge}}
}_{\text{fringing-field contribution}}
\\
\end{aligned}
\label{eq:cyl_C_total}
\end{equation}
where the first term corresponds to the dominant vertical capacitive coupling between the graphene and PEC ground plane, while the remaining terms capture lateral electromagnetic coupling and finite-edge effects associated with the isolated annular graphene geometry as shown in Figs.~\ref{fig:flat_reflector_configuration},~\ref{fig:flat_reflector_sectors}, and ~\ref{fig:flat_reflector_cross_section_1}.
The component of $C_{(n,k)}^{\mathrm{ox}}$ represents the oxide-stack capacitance between the $(n,k)$-th graphene sector and the PEC plane. In the cylindrical geometry, it is the capacitance associated with the dielectric layers separating the graphene from the PEC plane.

Electrically, the SiO$_2$ and hBN layers form a series dielectric stack between the graphene layer and the PEC ground plane, and their combined capacitive contribution is therefore determined by their respective electrical thicknesses. Consequently, their combined capacitive response is governed by the sum of their effective electrical thicknesses of $t/\epsilon_r$ rather than their physical dimensions alone as follows,
\begin{equation}
t_{\mathrm{eq}}
=
\frac{t_{\mathrm{SiO_2}}}
{\epsilon_{r,\mathrm{SiO_2}}}
+
\frac{t_{\mathrm{hBN}}}
{\epsilon_{r,\mathrm{hBN}}}
\label{eq:cyl_teq}
\end{equation}
For the proposed dielectric stack, as illustrated in Figs.~\ref{fig:flat_reflector_configuration}, \ref{fig:flat_reflector_sectors}, and \ref{fig:flat_reflector_cross_section_1},
\begin{equation}
t_{\mathrm{eq}}
=
\frac{1~\mu\mathrm{m}}{3.85}
+
\frac{0.1~\mu\mathrm{m}}{4.5}
\approx
0.282~\mu\mathrm{m}
\label{eq:cyl_teq_value}
\end{equation}

The quantum capacitance captures the finite electronic density of states of graphene and its contribution to the overall capacitance of the graphene element, as given by,
\begin{equation}
C_{(n,k)}^{Q}
=
A_{(n,k)}
\frac{2e^3|\mu^{(eV)}_{c,(n,k)}|}
{\pi\hbar^2 v_F^2}
\label{eq:quantum_capacitance}
\end{equation}
where $e$, $\hbar$, and $v_F$ refer to the elementary charge, the reduced Planck constant, and the Fermi velocity, respectively.  
The area of $A_{(n,k)}$ determines the total quantum capacitance of each graphene annular sectors. 
The chemical potential of $\mu_c$, expressed as an electronic energy, represents the tunable Fermi level of graphene and is controlled by the bias voltage applied to each graphene sector.

The Fermi velocity of $v_F$, typically $10^6$~m/s, characterizes the carrier dynamics in graphene. The factor of $2$ accounts for the spin and valley degeneracy of graphene.

The geometrical capacitance of $C_{(n,k)}^{\mathrm{geo}}$ accounts for the electrostatic coupling between the graphene element and the PEC through the SiO$_2$-hBN dielectric stack and is expressed as follows,
\begin{equation}
C_{(n,k)}^{\mathrm{geo}}
=
\frac{\epsilon_0 A_{(n,k)}}
{t_{\mathrm{eq}}}
=
\frac{
\epsilon_0
G_{(n,k),\rho}
G_{(n,k),\phi}
}{
\displaystyle
\frac{t_{\mathrm{SiO_2}}}{\epsilon_{r,\mathrm{SiO_2}}}
+
\frac{t_{\mathrm{hBN}}}{\epsilon_{r,\mathrm{hBN}}}
}
\label{eq:cyl_C_geo}
\end{equation}
where $G_{(n,k),\rho}$ and $G_{(n,k),\phi}$ denote the effective geometrical dimensions of the $(n,k)$-th element in the radial and azimuthal directions, respectively. 
Their product defines the effective area of the graphene annular sectors as
$A_{(n,k)}^{\mathrm{eff}}
=
G_{(n,k),\rho}G_{(n,k),\phi}$.
Because the oxide (or geometry) capacitance of $C_{(n,k)}^{\mathrm{geo}}$ is only associated with the dielectric (oxide) layers between the graphene sectors and the PEC ground plane, it characterizes the electrostatic coupling through the hBN--SiO$_2$ dielectric stack. This capacitance depends on the effective area of $A_{(n,k)}$ for the graphene annular sectors and the equivalent dielectric thickness of $t_{\mathrm{eq}}$ for the multilayer stack, as expressed by, 
An increase in the graphene area $A_{(n,k)}$ results in a higher $C_{(n,k)}^{\mathrm{geo}}$, whereas an increase in the equivalent dielectric thickness $t_{\mathrm{eq}}$ reduces the capacitance. 
The capacitance $C_{(n,k)}^{\mathrm{geo}}$ represents the contribution of the hBN--SiO$_2$ dielectric stack to the overall capacitance of graphene sectors and is modeled separately from the graphene-to-PEC capacitance, which incorporates both the geometrical and quantum capacitance contributions.

Furthermore, because the adjacent graphene sectors are separated by a finite radial gap of $g_{(n,k),\rho} > 0$, despite their
electrical and physical isolation, fringing electric fields extend across the gap, and then induce capacitive coupling between neighboring sectors, Figs.~\ref{fig:flat_reflector_configuration}, \ref{fig:flat_reflector_sectors}, and \ref{fig:flat_reflector_cross_section_1}.
Hence, for the first-order approximation, the fringing capacitance associated with the radial gap can be obtained by,
\begin{equation}
C_{(n,k),\rho}^{\mathrm{gap}}
\approx
\epsilon_0\epsilon_{\mathrm{eff}}
\frac{
A_{\rho}^{\mathrm{edge}}
}{
g_{(n,k),\rho}
}
F_\rho
\label{eq:cyl_Cgap_rho}
\end{equation}
where $A_{\rho}^{\mathrm{edge}}$ is the effective fringing-field area
and $F_\rho$ is a geometry-dependent correction factor.
Similarly, the finite azimuthal separation of $g_{(n,k),\phi}>0$ induces fringing-field coupling between adjacent sectors along the $\phi$-direction. The corresponding azimuthal gap capacitance is approximated as,
\begin{equation}
C_{(n,k),\phi}^{\mathrm{gap}}
\approx
\epsilon_0\epsilon_{\mathrm{eff}}
\frac{
A_{\phi}^{\mathrm{edge}}
}{
g_{(n,k),\phi}
}
F_\phi
\label{eq:cyl_Cgap_phi}
\end{equation}

Furthermore, near the finite boundaries of each graphene sector, the electric field is not completely confined between the graphene layer and the PEC ground plane. A portion of the field extends beyond the sector edges, resulting in fringing fields and an additional edge-capacitance contribution. This contribution is represented by $C_{(n,k)}^{\mathrm{edge}}$ in Eq.\ref{eq:cyl_C_total}.
\subsection{Equivalent RLC Model}
\begin{figure}[htbp]
    \centering
    \includegraphics[
        width=1\columnwidth,
        height=0.35\columnwidth
    ]{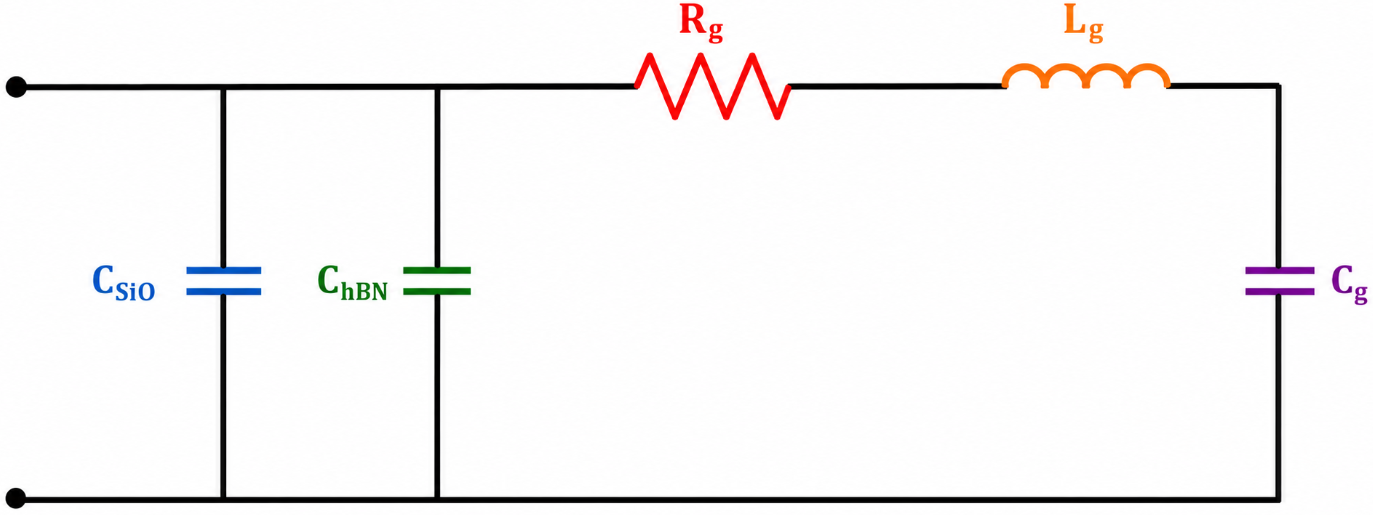}
    \caption{Equivalent RLC circuit modeling the unit cell of our proposed planar reflector, consistent with Fig.~\ref{fig:flat_reflector_configuration}.}
    \label{fig:flat_reflector_rlc_model}
\end{figure}
The equivalent impedance of the $(n,k)$-th graphene sector is represented by the RLC model, as illustrated in Fig.~\ref{fig:flat_reflector_rlc_model}, and is expressed as follows, 
\begin{equation}
\begin{aligned}
Z_{(n,k)}(\omega,V_{g,(n,k)})
={}&
R_{(n,k)}(V_{g,(n,k)})
\\
&+
j\omega L_{(n,k)}(V_{g,(n,k)})
+
\frac{1}{j\omega C_{(n,k)}}
\end{aligned}
\label{eq:cyl_Z_RLC}
\end{equation}
where the total inductance is decomposed into kinetic and geometrical contributions as follows,
\begin{equation}
L_{(n,k)}
=
L_{(n,k)}^{\mathrm{kin}}
+
L_{(n,k)}^{\mathrm{geo}}
\label{eq:cyl_L_total}
\end{equation}
and the total capacitance is expressed as the sum of the graphene-to-PEC, inter-sector gap, and edge-fringing capacitances, as defined in Eq.\ref{eq:cyl_C_total}. 
Accordingly, the resonance frequency of the $(n,k)$-th RLC unit cell is determined by its total inductance and capacitance as follows,
\begin{equation}
f_{0,(n,k)}
=
\frac{1}
{2\pi\sqrt{L_{(n,k)}C_{(n,k)}}}
\label{eq:cyl_resonance_frequency}
\end{equation}
Since the graphene kinetic inductance approximately scales as,
\begin{equation}
L_{(n,k)}^{\mathrm{kin}}
\propto
\frac{1}{\mu_{c,(n,k)}}
\label{eq:cyl_L_mu_relation}
\end{equation}
an increase in the chemical potential reduces the kinetic inductance, and thereby shifts the resonance frequency upward, which is consequently accompanied by providing direct electrical tunability of the RLC unit-cell response,
\begin{equation}
\mu_c\uparrow
\quad\Longrightarrow\quad
L_k\downarrow
\quad\Longrightarrow\quad
f_0\uparrow
\label{eq:cyl_tuning_relation}
\end{equation}
This behavior enables independent electrical tuning of the electromagnetic response of each electrically isolated graphene sector.

It is worth noting that the graphene sectors are electrically isolated from one another, allowing the chemical potential of each sector to be independently controlled through its corresponding gate voltage,
\begin{equation}
\mu_{c,(n,k)}
=
\mu_c\!\left(V_{g,(n,k)}\right)
\label{eq:cyl_mu_voltage}
\end{equation}
The corresponding chemical potential of the $(n,k)$-th graphene sector can be approximated as, 
\begin{equation}
\mu_{c,(n,k)}
=
\hbar v_F
\sqrt{
\frac{\pi C_{\mathrm{geo}}'}
{e}
\left|V_{g,(n,k)}-V_D\right|
}
\label{eq:chemical_potential_gate_voltage}
\end{equation}
where $C_{\mathrm{geo}}'$, $V_{g,(n,k)}$, and $V_D$ denote the effective gate-oxide capacitance per unit area, the gate voltage applied to the $(n,k)$-th graphene sector, and the Dirac-point voltage, respectively.
From Eqs.\ref{eq:cyl_C_total} and \ref{eq:cyl_teq}, the effective gate-oxide capacitance per unit area is given by,
\begin{equation}
\frac{C^{\mathrm{geo}}}{A}
=
\frac{\epsilon_0}
{
\displaystyle
\frac{t_{\mathrm{SiO_2}}}
{\epsilon_{r,\mathrm{SiO_2}}}
+
\frac{t_{\mathrm{hBN}}}
{\epsilon_{r,\mathrm{hBN}}}
}
\label{eq:cyl_Cox_area_equiv}
\end{equation}
Substituting the corresponding material parameters into Eq.\ref{eq:cyl_Cox_area_equiv} gives,
\begin{equation}
\frac{C^{\mathrm{geo}}}{A}
=
\frac{\epsilon_0}
{
\displaystyle
\frac{1\times10^{-6}}{3.85}
+
\frac{0.1\times10^{-6}}{4.5}
}
\label{eq:cyl_Cox_numeric}
\end{equation}
Accordingly, the effective oxide capacitance per unit area is approximately,
\begin{equation}
C_{\mathrm{geo}}'
=
\frac{C^{\mathrm{geo}}}{A}
\approx
3.14\times10^{-5}~\mathrm{F/m^2}
\label{eq:cyl_Cox_prime}
\end{equation}
For a more accurate electrostatic model, particularly at low chemical potentials, the additional capacitance contributions in Eq.\ref{eq:cyl_C_total} should also be taken into account.
\section{Array Factor of the Planar Reflector}
The reflected contribution of each graphene sector is governed by both its local reflection coefficient and its physical area. 
A larger sector intercepts a greater portion of the incident electromagnetic wavefront, consequently, contributes more significantly to the total reflected field. 
Thus, the physical sector area, horn illumination, and local graphene reflection response collectively determine the complex contribution of the $(n,k)$-th annular sector prior to incorporating the far-field propagation phase into the reflector array factor illustrated in Fig.~\ref{fig:flat_reflector_configuration}, as given by the following expression,
\begin{equation}
\begin{split}
W_{(n,k)}
&= \\
&\underbrace{\Delta S_{(n,k)}}_{\text{sector area}}
\underbrace{F^{\mathrm{inc}}_{(n,k)}}_{\text{horn illumination}}
\underbrace{\Gamma_{(n,k)}}_{\text{graphene reflection response}}
\end{split}
\label{eq:sector_weight}
\end{equation}
where $\Delta S_{(n,k)}$ denotes the physical area of the $(n,k)$-th graphene sector, $F^{\mathrm{inc}}_{(n,k)}$ characterizes the corresponding horn-illumination factor, and $\Gamma_{(n,k)}$ corresponds to the local graphene reflection coefficient.

Beam-steering synthesis is achieved by enforcing phase uniformity across the scattered wavefront.
The chemical potential $\mu_{c,(n,k)}$ of each sector is tuned such that the phase of its reflection coefficient $S_{11}$ matches the required local reflection phase, thereby maintaining phase uniformity across the aperture. 
Hence, the primary objective is to satisfy the constant-phase condition across the aperture,
\begin{equation}\Phi^{\mathrm{inc}}_{(n,k)}+\Phi^{\mathrm{g}}_{(n,k)}+\Phi^{\mathrm{prop}}_{(n,k)}=\mathrm{C}
\label{eq:constant_phase_condition}
\end{equation}
where $\Phi^{\mathrm{inc}}_{(n,k)}$ represents the phase accumulated by the incident field from the feedhorn to the $(n,k)$-th graphene sector, $\Phi^{\mathrm{g}}_{(n,k)}$ corresponds to the reflection phase introduced by the graphene sector, and $\Phi^{\mathrm{prop}}_{(n,k)}$ accounts for the propagation phase from the annular sector toward the desired far-field direction, and $C$ represents an arbitrary constant phase.
Enforcing this condition across the aperture aligns the phases of the fields reflected by the individual sectors toward the desired far-field direction, enabling their constructive interference and thereby establishing the specified beam-steering direction.
\begin{figure}[htbp]
    \centering
    \includegraphics[
        width=1\columnwidth,
        height=0.15\columnwidth
    ]{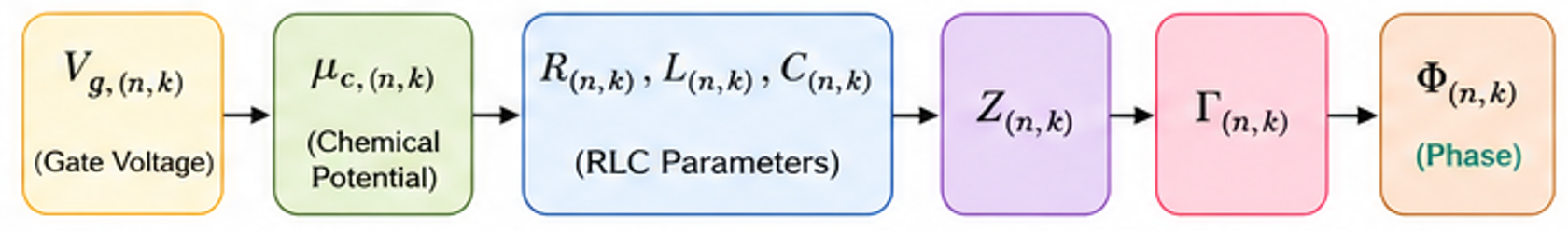}
    \caption{Illustration of the voltage-controlled tuning mechanism for the $(n,k)$-th graphene sector, showing the sequential relationship from the applied gate voltage $V_{g,(n,k)}$ to the chemical potential $\mu_{c,(n,k)}$, equivalent RLC parameters $R_{(n,k)}$, $L_{(n,k)}$, and $C_{(n,k)}$, impedance $Z_{(n,k)}$, reflection/transmission coefficient $\Gamma_{(n,k)}$, and the resulting phase response $\Phi_{(n,k)}$ for the planar reflector consistent with Fig.~\ref{fig:flat_reflector_configuration}.}
    \label{fig:V_to_Ph}
\end{figure}
Consequently, as illustrated in Fig.~\ref{fig:V_to_Ph}, this section presents the phase-synthesis procedure to determine the required $S_{11,(n,k)}$ phase by tuning the chemical potential of each annular graphene sector. 
This approach enables dynamic beam steering toward the desired (or required) direction while maintaining phase uniformity across the aperture.
\subsection{Annular-Sector Geometry}
Since the area of an annular sector is not uniformly distributed along the radial component of the polar coordinate system, its geometric centroid is shifted toward the outer radius and generally differs from the arithmetic radial midpoint. Accordingly, the area centroid provides a more physically representative equivalent location for the entire graphene sector, whereas the arithmetic mean of the inner and outer radii serves only as a first-order approximation.
In the polar coordinate system, the radial component of the area centroid of the $(n,k)$-th annular sector is defined as the area-weighted average of $\rho$ over the sector surface,
\begin{equation}
\rho_{(n,k)}^{(c)}
=
\frac{\displaystyle\int_{S_{(n,k)}} \rho\, dS}
{\displaystyle\int_{S_{(n,k)}} dS}
\label{eq:radial_centroid_definition}
\end{equation}
where $S_{(n,k)}$ denotes the area of the $(n,k)$-th annular sector.
Accordingly, the radial component of the area centroid of the $(n,k)$-th annular sector is obtained as,
\begin{equation}
\rho_{(n,k)}^{(c)}
=
\frac{2}{3}
\frac{\rho_{n,\mathrm{out}}^{3}-\rho_{n,\mathrm{in}}^{3}}
{\rho_{n,\mathrm{out}}^{2}-\rho_{n,\mathrm{in}}^{2}}
\end{equation}
This expression defines the radial component of the area-weighted mean of the $(n,k)$-th annular sector, which is particularly useful when representing the sector in a radial phase model. For sufficiently narrow sectors, the arithmetic midpoint closely approximates this geometric centroid. Therefore, the physical area of the $(n,k)$-th annular graphene sector is expressed as, 
\begin{equation}
\Delta S_{(n,k)}
=
\frac{1}{2}
\left(
\rho_{n,\mathrm{out}}^{2}
-
\rho_{n,\mathrm{in}}^{2}
\right)
\Delta\phi_{(n,k)}
\end{equation}
where $\Delta\phi_{(n,k)}$ denotes the angular width of the sector in radians.

As a result, since each graphene sector occupies a finite annular area rather than a discrete spatial point, a single representative radial component should be established to evaluate the incident and reflected phases. 
Although the arithmetic radial midpoint provides a computationally simple approximation, the radial component of the weighted area explicitly accounts  for the nonuniform distribution of the physical area along the radial direction in the polar coordinate system. 
Consequently, the geometric centroid provides a more accurate equivalent location for annular sectors with substantial radial dimensions. 
The physical area of each annular sector is intrinsically incorporated into the amplitude coefficients of the array factor to account for the variation in the effective aperture contribution of the graphene elements.
\subsection{Phase Synthesis for Dynamic Beam Steering}
The required reflection-phase distribution is determined by enforcing phase coherence among the fields reflected from all annular graphene sectors toward the desired far-field direction. 
Consequently, the total phase contribution associated with the $(n,k)$-th sector should satisfy the constant-phase condition given by in Eq.\ref{eq:constant_phase_condition}.

The spatial position of the $(n,k)$-th annular sector is defined in the cylindrical coordinate system on the $z = 0$ plane, with its position vector given by,
\begin{equation}
\mathbf{r}_{(n,k)} = \rho_{(n,k)}\cos\varphi_{(n,k)}\hat{\mathbf{x}} + \rho_{(n,k)}\sin\varphi_{(n,k)}\hat{\mathbf{y}}
\label{eq:sector_position_vector}
\end{equation}
The physical distance of $\mathbf{d}^{\mathrm{inc}}_{(n,k)}$ between the feedhorn phase center located at $\mathbf{r}_{\mathrm{F}} = (\rho_{\mathrm{F}}, \varphi_{\mathrm{F}}, z_{\mathrm{F}})$ and the center of the $(n,k)$-th graphene annular sector at $\mathbf{r}_{(n,k)} = (\rho_{(n,k)}, \varphi_{(n,k)}, 0)$ is expressed in vector form as,
\begin{equation} 
\mathbf{d}^{\mathrm{inc}}_{(n,k)} = \left\Vert{} \mathbf{r}_{(n,k)} - \mathbf{r}_{\mathrm{F}} \right\Vert{} 
\label{eq:vector_distance} 
\end{equation}
By expanding Eq.\ref{eq:vector_distance} in cylindrical coordinate system, the explicit distance simplifies to,
\begin{equation} 
\mathbf{d}^{\mathrm{inc}}_{(n,k)} = \sqrt{ \rho_{(n,k)}^{2} + \rho_{\mathrm{F}}^{2} - 2\rho_{(n,k)}\rho_{\mathrm{F}} \cos\left( \varphi_{(n,k)}-\varphi_{\mathrm{F}} \right) + z_{\mathrm{F}}^{2} } 
\label{eq:explicit_distance} 
\end{equation}
It is worth mentioning that the proposed hemispherical transmitarray antenna in Fig.~\ref{fig:hemispherical_transmitarray_2} intrinsically provides nearly uniform illumination across the aperture lens due to the alignment between the feedhorn-incident wave vector and the position vectors of the graphene sectors\cite{komeylian2026graphenebasedhemisphericaltransmitarrayantenna}. In contrast to the hemispherical transmitarray antenna, the incident-field factor of $F^{\mathrm{inc}}_{(n,k)}$ of the planar reflector in Fig.~\ref{fig:flat_reflector_configuration} varies spatially and cannot be assumed uniform.

The incident-field amplitude at each $(n,k)$-th graphene sector depends on the angular radiation pattern of the feedhorn, spherical-wave distance spreading, local incidence angle, and the polarization mismatch between the incident field and the local sector response, and the local incidence angle. 
\begin{equation}
\begin{aligned}
F^{\mathrm{inc}}_{(n,k)}
\left(
\rho^{\mathrm{inc}}_{(n,k)},
\varphi^{\mathrm{inc}}_{(n,k)}
\right)
&=
A_{\mathrm{ref}}
\frac{\left|\mathbf{r}_{\mathrm{F}}\right|}
     {\left|\mathbf{d}^{\mathrm{inc}}_{(n,k)}\right|}     
F_{\mathrm{horn}}
\left(
\theta^{\mathrm{inc}}_{(n,k)},
\phi^{\mathrm{inc}}_{(n,k)}
\right)
\\
&\quad\times
C_{\mathrm{pol},(n,k)}
C_{\mathrm{obl},(n,k)}
e^{-j k_0 d^{\mathrm{inc}}_{(n,k)}}
\label{eq:incident_field}
\end{aligned}
\end{equation}
where,
\begin{equation}
\begin{aligned}
C_{\mathrm{pol},(n,k)}
&=
\left|
\hat{\mathbf{p}}_{\mathrm{horn}}
\cdot
\hat{\mathbf{p}}_{\mathrm{loc},(n,k)}
\right|,
\quad
C_{\mathrm{obl},(n,k)}
=
\left|
\hat{\mathbf{k}}^{\mathrm{inc}}_{(n,k)}
\cdot
\hat{\mathbf{n}}_{(n,k)}
\right|
\end{aligned}
\end{equation}
The normalized field pattern of the feedhorn of $F_{\mathrm{horn}}(\theta^{\mathrm{inc}}{(n,k)},\phi^{\mathrm{inc}}{(n,k)})$ characterizes the angular variation of the incident illumination, whereas $\frac{\left|\mathbf{r}_{\mathrm{F}}\right|}
     {\left|\mathbf{d}^{\mathrm{inc}}_{(n,k)}\right|}$ 
describes the variation of the spherical-wave amplitude due to free-space propagation.
Furthermore, $C_{\mathrm{pol},(n,k)}$ and $C_{\mathrm{obl},(n,k)}$ account for the polarization mismatch and local obliquity of the incident wave, respectively.
This formulation thus enables an accurate sector-dependent representation of the incident field for the quantitative evaluation of the proposed reflector. 

Assuming an $e^{j\omega t}$ time-harmonic convention, the incident-wave phase at the $(n,k)$-th sector is given by,
\begin{equation}
\Phi^{\mathrm{inc}}_{(n,k)} = - k_0 d^{\mathrm{inc}}_{(n,k)}
\label{eq:incident_phase}
\end{equation}
where $k_0 = \frac{2\pi}{\lambda_0} = \omega\sqrt{\mu_0\epsilon_0}$ refers to the free-space wavenumber. 
Let $\hat{\mathbf{s}}_0$ represent the unit vector pointing toward the desired far-field direction $(\Theta_0,\Phi_0)$, expressed as,
\begin{equation}
\hat{\mathbf{s}}_0 = \sin\Theta_0\cos\Phi_0\hat{\mathbf{x}} + \sin\Theta_0\sin\Phi_0\hat{\mathbf{y}} + \cos\Theta_0\hat{\mathbf{z}}
\label{eq:desired_direction}
\end{equation}
The propagation phase associated with the path from the $(n,k)$-th sector toward the desired far-field direction is given by, 
\begin{equation}
\Phi_{\mathrm{prop},(n,k)} = k_0\hat{\mathbf{s}}_0 \cdot \mathbf{r}_{(n,k)}
\label{eq:propagation_phase}
\end{equation}
Substituting Eq.\ref{eq:incident_phase} and Eq.\ref{eq:propagation_phase} into Eq.\ref{eq:constant_phase_condition} yields the required local reflection phase of the $(n,k)$-th graphene annular sector as,
\begin{equation}
\psi^{\mathrm{req}}_{(n,k)} = \Phi_{(n,k)}^{\mathrm{g},\mathrm{req}} = C + k_0 \mathbf{d}^{\mathrm{inc}}_{(n,k)} - k_0\hat{\mathbf{s}}_0 \cdot \mathbf{r}_{(n,k)}
\label{eq:required_graphene_phase}
\end{equation}
In the cylindrical coordinate system, the projection of the position vector of the graphene sector onto the desired propagation direction is given by,
\begin{equation}
\hat{\mathbf{s}}_0 \cdot \mathbf{r}_{(n,k)} = \rho_{(n,k)}\sin\Theta_0\cos(\Phi_0 - \varphi_{(n,k)})
\label{eq:desired_direction_projection}
\end{equation}
Therefore, the required reflection phase reduces to,
\begin{equation}
\psi^{\mathrm{req}}_{(n,k)} = C + k_0 \mathbf{d}^{\mathrm{inc}}_{(n,k)}- k_0 \rho_{(n,k)}\sin\Theta_0\cos(\Phi_0 - \varphi_{(n,k)})
\label{eq:required_reflection_phase}
\end{equation}
For the $(n,k)$-th graphene sector, the complex reflection coefficient is characterized by, 
\begin{equation}
\Gamma_{(n,k)} = S_{11,(n,k)} = |\Gamma_{(n,k)}| e^{j\psi_{(n,k)}}
\label{eq:graphene_reflection_coefficient}
\end{equation}
where 
\begin{equation}
\psi_{(n,k)} = \angle S_{11,(n,k)}
\label{eq:graphene_reflection_phase}
\end{equation}
Accordingly, the chemical potential $\mu_{c,(n,k)}$ of each independently controlled graphene sector is tuned such that the phase of its reflection coefficient matches the required local reflection phase, thereby satisfying the phase-coherence condition across the reflector aperture,
\begin{equation}
\begin{aligned}
\angle S_{11,(n,k)}
\left(
\mu_{c,(n,k)},f;\text{geometrical parameters}
\right)
&\approx
\psi^{\mathrm{req}}_{(n,k)}
\\[-1mm]
&\pmod{2\pi}
\end{aligned}
\label{eq:s11_phase_matching}
\end{equation}
The following discussion establishes the constraints for determining the allowable graphene chemical-potential range required for dynamic beam steering.
\begin{equation}
\mathcal{F}_{(n,k)}
=
\left\{
\begin{aligned}
\mu_c: \;& \mu_{c,\min} \leq \mu_c \leq \mu_{c,\max},\\
&\left|S_{11,(n,k)}(\mu_c)\right|
\geq S_{11,\min}
\end{aligned}
\right\}
\label{eq:feasible_chemical_potential_set}
\end{equation}
The set of $\mathcal{F}_{(n,k)}$ in Eq.\ref{eq:feasible_chemical_potential_set} defines the physically realizable chemical potentials that yield a sufficient reflection magnitude under the two given constraints. 
The first constraint restricts the chemical potential to its physically realizable range, thereby ensuring compatibility with the practical graphene biasing mechanism.
The second constraint enforces a minimum allowable reflection magnitude for each graphene sector, ensuring that the reflected field maintains sufficient intensity for robust beam steering.
\begin{equation}
\mu_{c,(n,k)}^{\star}
=
\operatorname*{arg\,min}_{\mu_c \in \mathcal{F}_{(n,k)}}
\left|
\operatorname{wrap}
\left[
\angle S_{11,(n,k)}(\mu_c)
-
\psi_{(n,k)}^{\mathrm{req}}
\right]
\right|
\label{eq:chemical_potential_optimization}
\end{equation}
It is worth mentioning that the optimization process does not merely select the chemical potential yielding the closest phase. Instead, it enforces a hierarchical selection criterion of Eq.\ref{eq:optimization_flow}: it first verifies physical realizability and sufficient reflection magnitude, and subsequently minimizes the phase error. This constrained approach makes the formulation uniquely suited for practical, bias-controlled planar reflector.
\begin{equation}
\text{Feasible states in }  \mathcal{F}_{(n,k)}
\;\longrightarrow\;
\min_{\mu_c \in \mathcal{F}_{(n,k)}} \epsilon_{\phi,(n,k)}
\;\longrightarrow\;
\mu_{c,(n,k)}^{\star}
\label{eq:optimization_flow}
\end{equation}
The chemical potential is selected as the physically realizable value that minimizes the phase error while satisfying the required reflection-magnitude constraint. 
To account for the physical limitations of the graphene-based phase modulation, the feasible chemical-potential domain is first mapped to the corresponding set of physically attainable phase states. Specifically, for each $(n,k)$-th graphene sector, every feasible value of the chemical potential $\mu_c \in \mathcal{F}_{(n,k)}$ yields a corresponding phase response of the reflection coefficient $S_{11,(n,k)}(\mu_c)$. The corresponding set of phase values defines the attainable-phase set $\Psi_{(n,k)}^{\mathrm{att}}$.
\begin{equation}
\psi_{(n,k)}^{\mathrm{att}}
=
\operatorname*{arg\,min}_{\psi \in \Psi_{(n,k)}^{\mathrm{att}}}
\left|
\operatorname{wrap}
\left(
\psi
-
\psi_{(n,k)}^{\mathrm{req}}
\right)
\right|
\label{eq:attainable_phase}
\end{equation}
This mapping establishes the relationship between the allowable graphene chemical potential and the phase states that can be physically realized by the $(n,k)$-th annular sector.
Moreover, the wrapping operation accounts for the $2\pi$-periodicity of phase and therefore evaluates the minimum angular difference between the required and attainable phases. This formulation ensures that the phase synthesis remains consistent with the physical operating range of the graphene elements rather than assuming arbitrary phase realizability.

\begin{equation}
\Psi_{(n,k)}^{\mathrm{att}}
=
\left\{
\angle S_{11,(n,k)}(\mu_c)
\,:\,
\mu_c \in \mathcal{F}_{(n,k)}
\right\}
\label{eq:attainable_phase_set}
\end{equation}
Among the physically attainable phase states, the phase that most closely matches the required phase is selected. 
Specifically, $\psi_{(n,k)}^{\mathrm{req}}$ in Eq.\ref{eq:s11_phase_matching} denotes the phase required by the beam steering formulation, whereas $\psi_{(n,k)}^{\mathrm{att}}$ represents the closest physically realizable phase within the attainable-phase set $\Psi_{(n,k)}^{\mathrm{att}}$. 
This selection minimizes the phase error while ensuring that the selected phase remains consistent with the physical operating constraints of the graphene annular sector.

Consequently, the target beam direction determines the necessary spatial reflection-phase distribution across the reflector aperture, while the finite-distance feedhorn introduces a spatial variation of the incident phase delay characterized by $\mathbf{d}^{\mathrm{inc}}_{(n,k)}$. Therefore, the chemical potential of each graphene sector is then independently adjusted to produce the corresponding $S_{11,(n,k)}$ phase. Satisfying local phase-matching condition across the aperture aligns the reflected fields toward the desired far-field direction, thereby enabling electronically reconfigurable beam steering.
\subsection{Array factor of the Planar Reflector}
The array factor of the planar reflector in Fig.~\ref{fig:flat_reflector_configuration} is formulated by accounting for the contribution of the graphene apex sector and the surrounding graphene annular sectors as follow, 
\begin{equation}
\begin{aligned}
AF^{\mathrm{ref}}{(\Theta,\Phi)}
={}&
\underbrace{
W_0
e^{j k_0
\hat{\mathbf{s}}(\Theta,\Phi)\cdot\mathbf{r}_0}
}_{\text{graphene apex sector}}
\\
&+
\underbrace{
\sum_{n=1}^{N_r}
\sum_{k=1}^{N_n}
W_{(n,k)}
e^{j k_0
\hat{\mathbf{s}}(\Theta,\Phi)\cdot\mathbf{r}_{(n,k)}}
}_{\text{$(n,k)$-th graphene annular sector}} 
\end{aligned}
\label{eq:reflector_array_factor_general}
\end{equation}
where $k_0=2\pi/\lambda_0$, $\mathbf{r}_0$, $\mathbf{r}_{(n,k)}$ denote the free-space wavenumber, the position vector of the graphene apex sector, and the position vector of the $(n,k)$-th graphene annular sector, respectively. 
Accordingly, the coefficients of $W_0$ and $W_{(n,k)}$ refer to the corresponding excitation amplitudes, respectively. 

In what follows, several performance metrics are introduced to quantitatively assess the beam-steering performance of the planar reflector. These metrics include the elevation and azimuthal pointing errors, relative gain, reflection efficiency, and sidelobe level. The beam-pointing accuracy is quantified separately in the elevation and azimuthal directions. The elevation pointing error is defined as,
\begin{equation}
\delta_{\Theta}
=
\left|
\Theta_{\mathrm{pk}}
-
\Theta_{0}
\right|
\label{eq:elevation_pointing_error}
\end{equation}
where $\Theta_{\mathrm{pk}}$ represents the elevation angle corresponding to the peak of the reflected beam. Similarly, the azimuthal pointing error is defined as,
\begin{equation}
\delta_{\Phi}
=
\min_{m\in\mathbb{Z}}
\left|
\Phi_{\mathrm{pk}}
-
\Phi_{0}
+
2\pi m
\right|
\label{eq:azimuth_pointing_error}
\end{equation}
where $\Phi_{\mathrm{pk}}$ denotes the actual peak azimuth angle corresponding to the peak of the reflected beam. The modulo-$2\pi$ formulation accounts for the periodic nature of the azimuthal coordinate.
The normalized array-factor gain to quantify the relative peak response is defined as,
\begin{equation}
G^{\mathrm{rel}}
=
20\log_{10}
\left[
\frac{
\left|
AF^{\mathrm{ref}}
(\Theta_{\mathrm{pk}},\Phi_{\mathrm{pk}})
\right|
}{
\left|
AF^{\mathrm{ref}}
\right|_{\max}
}
\right]
\label{eq:relative_gain}
\end{equation}
The relative array factor gain of $G_{\mathrm{rel}}$ quantifies the normalized peak response of the planar reflector at the actual beam-steering direction. 
This normalization provides a convenient metric for evaluating the relative beamforming response independently of the absolute excitation amplitude.

Since the planar reflector shown in Fig.~\ref{fig:flat_reflector_configuration} is a passive aperture illuminated by a port-fed antenna, its efficiency is defined with respect to the power incident on the reflecting surface, rather than the accepted input power at the feed port used in the conventional definition of antenna efficiency.
The reflector efficiency therefore quantifies the fraction of the incident power that is re-directed into the reflected field. 
Specifically, it is evaluated as the ratio of reflected power to incident power. The reflected power is obtained by integrating the reflected far-field power density over the observation region, while the incident power is calculated from the corresponding incident field.
This definition separates the electromagnetic losses of the planar reflector from those in the feed and illumination path. 
These external losses, such as input-port mismatch, feed-network loss, and transmitting antenna dissipation, are accounted for separately when evaluating overall system efficiency. 
Accordingly, the reflector efficiency is defined as,
\begin{equation}
\eta_{\mathrm{beam}}
=
\frac{P_{\mathrm{ref,desired}}}{P_{\mathrm{inc}}}
=
\frac{P_{\mathrm{beam}}}{P_{\mathrm{inc}}}
\end{equation}
Indeed $\eta_{\mathrm{beam}}$ measures how much of the incident power is actually delivered into the desired reflected beam, rather than the total power reflected in all directions.
where
\begin{equation}
P_{\mathrm{beam}}
=
\int_{\Omega_{\mathrm{beam}}}
S_{\mathrm{ref}}(\Theta,\Phi)\,d\Omega
\label{eq:reflected_power}
\end{equation}
\begin{equation}
P_{\mathrm{inc}}
=
\int_{\Omega_{\mathrm{inc}}}
S_{\mathrm{inc}}(\Theta,\Phi)\,d\Omega
\label{eq:incident_power}
\end{equation}
where $P_{\mathrm{beam}}$ denotes the portion of the total reflected
power contained within the prescribed angular region
$\Omega_{\mathrm{beam}}$ corresponding to the desired beam, where $\Omega_{\mathrm{beam}}\subseteq\Omega_{\mathrm{ref}}$. Also, $\Omega_{\mathrm{inc}}$ refers to the angular region occupied by
the incident field over the reflecting aperture.  
For far-field quantities, the power-density terms can be written as,
\begin{equation}
S_{\mathrm{ref}}(\Theta,\Phi)
\propto
\left|
\mathbf{E}_{\mathrm{element}}^{\mathrm{ref}}(\Theta,\Phi)
AF^{\mathrm{ref}}(\Theta,\Phi)
\right|^{2}
\end{equation}
\begin{equation}
S_{\mathrm{inc}}(\Theta,\Phi)
\propto
\left|
\mathbf{E}_{\mathrm{element}}^{\mathrm{inc}}(\Theta,\Phi)
AF^{\mathrm{inc}}(\Theta,\Phi)
\right|^{2}
\end{equation}
The common proportionality factor associated with the far-field power normalization is identical for the incident and reflected fields and therefore cancels when evaluating the reflector efficiency as follows,
\begin{equation}
\eta_{\mathrm{ref}}
=
\frac{
\displaystyle
\int_{\Omega_{\mathrm{beam}}}
\left|
\mathbf{E}_{\mathrm{element}}^{\mathrm{ref}}(\Theta,\Phi)
AF^{\mathrm{ref}}(\Theta,\Phi)
\right|^{2}
\,d\Omega
}{
\displaystyle
\int_{\Omega_{\mathrm{inc}}}
\left|
\mathbf{E}_{\mathrm{element}}^{\mathrm{inc}}(\Theta,\Phi)
AF^{\mathrm{inc}}(\Theta,\Phi)
\right|^{2}
\,d\Omega
}
\end{equation}
where the quantities of $\mathbf{E}_{\mathrm{element}}^{\mathrm{ref}}(\Theta,\Phi)$ and $\mathbf{E}_{\mathrm{element}}^{\mathrm{inc}}(\Theta,\Phi)$ represent the corresponding element radiation patterns, whereas $AF^{\mathrm{ref}}(\Theta,\Phi)$ and $AF^{\mathrm{inc}}(\Theta,\Phi)$ denote the associated array factors. 
The term of $d\Omega=\sin\Theta\,d\Theta\,d\Phi$ represents the differential solid angle.

First, the illuminating feedhorn antenna transfers the power accepted at its input port to the reflecting aperture, yielding a feed efficiency defined as,
\begin{equation}
\eta_{\mathrm{feed}}
=
\frac{P_{\mathrm{inc}}}{P_{\mathrm{acc}}}
\label{eq:feed_efficiency}
\end{equation}
where $P_{\mathrm{acc}}$ denotes the power accepted at the input port
of the illuminating antenna and $P_{\mathrm{inc}}$ denotes the power
incident on the reflecting aperture. 
Then, the planar reflector converts the incident power into the reflected field, with its performance characterized by the reflector efficiency of $\eta _{\mathrm{ref}}$,
\begin{equation}
\eta_{\mathrm{ref}}
=
\frac{P_{\mathrm{ref}}}{P_{\mathrm{inc}}}
\label{eq:total_reflector_efficiency}
\end{equation}
Accordingly, the corresponding power-flow relationships are
\begin{equation}
P_{\mathrm{acc}}
\;\xrightarrow{\;\eta_{\mathrm{feed}}\;}\;
P_{\mathrm{inc}}
\;\xrightarrow{\;\eta_{\mathrm{ref}}\;}\;
P_{\mathrm{ref,total}}
\label{eq:power_flow_total}
\end{equation}
\begin{equation}
P_{\mathrm{acc}}
\;\xrightarrow{\;\eta_{\mathrm{feed}}\;}\;
P_{\mathrm{inc}}
\;\xrightarrow{\;\eta_{\mathrm{beam}}\;}\;
P_{\mathrm{ref,desired}}
\label{eq:power_flow_beam}
\end{equation}
Thus, the end-to-end system efficiency with respect to the total reflected power is
\begin{equation}
\eta_{\mathrm{sys}}
=
\frac{P_{\mathrm{ref,total}}}{P_{\mathrm{acc}}}
=
\eta_{\mathrm{feed}}\eta_{\mathrm{ref}}
\label{eq:system_efficiency_product}
\end{equation}
whereas the corresponding efficiency for the desired reflected beam
is given by,
\begin{equation}
\eta_{\mathrm{sys,beam}}
=
\frac{P_{\mathrm{ref,desired}}}{P_{\mathrm{acc}}}
=
\eta_{\mathrm{feed}}\eta_{\mathrm{beam}}
\label{eq:system_beam_efficiency_product}
\end{equation}

The sidelobe level (SLL) is used to quantify the maximum undesired radiation relative to the main beam and is given by,
\begin{equation}
\mathrm{SLL}
=
20\log_{10}
\left[
\frac{
\displaystyle
\max_{(\Theta,\Phi)\in
\Omega_{\mathrm{sidelobe}}}
\left|
AF^{\mathrm{ref}}(\Theta,\Phi)
\right|
}{
\left|
AF^{\mathrm{ref}}
(\Theta_{\mathrm{pk}},\Phi_{\mathrm{pk}})
\right|
}
\right]
\label{eq:sidelobe_level}
\end{equation}
where $\Omega_{\mathrm{sidelobe}}$ represents the angular region excluding the main-beam region. A lower SLL indicates better suppression of undesired radiation outside the main beam.

Finally, dynamic beam steering is implemented by mapping the desired beam direction to the required reflection phase and subsequently to the graphene chemical potential. The overall synthesis procedure is represented by,
\begin{equation}
(\Theta_0,\Phi_0)
\longrightarrow
\psi^{\mathrm{req}}_{(n,k)}
\longrightarrow
\angle S_{11,(n,k)}
\longrightarrow
\mu_{c,(n,k)}
\label{eq:design_flow}
\end{equation}
Here, $(\Theta_0,\Phi_0)$ specifies the desired beam direction, $\psi^{\mathrm{req}}_{(n,k)}$ is the required reflection phase for the $(n,k)$-th graphene sector, $\angle S_{11,(n,k)}$ represents the phase obtained from the electromagnetic response of that sector, and $\mu_{c,(n,k)}$ is the corresponding graphene chemical potential. Thus, the desired beam direction is translated into a spatial distribution of graphene bias conditions, enabling dynamic beam steering of the planar reflector.
\section{Broadband Beam-Steering Performance of the Hemispherical Transmitarray Antenna}
\begin{figure}[htbp]
\centering
\subfloat[]{%
    \includegraphics[
        width=\columnwidth
    ]{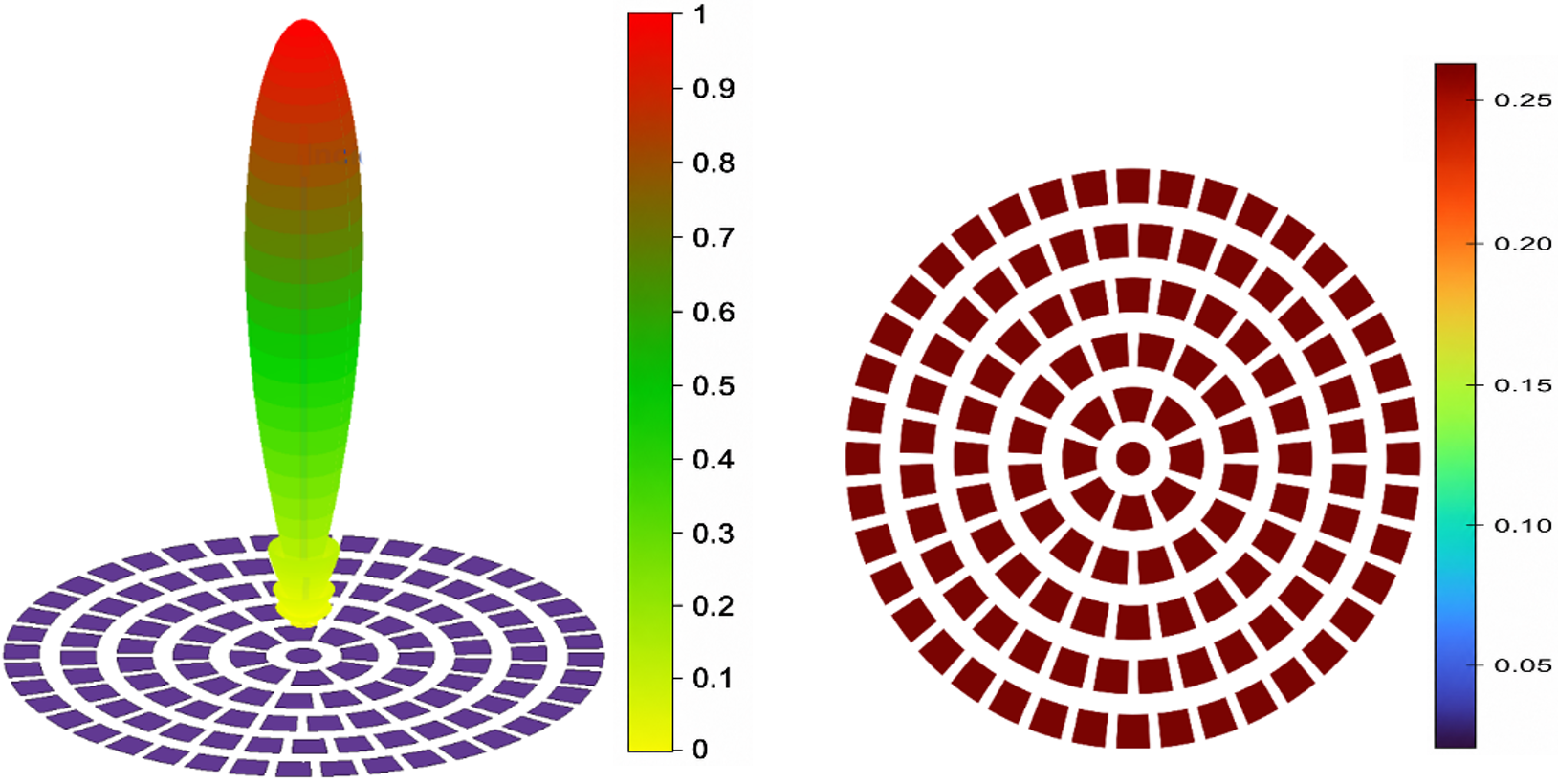}%
}
\vspace{1mm}

\subfloat[]{%
    \includegraphics[
        width=\columnwidth
    ]{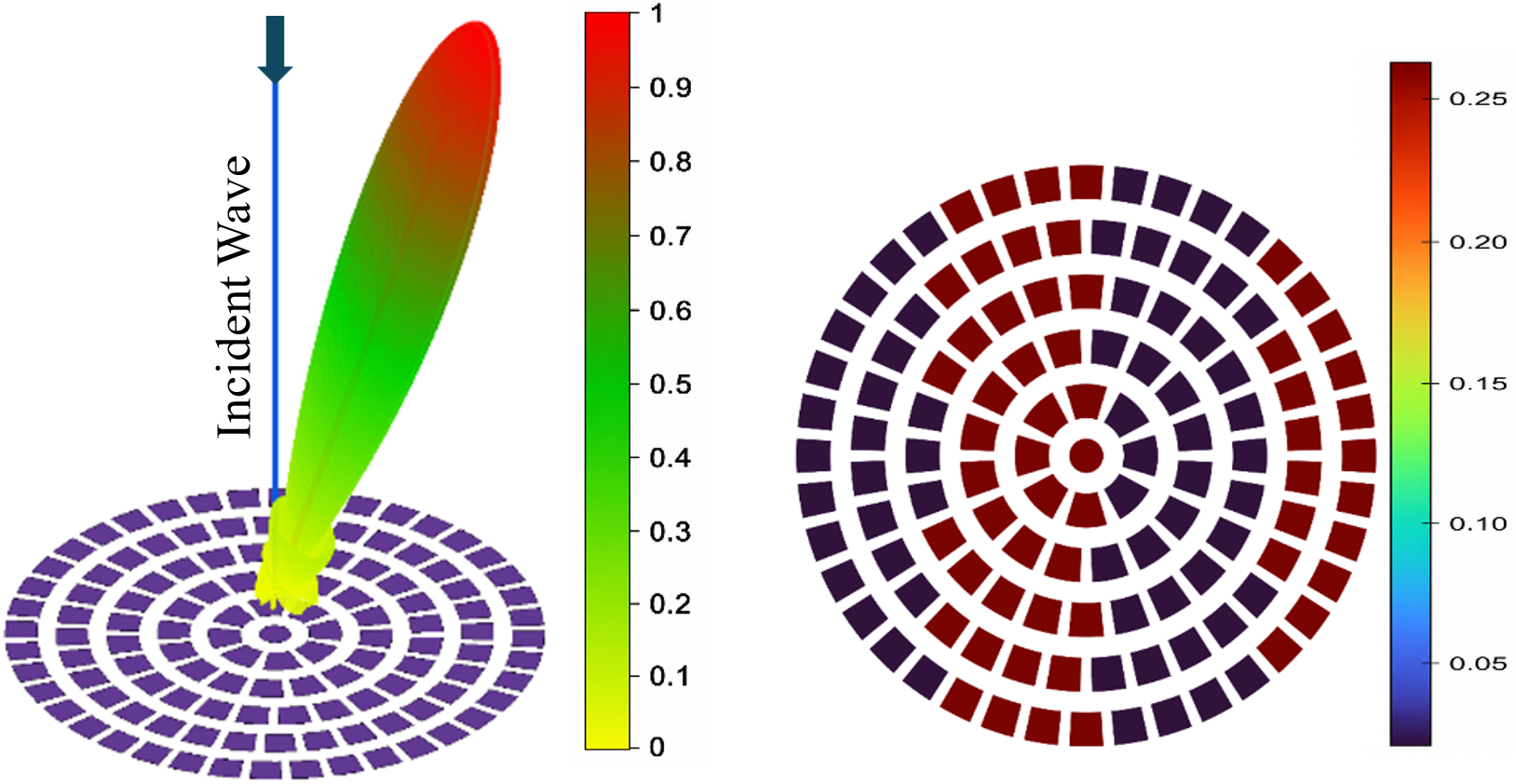}%
}
\vspace{1mm}

\subfloat[]{%
    \includegraphics[
        width=\columnwidth
    ]{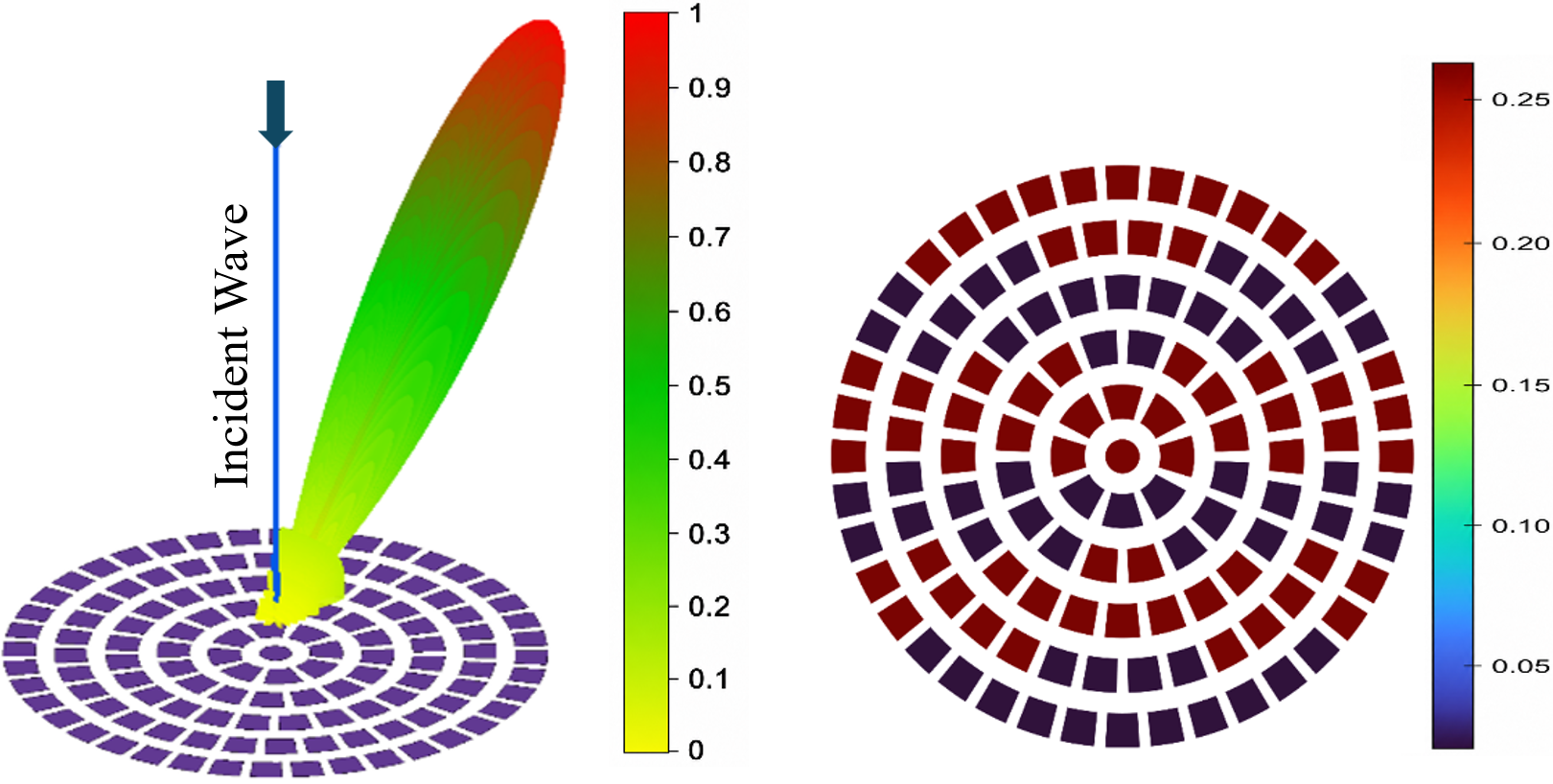}%
}
\vspace{1mm}
\subfloat[]{%
    \includegraphics[
        width=\columnwidth
    ]{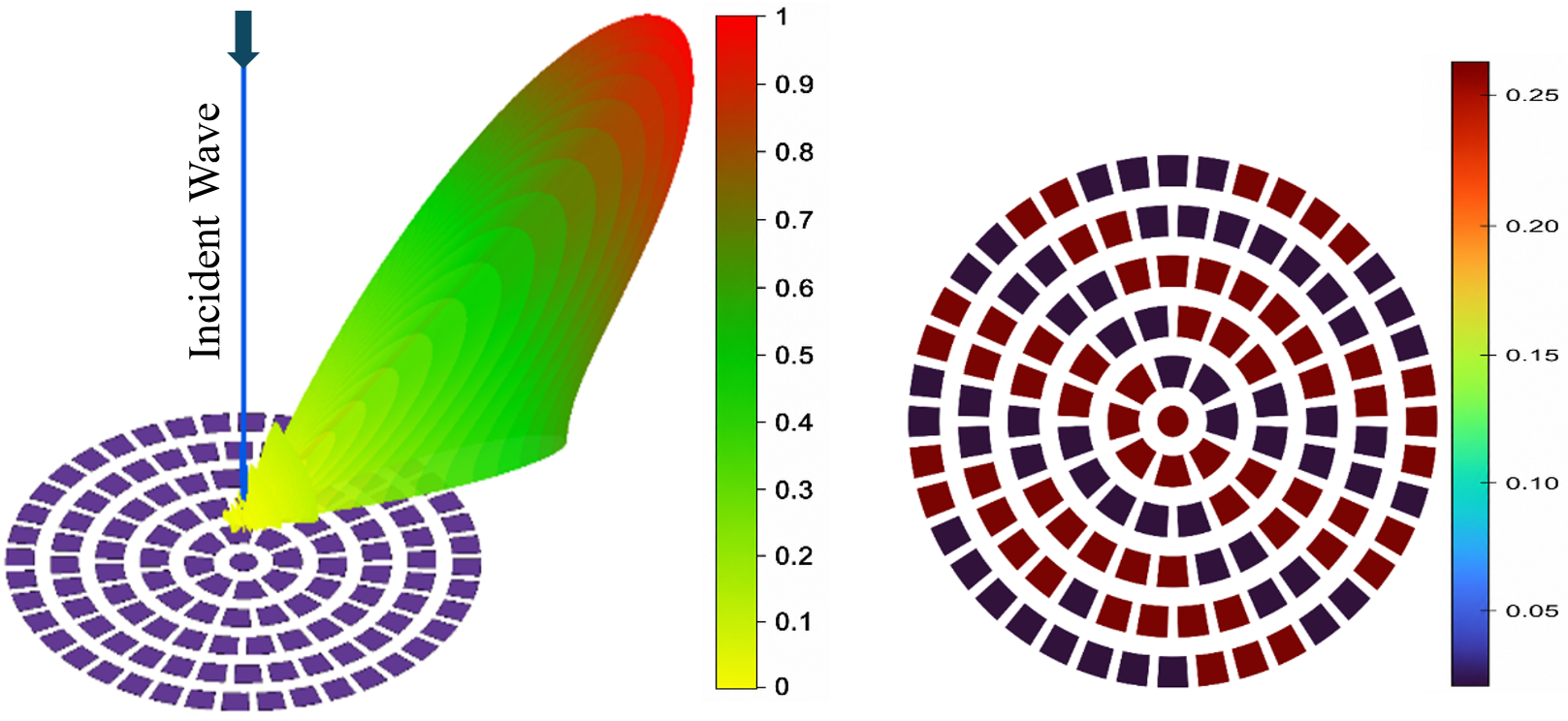}%
}
\caption{The vertically incident waves are generated by a circular feedhorn with an aperture diameter of 5$mm$ positioned approximately $41.7\lambda_0$ above the surface of the proposed flat reflector, as illustrated in Fig.~\ref{fig:flat_reflector_configuration}.
The reflected-wave data were obtained directly from CST Studio Suite for representative bias-voltage distributions applied to the graphene sectors. The obtained raw CST data were subsequently exported to MATLAB for post-processing and visualization. We obtained the results across the 200–300 GHz frequency range.
Each bias-voltage distribution applied to the individual graphene sectors is presented on the left-hand side, while the corresponding reflected-wave response is shown on the right-hand side. The reflected-wave angles corresponding to the specified bias-voltage distributions applied to the graphene sectors are given by the following values: (a) $(\theta_{\mathrm{r}},\phi_{\mathrm{r}})=(0^\circ,0^\circ)$, (b) $(\theta_{\mathrm{r}},\phi_{\mathrm{r}})=(25^\circ,0^\circ)$, (c) $(\theta_{\mathrm{r}},\phi_{\mathrm{r}})=(45^\circ,90^\circ)$, and (d) $(\theta_{\mathrm{r}},\phi_{\mathrm{r}})=(60^\circ,60^\circ)$.}
\label{fig:flat_reflector_four_parts_1}
\end{figure}
\begin{figure}[htbp]
\centering
\subfloat[]{%
    \includegraphics[
        width=\columnwidth
    ]{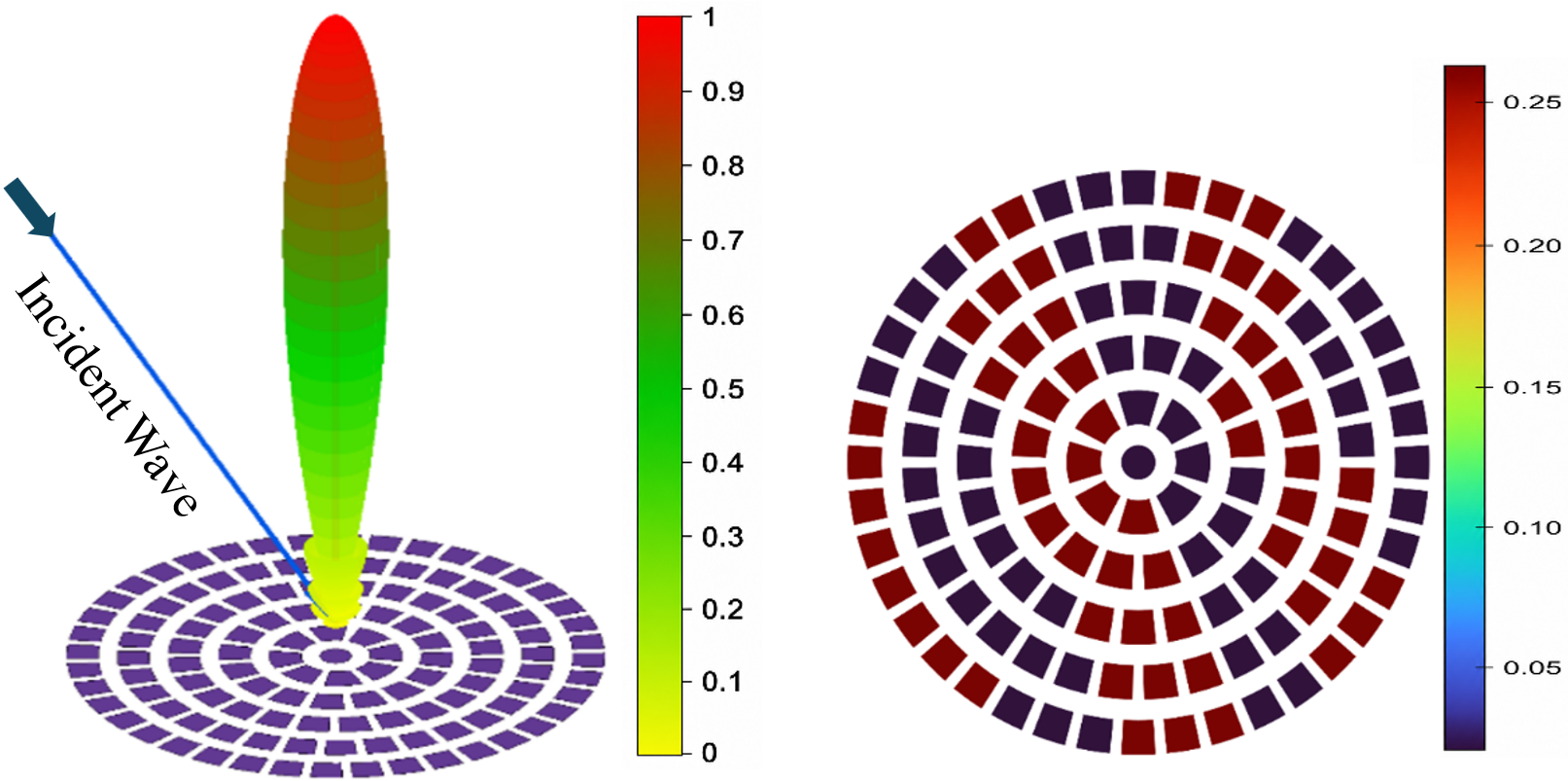}%
}
\vspace{1mm}
\subfloat[]{%
    \includegraphics[
        width=\columnwidth
    ]{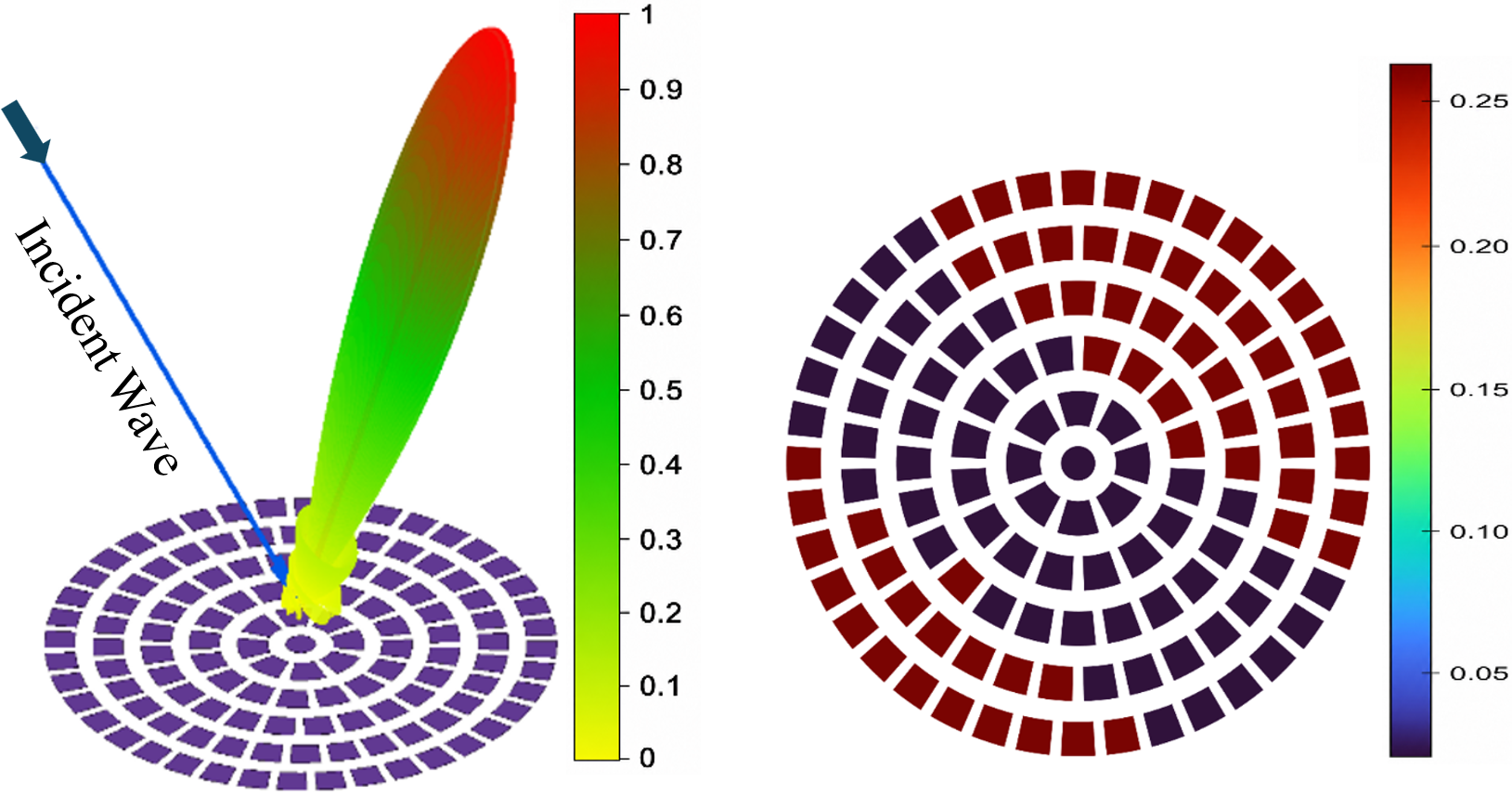}%
}
\vspace{1mm}

\subfloat[]{%
    \includegraphics[
        width=\columnwidth
    ]{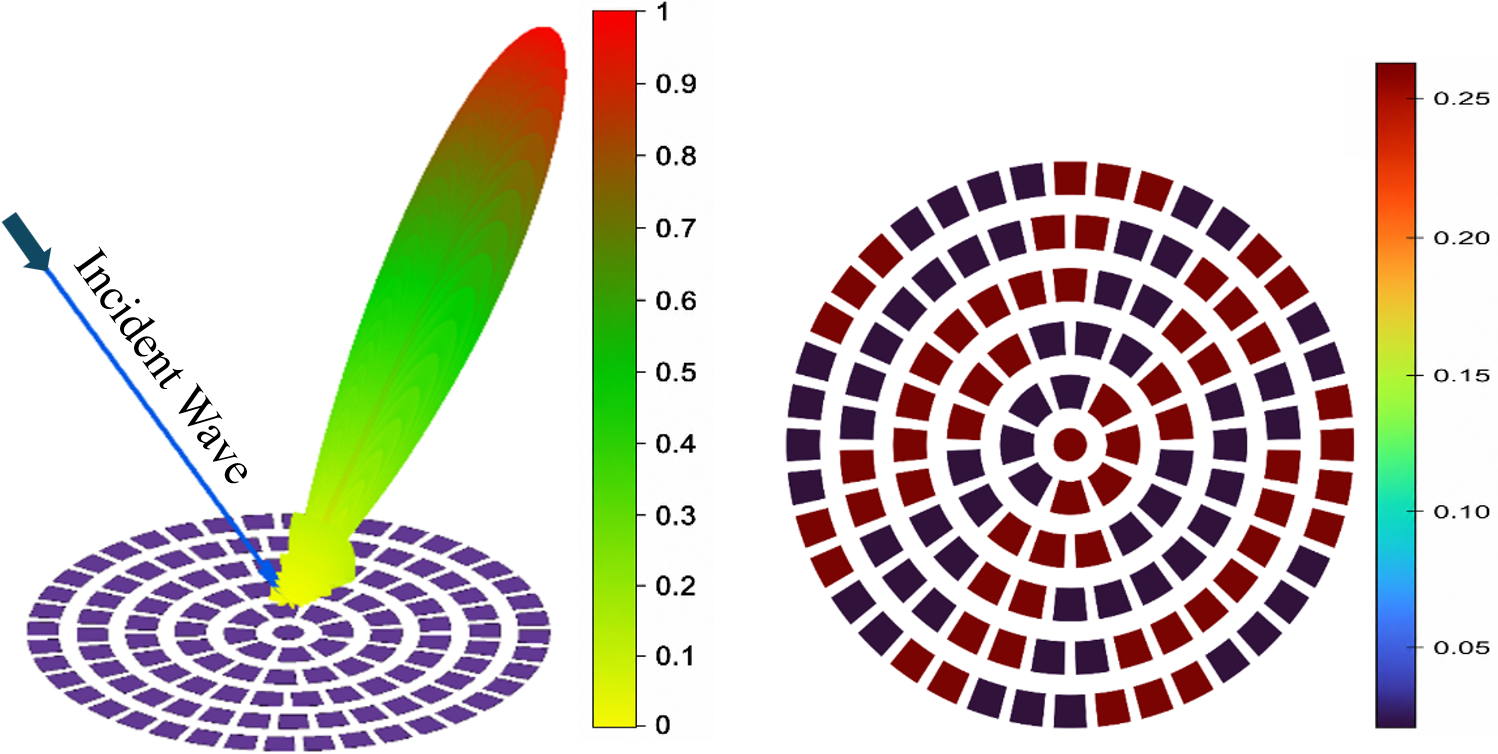}%
}
\vspace{1mm}

\subfloat[]{%
    \includegraphics[
        width=\columnwidth
    ]{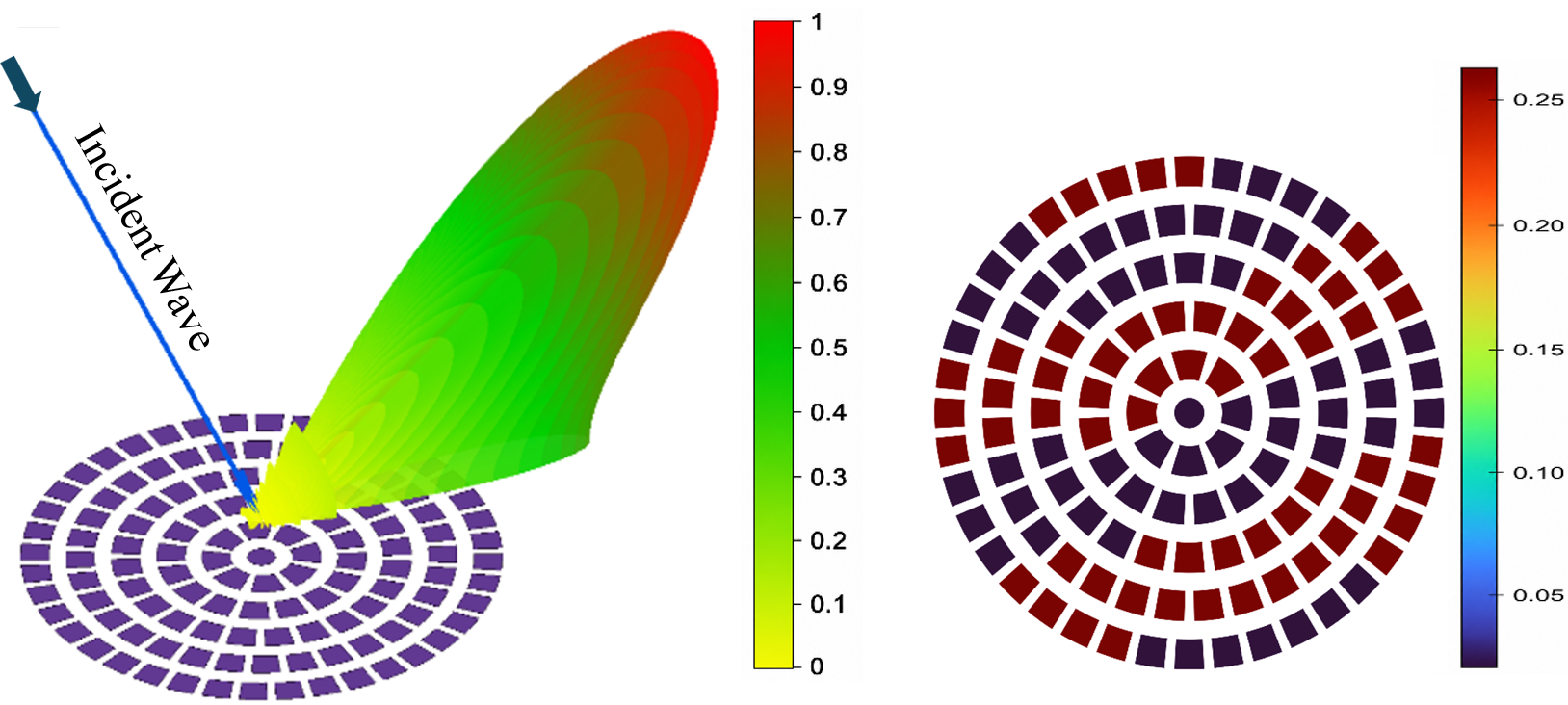}%
}
\caption{The incident waves are generated by a circular feedhorn with an aperture diameter of $5~\mathrm{mm}$, positioned approximately $41.7\lambda_0$ from the surface of the proposed flat reflector, as illustrated in Fig.~\ref{fig:flat_reflector_configuration}.
The feedhorn is oriented to illuminate the reflector with an obliquely incident wave with $(\theta_{\mathrm{inc}},\phi_{\mathrm{inc}})=(40^\circ,20^\circ)$ rather than a normally incident wave. 
The reflected-wave data were obtained directly from CST Studio Suite for representative bias-voltage distributions applied to the graphene sectors. The obtained raw CST data were subsequently exported to MATLAB for post-processing and visualization. We obtained the results across the $200$--$300$ GHz frequency range.
For each bias-voltage distribution applied to the individual graphene sectors is presented on the left-hand side, while the corresponding reflected-wave response is shown on the right-hand side. The reflected-wave directions corresponding to the specified bias-voltage distributions are as follows: (a) $(\theta_{\mathrm{r}},\phi_{\mathrm{r}})=(0^\circ,0^\circ)$, (b) $(\theta_{\mathrm{r}},\phi_{\mathrm{r}})=(25^\circ,0^\circ)$, (c) $(\theta_{\mathrm{r}},\phi_{\mathrm{r}})=(45^\circ,90^\circ)$, and (d) $(\theta_{\mathrm{r}},\phi_{\mathrm{r}})=(60^\circ,60^\circ)$.}
\label{fig:flat_reflector_four_parts_2}
\end{figure}
\begin{table}[htbp]
\caption{Performance evaluation of the proposed planar reflector in Fig.~\ref{fig:flat_reflector_configuration} for representative bias-voltage distributions applied to the graphene sectors, in terms of the reflected beam characteristics and HPBW. We have obtained the results across the $200$--$350$ GHz frequency range.}
\label{tab:beam_steering_performance}
\centering
\footnotesize
\setlength{\tabcolsep}{2.5pt}
\renewcommand{\arraystretch}{1.15}
\begin{tabular}{c c c c c}
\hline
$\theta_{\mathrm{inc}}$ &
$\phi_{\mathrm{inc}}$ &
$\theta_{\mathrm{ref}}$ &
$\phi_{\mathrm{ref}}$ &
HPBW \\
(deg.) & (deg.) & (deg.) & (deg.) & (deg.) \\
\hline
$0^\circ$  & $0^\circ$  & $60^\circ$  & $220^\circ$  & $31.9^\circ$ \\
$0^\circ$  & $0^\circ$  & $25^\circ$ & $0^\circ$  & $14.94^\circ$ \\
$0^\circ$  & $0^\circ$  & $45^\circ$ & $90^\circ$ & $19.44^\circ$ \\
$0^\circ$  & $0^\circ$  & $60^\circ$ & $300^\circ$ & $34.71^\circ$ \\
\hline
\end{tabular}
\end{table}
\begin{figure}[htbp]
\centering
\subfloat[]{%
    \includegraphics[
        width=\columnwidth
    ]{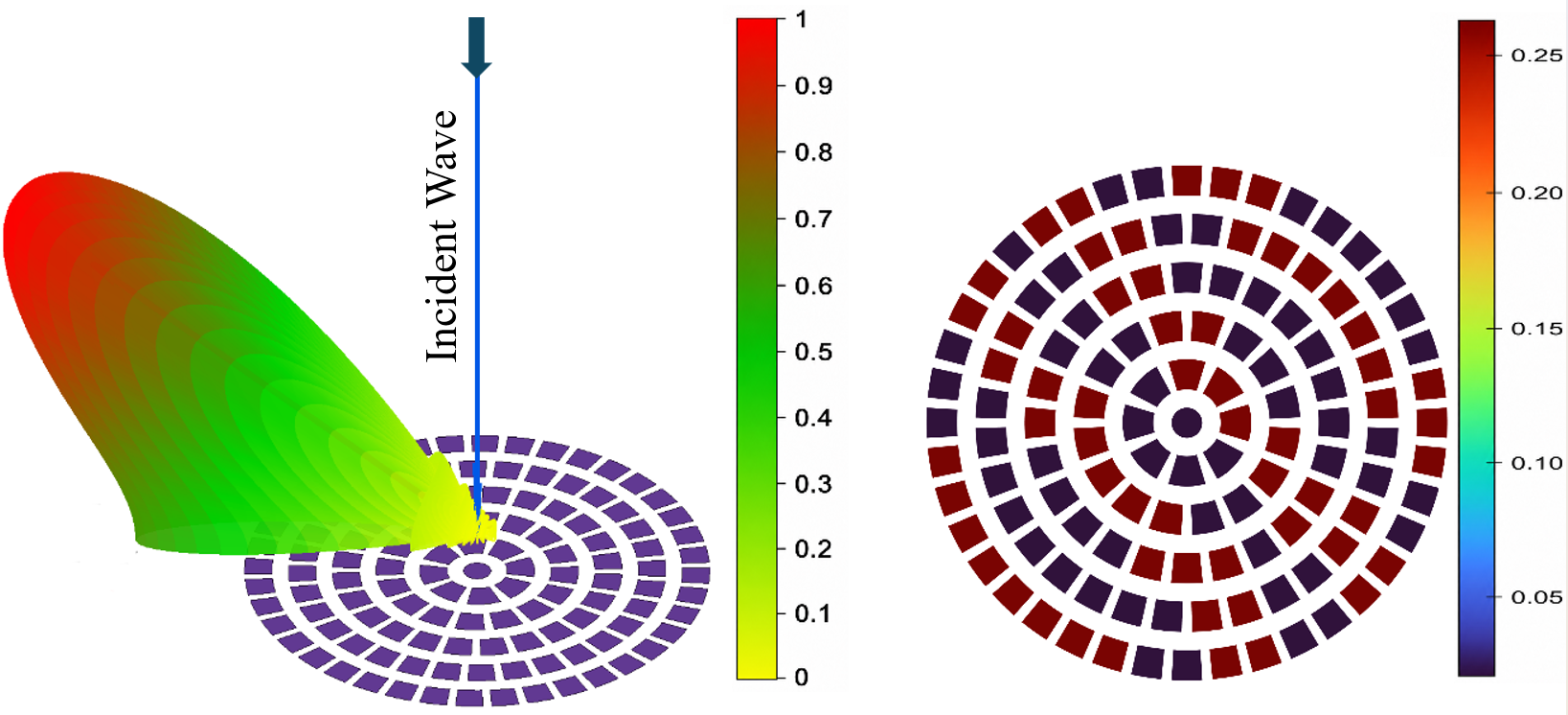}%
}
\vspace{1mm}
\subfloat[]{%
    \includegraphics[
        width=\columnwidth
    ]{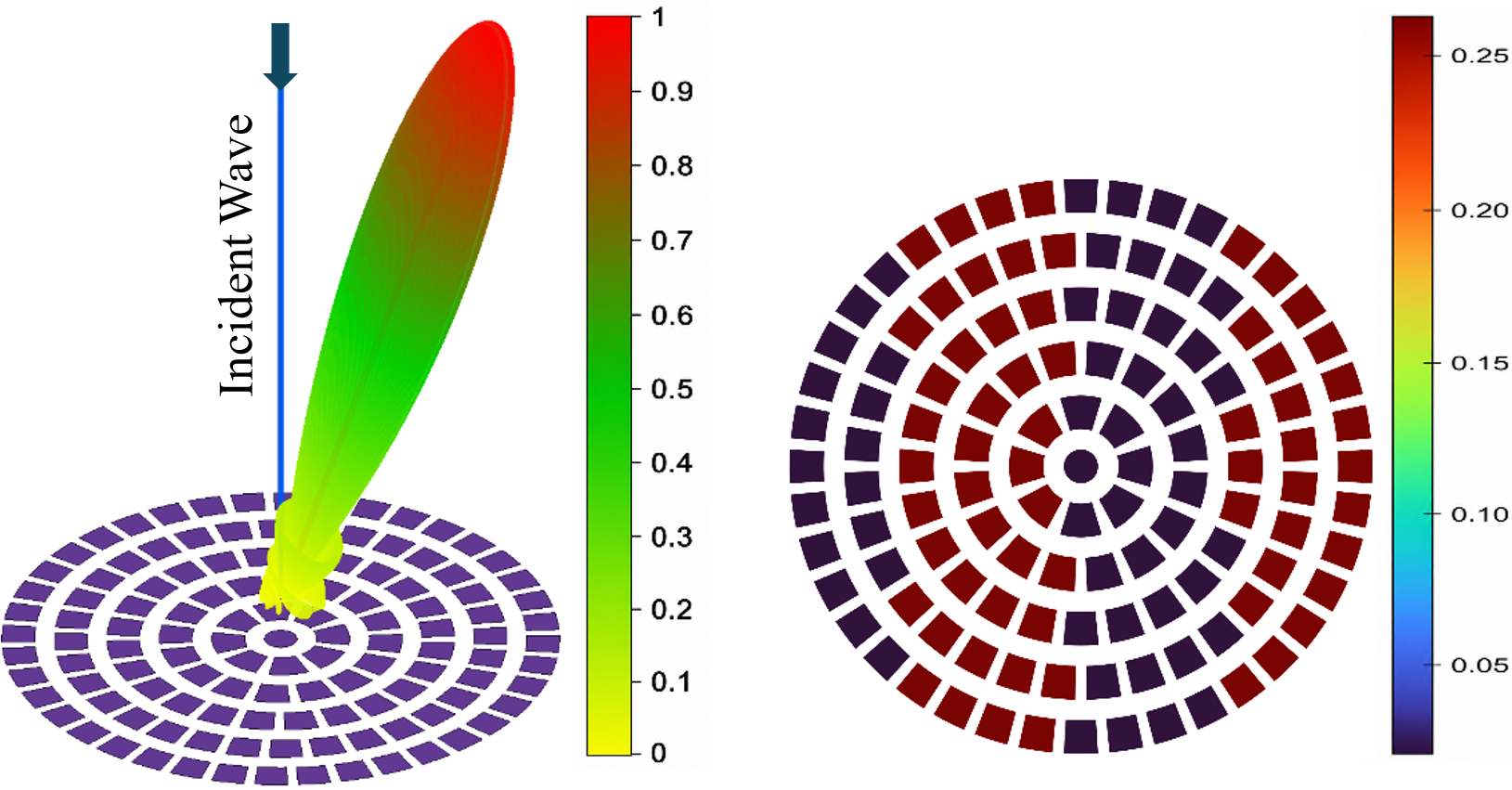}%
}
\vspace{1mm}

\subfloat[]{%
    \includegraphics[
        width=\columnwidth
    ]{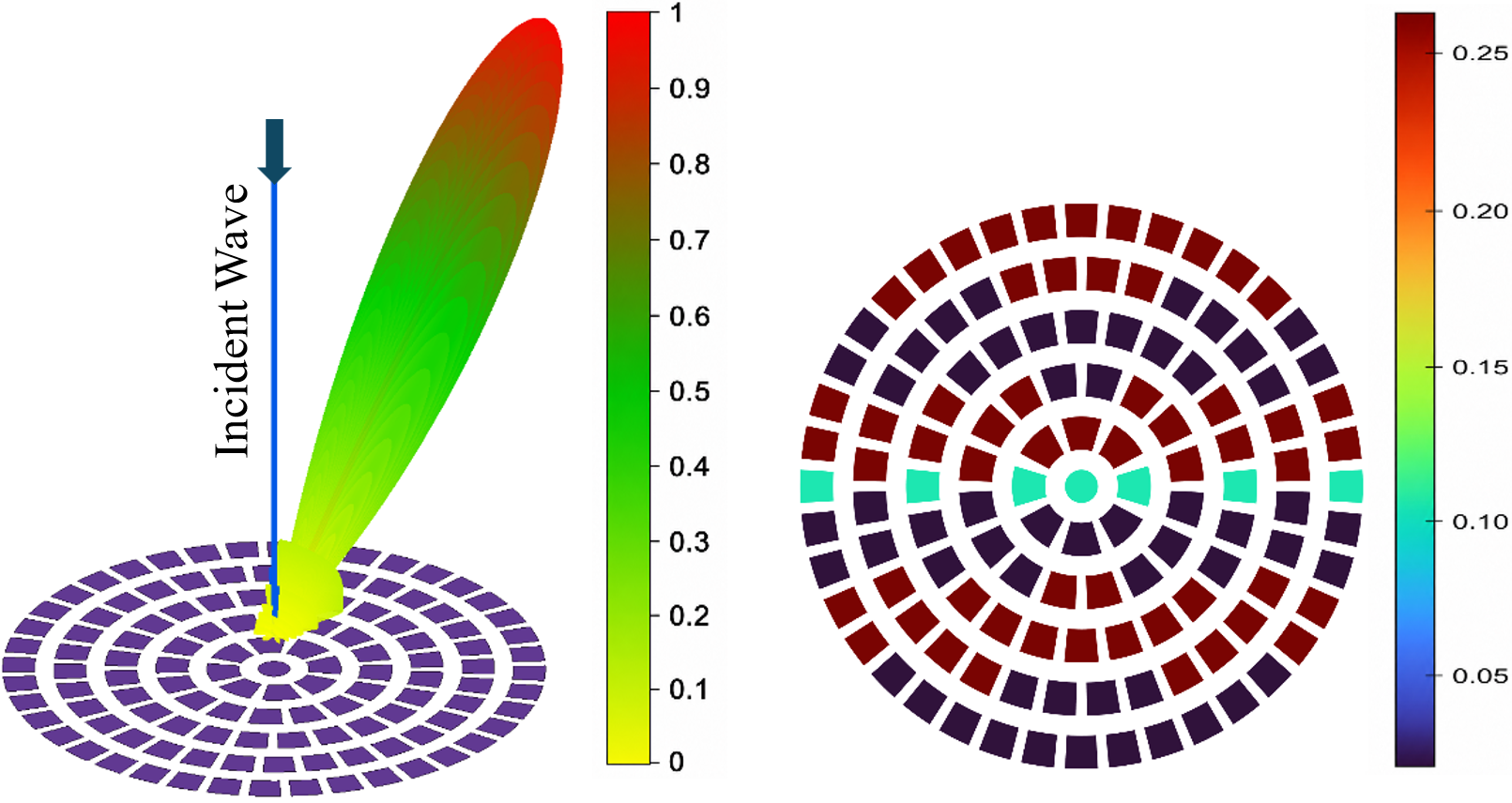}%
}
\vspace{1mm}

\subfloat[]{%
    \includegraphics[
        width=\columnwidth
    ]{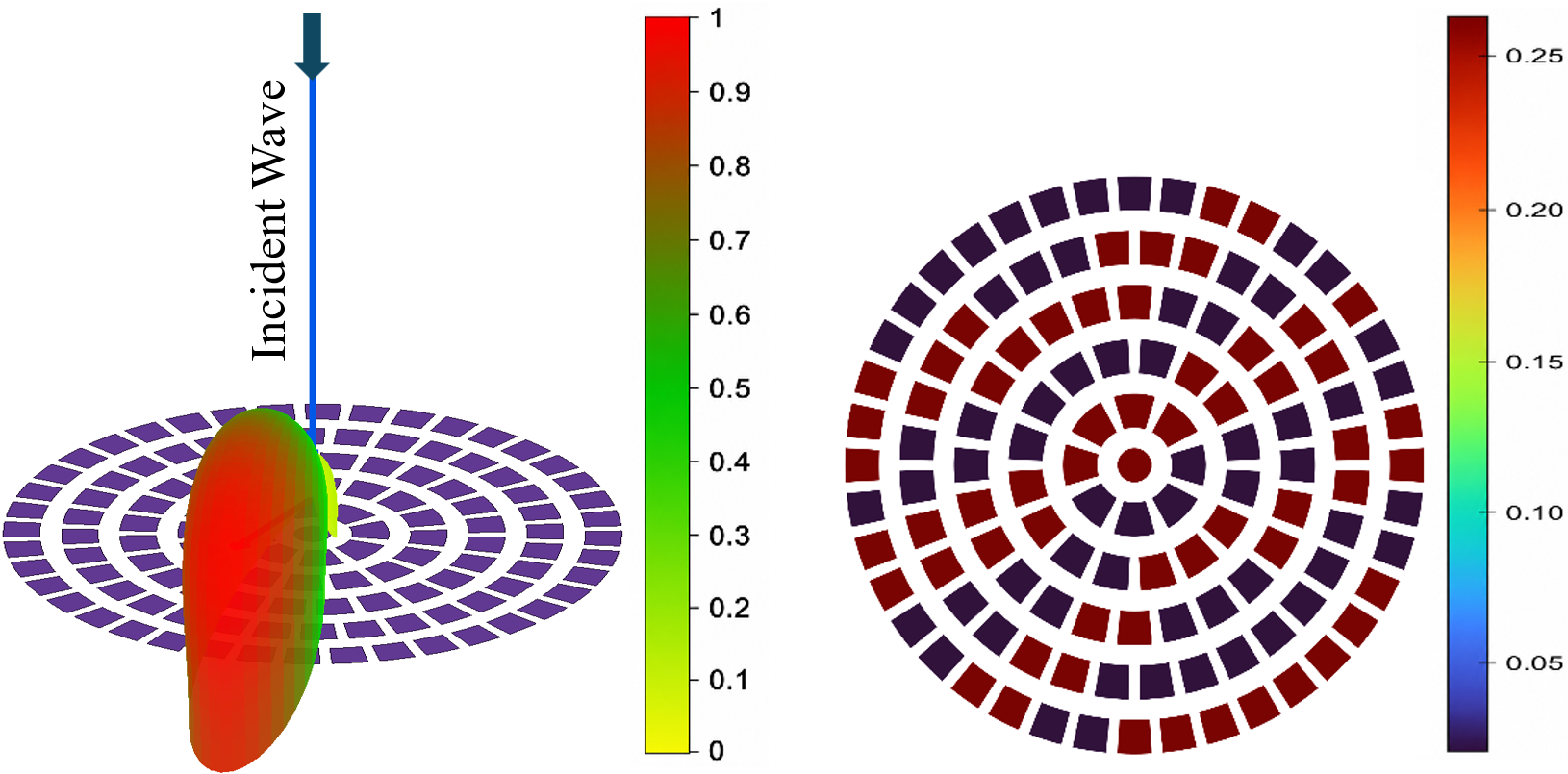}%
}
\caption{The vertically incident waves are generated by a circular feedhorn with an aperture diameter of 5$mm$ positioned approximately $41.7\lambda_0$ above the surface of the proposed flat reflector, as illustrated in Fig.~\ref{fig:flat_reflector_configuration}.
The reflected-wave data were obtained directly from CST Studio Suite for representative bias-voltage distributions applied to the graphene sectors. The obtained raw CST data were subsequently exported to MATLAB for post-processing and visualization. We obtained the results across the $200$--$350$ GHz frequency range. Each bias-voltage distribution applied to the individual graphene sectors is presented on the left-hand side, while the corresponding reflected-wave response is shown on the right-hand side. The reflected-wave angles corresponding to the specified bias-voltage distributions applied to the graphene sectors are given by the following values: (a) $(\theta_{\mathrm{r}},\phi_{\mathrm{r}})=(60^\circ,220^\circ)$, (b) $(\theta_{\mathrm{r}},\phi_{\mathrm{r}})=(25^\circ,0^\circ)$, (c) $(\theta_{\mathrm{r}},\phi_{\mathrm{r}})=(45^\circ,90^\circ)$, and (d) $(\theta_{\mathrm{r}},\phi_{\mathrm{r}})=(70^\circ,300^\circ)$.}
\label{fig:flat_reflector_four_parts_3}
\end{figure}
The discussion in this section has focused on demonstrating a major advantage of the proposed hemispherical transmitarray antenna, shown in Fig.~\ref{fig:hemispherical_transmitarray_2}, over its planar reflector counterpart, shown in Fig.~\ref{fig:flat_reflector_configuration}, particularly in terms of beam-steering range and angular coverage over the $200$--$300$~GHz frequency range.
To further validate the consistency of its beam-steering capability, the planar reflector is evaluated over an extended frequency range up to $350$~GHz.
\subsection{Broadband Beam-Steering Performance from $200$ to $300$ GHz}
The results in Figs. \ref{fig:flat_reflector_four_parts_1} and \ref{fig:flat_reflector_four_parts_2} demonstrate that each desired beam-steering direction requires a unique spatial distribution of bias voltages across the independently controlled graphene annular sectors.
As shown on the right side of each part, tuning the chemical potential of the graphene annular sectors induces a spatial variation in the reflection phase, which subsequently dictates the direction of the reflected beam shown on the left of each part.
This direct correspondence between the bias-voltage distribution and the reflected-beam direction confirms the capability of the proposed graphene-based reflector to achieve electronically reconfigurable beam steering.

For the normally incident case in Fig.~\ref{fig:flat_reflector_four_parts_1}, the reflected beam can be progressively steered from broadside toward large angular directions by appropriately tailoring the bias-voltage distribution.
However, as the steering angle approaches $\theta_{\mathrm{r}}\approx60^\circ$, the reflected beam pattern degrades, and the power concentration around the target direction decreases substantially. 
These observations indicate that the practical beam-steering range of the proposed reflector is limited to approximately $\theta_{\mathrm{r}}\le60^\circ$ in the elevation direction.

A comparison of Figs.~\ref{fig:flat_reflector_four_parts_1} and \ref{fig:flat_reflector_four_parts_2} further demonstrates that the beam-steering performance of the planar reflector is significantly vulnerable to the incidence angle.
Under normal incidence, the reflector maintains more symmetric and stable phase distribution across its aperture, enabling the reflected beam to obtain  wider steering angles.

In contrast, oblique illumination with $(\theta_{\mathrm{inc}},\phi_{\mathrm{inc}})=(40^\circ,20^\circ)$ introduces additional spatial phase variation across the planar reflector.
This additional phase variation modifies the phase distribution required for accurate beam steering, resulting in a progressive degradation of the reflected-wave pattern at large steering angles.
Thus, the reflected-wave pattern progressively degrades as the steering angle increases, thereby limiting the maximum obtainable steering angle in the $\theta$ direction under oblique incidence. Hence, under normal incidence, the planar reflector achieves its maximum steering range in the $\theta$ direction.

In comparison, the proposed hemispherical transmitarray antenna provides substantially wider angular beam steering, extending approximately from $-78^\circ$ to $+78^\circ$\cite{komeylian2025high,komeylian2026graphenebasedhemisphericaltransmitarrayantenna,komeylian2026activehemisphericalmetasurfacetransmitarray}.
The hemispherical configuration substantially reduces the spatial phase nonuniformity associated with feed illumination and maintains more favorable phase coherence among the radiating elements at large scan angles.
Owing to the alignment of the incident wave from the feedhorn with the local position vectors of the graphene sectors, the hemispherical transmitarray maintains an approximately uniform phase difference across the entire aperture lens.
In contrast, the graphene annular sectors of the planar reflector experience significant phase variations across the aperture due to the misalignment between the incident wave from the feed horn and the position vectors of the individual sectors, resulting in a nonuniform aperture phase distribution.
This comparison therefore demonstrates that the hemispherical topology provides a substantially broader angular steering capability than its geometrically corresponding planar reflector counterpart, particularly for the three-dimensional extreme-angle beam steering.
\subsection{Broadband Beam-Steering Performance from $200$ to $350$ GHz}
As shown in Fig.~\ref{fig:flat_reflector_four_parts_3}, the proposed planar reflector in Fig.~\ref{fig:flat_reflector_configuration} exhibits wide operating bandwidth, over which consistent beam-steering performance is maintained, with reliable beam steering extending up to 350 GHz. 
However, under oblique illumination from the feedhorn, the planar reflector fails to maintain stable beam-steering performance across the $200$--$350$ GHz frequency range. 
\section{Enhancement of Half-Power Beamwidth in the Hemispherical Transmitarray Antenna}
Table~\ref{tab:radiation_performance} has summarized the variation in HPBWs of the proposed transmitarray antenna for different bias-voltage distributions applied to its graphene sectors, as illustrated in Fig.~\ref{fig:hemispherical_transmitarray_2}.
For reference, the standalone feedhorn exhibits an HPBW of $28.7^\circ$ at the operating frequency of $250~\mathrm{GHz}$.
As evident from Table~\ref{tab:radiation_performance}, the maximum HPBW of $41.8^\circ$ occurs at $(\theta,\phi)=(78^\circ,360^\circ)$, corresponding to the maximum beam-steering angles. Compared with the reference HPBW of the standalone feedhorn, this represents a small HPBW variation of $13.1^\circ$ relative to the feedhorn reference in Table~\ref{tab:beam_steering_performance_13}. 
Furthermore, the proposed hemispherical transmitarray antenna exhibits a smooth and predictable variation in HPBWs. 
The reference HPBW can be systematically tailored by appropriately adjusting the feedhorn and lens dimensions, thereby providing an additional degree of freedom for controlling HPBWs.

Table~\ref{tab:beam_steering_performance} has reported the variation of HPBWs for some representative bias-voltage distributions applied to the graphene sectors of the planar reflector consistent with Figs.~\ref{fig:flat_reflector_configuration}, and ~\ref{fig:flat_reflector_four_parts_3} across the frequency range of $200$--$350$ GHz.

\textbf{Planar reflector:} for the planar reflector, the HPBW obtained when a uniform bias voltage
is applied to all graphene annular sectors is adopted as the reference, yielding an HPBW of $5.85^\circ$. 
Table~\ref{tab:beam_steering_performance_13} highlights that the maximum HPBW of $32.33^\circ$ occurs at $(\theta_{r},\phi_{r})=(60^\circ,60^\circ)$, corresponding to the maximum reflected beam angles.
Thus, under normal incidence, the maximum HPBW increase relative to the uniform-bias reference case is obtained by,
\begin{equation}
\begin{aligned}
\Delta\mathrm{HPBW}_{\mathrm{P.R.N}}
&=31.19^\circ-6.75^\circ\
&=24.44^\circ
\end{aligned}
\end{equation}
Under oblique-wave incidence of $(\theta_{r},\phi_{r})=(40^\circ,20^\circ)$, the HPBW variation of the planar reflector is given by,
\begin{equation}
\begin{aligned}
\Delta\mathrm{HPBW}_{\mathrm{P.R.O}}
&=32.33^\circ-5.85^\circ\
&=26.48^\circ
\end{aligned}
\end{equation}
Hence, we can conclude that the proposed hemispherical transmitarray provides a more stable and controllable HPBW response than the flat reflector. In addition, compared to the hemispherical transmitarray, the flat reflector distributes the received power over a wider region, and thereby the power density, directivity, and overall gain of the flat reflected wave are reduced significantly.

\textbf{Hemispherical transmitarray antenna:}
The HPBW of the standalone feedhorn is considered as the reference beamwidth and is equal to $28.7^\circ$. 
As evident from Table~\ref{tab:radiation_performance}, the maximum HPBW of $41.8^\circ$ occurs at $(\theta,\phi)=(78^\circ,360^\circ)$, corresponding to the maximum beam-steering angles. Therefore, the maximum HPBW increase relative to the standalone feedhorn is estimated by, 
\begin{equation}
\Delta\mathrm{HPBW}_{\mathrm{T.A.}}
=41.8^\circ-28.7^\circ
=13.1^\circ
\end{equation}
These results correspond to reductions in HPBW variation of approximately $50.6\%$ under oblique-wave incidence and $46.4\%$ under normal-wave incidence, demonstrating a substantial enhancement in the beamwidth stability for the proposed transmitarray antenna, as shown in Fig.~\ref{fig:hemispherical_transmitarray_2}, compared with the planar reflector counterpart in Fig.~\ref{fig:flat_reflector_configuration}.
\begin{table}[htbp]
\caption{Radiation performance of the proposed hemispherical transmitarray antenna in Fig.~\ref{fig:hemispherical_transmitarray_2} for representative bias-voltage distributions applied to the graphene sectors and corresponding beam-steering directions $(\theta,\phi)$, characterized in terms of directivity, HPBW, and SLL. We have obtained the results across the 200–300 GHz frequency range.}
\label{tab:radiation_performance}
\centering
\footnotesize
\setlength{\tabcolsep}{4pt}
\renewcommand{\arraystretch}{1.15}
\begin{tabular}{c c c c}
\hline
$(\theta,\phi)$ & Directivity & HPBW & SLL \\
(deg.) & (dBi) & (deg.) & (dB) \\
\hline
$(0,0)$    & 14.72 & 21.6 & $-8.5$ \\
$(23,0)$   & 12.15 & 39.9 & $-9.4$ \\
$(22,150)$ & 11.96 & 39.6 & $-9.6$ \\
$(23,300)$ & 12.09 & 39.4 & $-9.8$ \\
$(50,0)$   & 13.29 & 22.7 & $-8.8$ \\
$(50,150)$ & 13.24 & 21.3 & $-7.7$ \\
$(76,360)$ & 11.38 & 41.8 & $-6.8$ \\
$(78,150)$ & 11.70 & 40.1 & $-9.2$ \\
$(78,300)$ & 11.43 & 37.4 & $-7.7$ \\
\hline
\end{tabular}
\end{table}
\begin{table}[htbp]
\caption{Performance evaluation of the proposed planar reflector in Fig.~\ref{fig:flat_reflector_configuration} for representative bias-voltage distributions applied to the graphene sectors, in terms of the reflected beam characteristics and HPBW. We have obtained the results across the 200–300 GHz frequency range.}
\label{tab:beam_steering_performance_13}
\centering
\footnotesize
\setlength{\tabcolsep}{2.5pt}
\renewcommand{\arraystretch}{1.15}
\begin{tabular}{c c c c c}
\hline
$\theta_{\mathrm{inc}}$ &
$\phi_{\mathrm{inc}}$ &
$\theta_{\mathrm{ref}}$ &
$\phi_{\mathrm{ref}}$ &
HPBW \\
(deg.) & (deg.) & (deg.) & (deg.) & (deg.) \\
\hline
$0^\circ$  & $0^\circ$  & $0^\circ$  & $0^\circ$  & $6.75^\circ$ \\
$0^\circ$  & $0^\circ$  & $25^\circ$ & $0^\circ$  & $14.94^\circ$ \\
$0^\circ$  & $0^\circ$  & $45^\circ$ & $90^\circ$ & $19.44^\circ$ \\
$0^\circ$  & $0^\circ$  & $60^\circ$ & $60^\circ$ & $31.19^\circ$ \\
$40^\circ$ & $20^\circ$ & $0^\circ$  & $0^\circ$  & $5.85^\circ$ \\
$40^\circ$ & $20^\circ$ & $25^\circ$ & $0^\circ$  & $14.1^\circ$ \\
$40^\circ$ & $20^\circ$ & $45^\circ$ & $90^\circ$ & $18.8^\circ$ \\
$40^\circ$ & $20^\circ$ & $60^\circ$ & $60^\circ$ & $32.33^\circ$ \\
\hline
\end{tabular}
\end{table}

As a result, we can conclude that the proposed hemispherical
transmitarray provides a more stable and controllable HPBW response than the flat reflector, Figs.~\ref{fig:flat_reflector_four_parts_2}--\ref{fig:flat_reflector_four_parts_3}. 
In addition, compared to the hemispherical transmitarray, the flat reflector distributes the received power over a wider region, and thereby the power density, directivity, and overall gain of the flat reflected wave
are reduced significantly\cite{mahony2012relationship,stegen1964gain}.
\section{Substantial Bandwidth Improvement with Fractal Concentric Circular Elements}
\begin{figure}[htbp]
    \centering
    \includegraphics[
        width=1\columnwidth,
        height=0.68\columnwidth
    ]{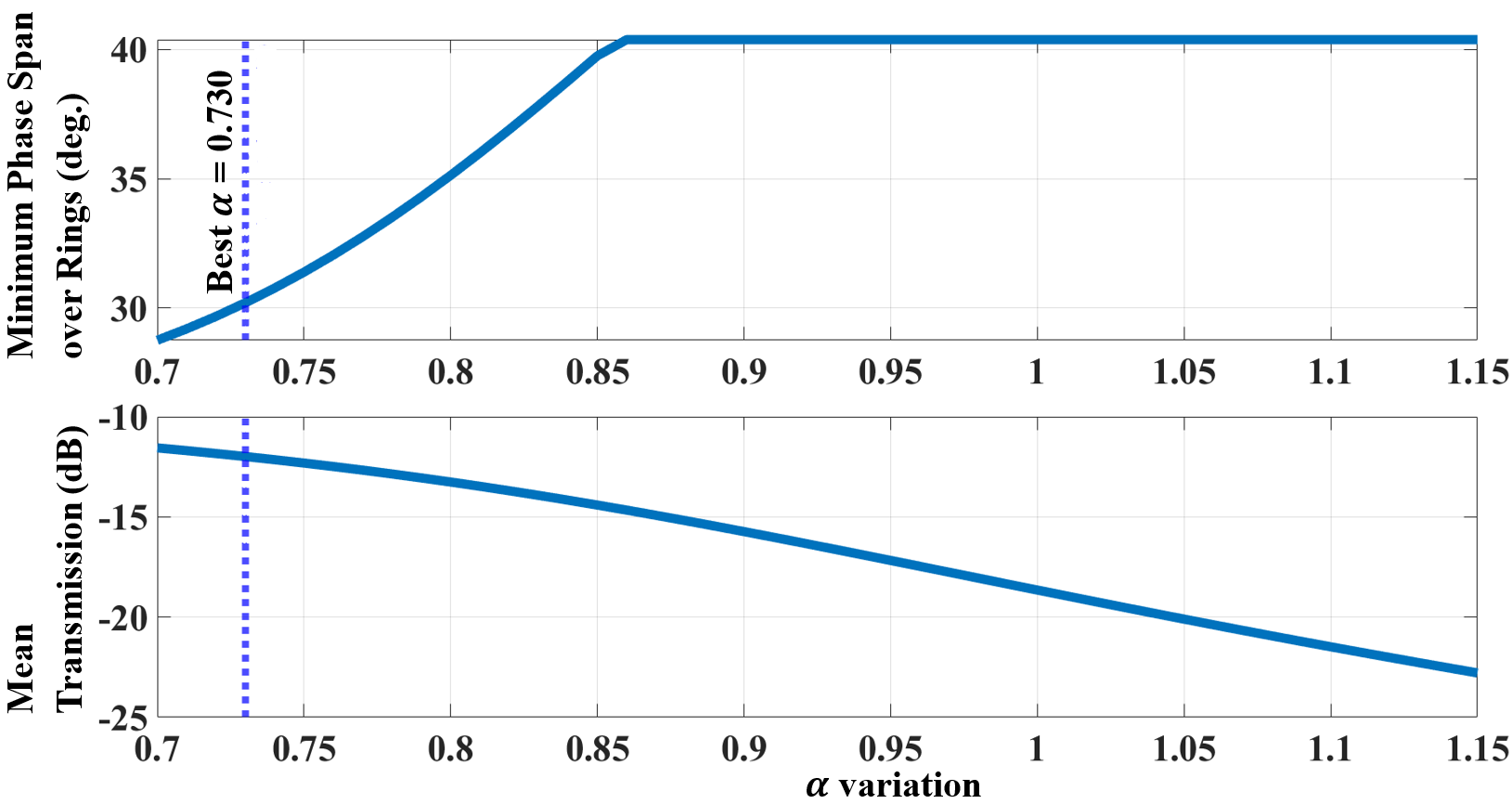}
    \caption{Optimization of the scaling factor $\alpha$ using a parametric sweep from $0.70$ to $1.15$: (a) minimum transmission-phase span among all concentric rings and (b) mean transmission magnitude as functions of $\alpha$, evaluated over the chemical potential range $0.1 \leq \mu_c \leq 1$ eV. The optimal $\alpha$ is selected to maximize the minimum available phase span while maintaining the mean transmission above $-12$ dB.}
    \label{fig:alpha}
\end{figure}
\begin{figure}[htbp]
    \centering
    \subfloat[]{%
        \includegraphics[
            width=1\columnwidth,
            height=0.51\columnwidth]
            {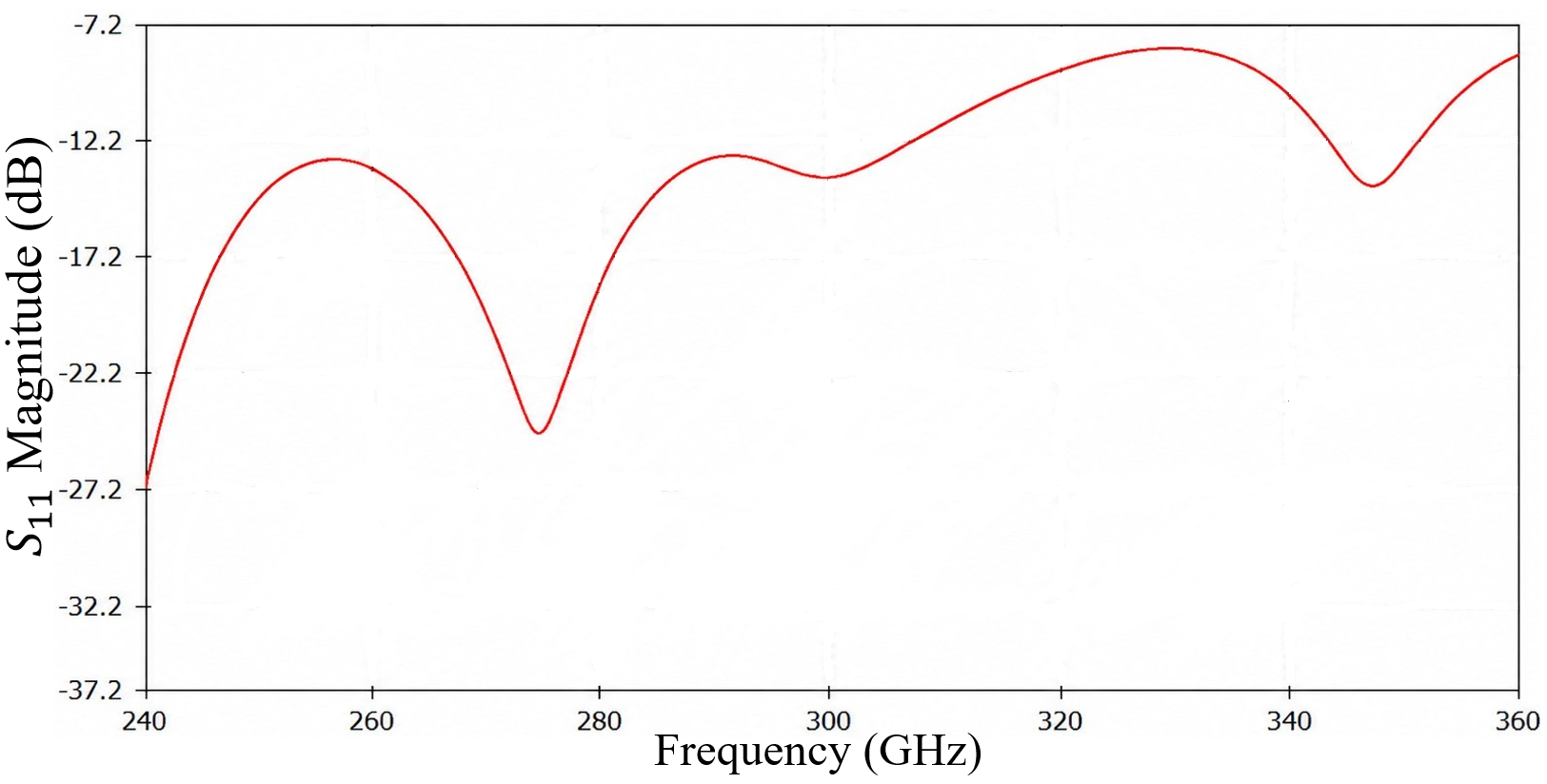}%
    }
    \hfill
    \subfloat[]{%
        \includegraphics[
            width=1\columnwidth,
            height=0.5\columnwidth]
            {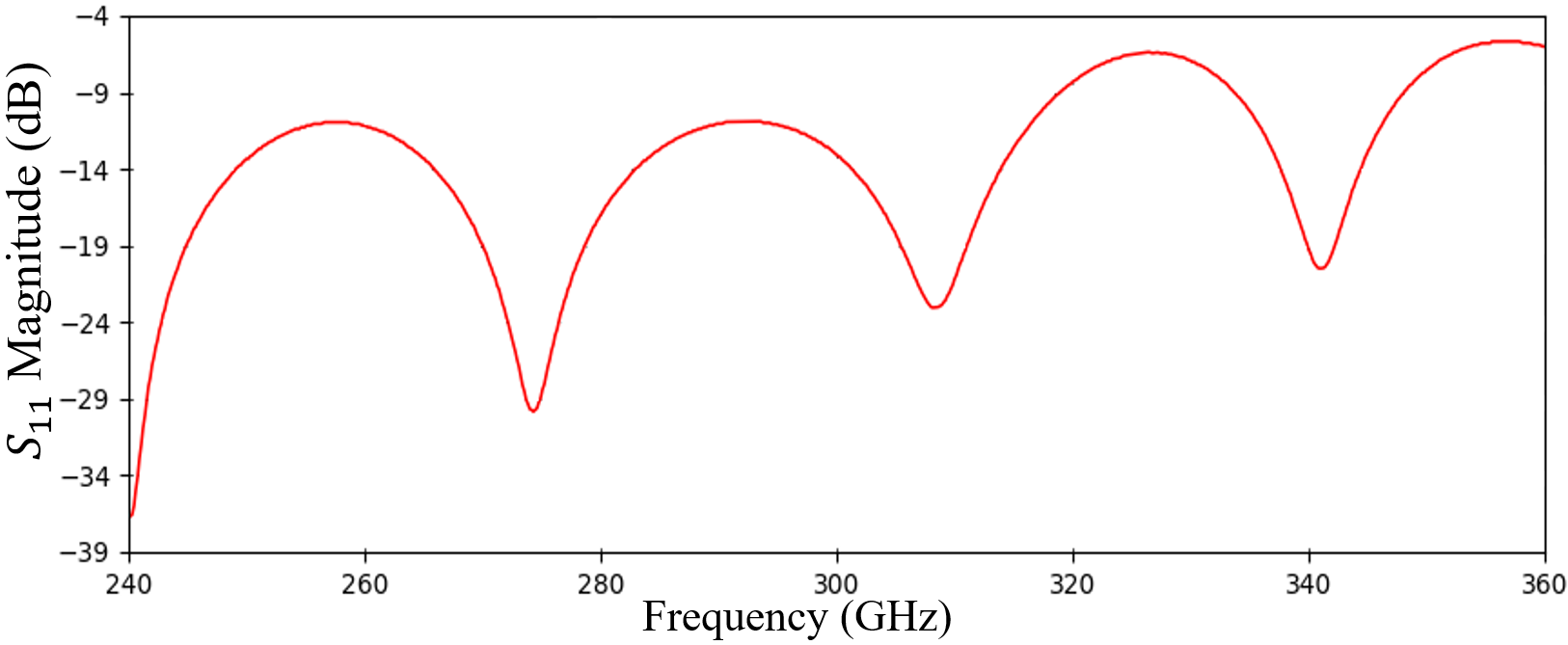}%
    }\\[2mm]
    \subfloat[]{%
       \includegraphics[
            width=1\columnwidth,
            height=0.5\columnwidth]
            {Figures/S11-T30-P60-2.png}
        }
\caption{Variation of $|S_{11}|$ versus frequency for our proposed transmitarray antenna over $240$--$360$ GHz frequency range with an operating frequency of $300$ GHz for different beam-steering directions of (a) $\theta=78^\circ$ and $\phi=200^\circ$, (b) $\theta=30^\circ$ and $\phi=60^\circ$, and (c) $\theta=0^\circ$ and $\phi=0^\circ$. The results presented herein are obtained from the hemispherical transmitarray antenna shown in Fig.~\ref{fig:hemispherical_transmitarray_2} without changing design dimensions.}
\label{fig:transmitarray_S11}
\end{figure}
\begin{figure}[htbp]
    \centering
    \subfloat[]{%
        \includegraphics[
            width=1\columnwidth,
            height=0.5\columnwidth]
            {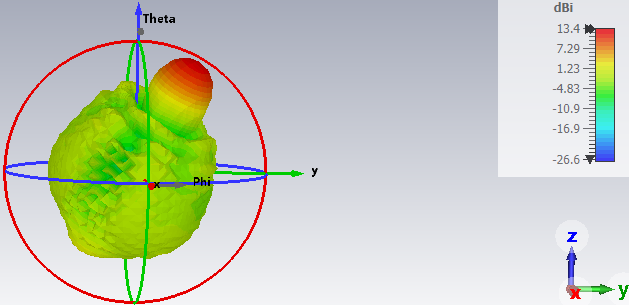}%
    }
    \hfill
    \subfloat[]{%
        \includegraphics[
            width=1\columnwidth,
            height=0.5\columnwidth]
            {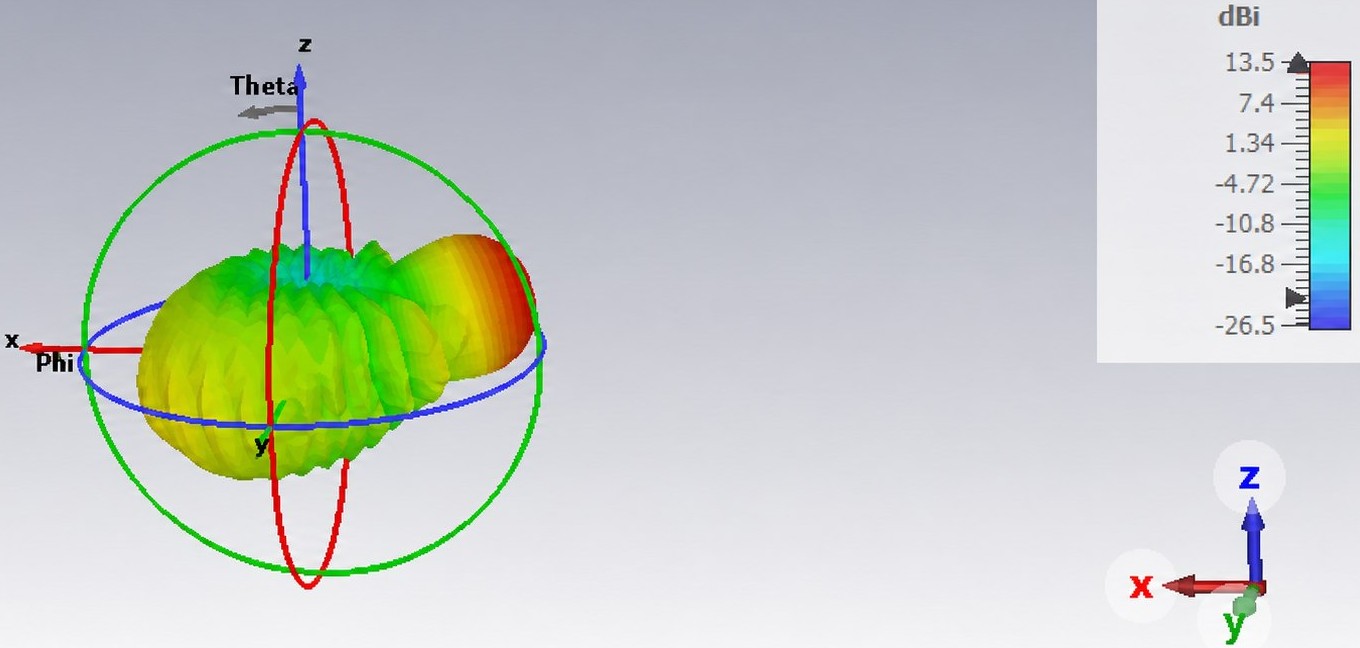}%
    }\\[2mm]
    \subfloat[]{%
       \includegraphics[
            width=1\columnwidth,
            height=0.5\columnwidth]
            {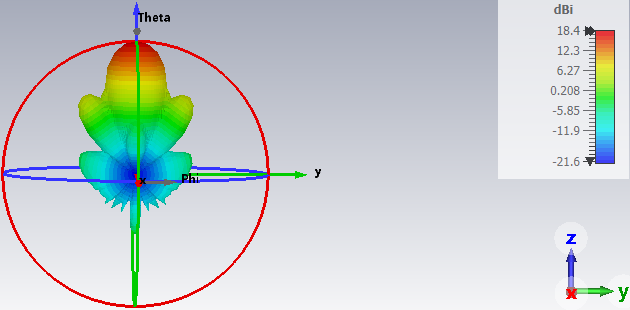}%
        }
\caption{Three-dimensional radiation pattern of our proposed transmitarray antenna over $240$--$360$ GHz frequency range with an operating frequency of 300 GHz for different beam-steering directions of (a) $\theta=78^\circ$ and $\phi=200^\circ$, (b) $\theta=30^\circ$ and $\phi=60^\circ$, and (c) $\theta=0^\circ$ and $\phi=0^\circ$. The results presented herein are obtained from the hemispherical transmitarray antenna shown in Fig.~\ref{fig:hemispherical_transmitarray_2} without changing design dimensions.}
\label{fig:transmitarray_Radiation}
\end{figure}
Planar reflectors exhibit substantial phase delays stemming from the spatial separation between the feed and reflector planes, which imposes a fundamental constraint on bandwidth. 
To overcome this limitation, the proposed hemispherical antenna in Fig.~\ref{fig:hemispherical_transmitarray_2} has employed electrically thin, subwavelength FCC elements whose dimensions are gradually scaled from one concentric ring to the next to realize the required spatial impedance distribution.
The scaling factor $\alpha$ was determined through a systematic parametric phase-range sweep. By setting the initial geometric parameters to $a=8$ and $\rho=1.45$, $\alpha$ was swept from $0.70$ to $1.15$ in increments of $0.01$, where $\rho$ denotes the growth factor governing the number of elements in each ring.
For each value of $\alpha$, the transmission-phase range of the unit cells was evaluated across a tunable chemical potential $\mu _{c}$ spanning $0.1$ to $1\text{ eV}$.
To ensure adequate phase-tuning capability across the entire aperture, the minimum phase span among all concentric rings was selected as the primary optimization metric.
The optimization primarily maximized the minimum available transmission-phase span while constraining the mean transmission to remain above $-12$~dB, thereby maintaining adequate $|S_{21}|$ while ensuring sufficient phase-tuning capability.
Ultimately, the script identifies the $\alpha$ configuration with the highest overall score.
For visualization, a high-resolution, two-panel plot was generated to illustrate the variation of the minimum phase span and mean transmission as functions of $\alpha$ in Fig.~\ref{fig:alpha}.

Parametric optimization identified an optimal scaling factor of $\alpha = 0.73$, which was subsequently rounded to $\alpha = 0.8$ to achieve a practical tradeoff between performance and implementation considerations.
Accordingly, the dimensions of the FCC elements are progressively scaled by a constant factor of 0.8 between successive concentric rings, as detailed in Table~I of\cite{komeylian2025high}.
This value aligns with the independent trial-and-error findings reported in \cite{komeylian2026graphenebasedhemisphericaltransmitarrayantenna,
komeylian2025high,
komeylian2026activehemisphericalmetasurfacetransmitarray}.
As illustrated in~\cite{komeylian2026graphenebasedhemisphericaltransmitarrayantenna,komeylian2025high,komeylian2026activehemisphericalmetasurfacetransmitarray}, the variation of reflection coefficient amplitude of $\lvert S_{11} \rvert$ in dB remains below $-10$~dB across the entire 
200--300~GHz frequency band. 
This performance corresponds to a fractional bandwidth of 40\%, validating the wideband impedance-matching capability of our proposed hemispherical transmitarray antenna.

In this work, to assess the consistency of the beam-steering performance over an extended frequency range, the operating frequency was swept from 250 GHz to 360 GHz while maintaining all geometrical dimensions of the proposed hemispherical transmitarray antenna in Fig.~\ref{fig:hemispherical_transmitarray_2} unchanged. In this case, the performance evaluation in the CST Studio environment in Figs.\ref{fig:transmitarray_S11} and \ref{fig:transmitarray_Radiation} demonstrate three distinct reflected-beam directions produced by three representative bias-voltage distributions across the graphene sectors of our proposed hemispherical transmitarray antenna.

It is evident in Fig.\ref{fig:transmitarray_S11} that the beam-steering performance remains stable over the $240$--$300$ GHz frequency range, while the reflection coefficient remains well below $-10$~dB. 
Beyond 300 GHz, the reflection coefficient degrades progressively, reaching approximately $-5$~dB at 360 GHz. Although a noticeable degradation in $S_{11}$ is observed above 300 GHz, the available incident power in this frequency range is relatively low compared with that at the lower frequencies\cite{song2011present}.
As shown in Fig.~\ref{fig:transmitarray_S11}, the proposed transmitarray antenna maintains stable beam-steering performance over the $240$--$300$~GHz frequency range, while the reflection coefficient remains below $-10$~dB. 
Above $300$~GHz, the reflection coefficient gradually degrades, reaching approximately $-5$~dB at $360$~GHz. 
Although the impedance-matching performance deteriorates beyond $300$~GHz, the incident power available in this frequency range is lower than that at lower frequencies~\cite{song2011present}. 
Consequently, the associated reduction in accepted power may have a limited impact on the overall system performance, depending on the source power spectrum and the system-level link budget.

For further verification of the stability of the beam-steering performance, the three-dimensional radiation patterns are presented in Fig.~\ref{fig:transmitarray_Radiation} to demonstrate consistent beam-steering capability with stable gains over the extended $240$--$360$-GHz frequency range.
\section{More Comparison of the Hemispherical Transmitarray Antenna and its Planar Reflector Counterpart}
This section further compares the proposed hemispherical transmitarray antenna with its planar reflector counterpart to elucidate the physical mechanisms underlying its improvement in the radiation performance and wide-angle beam-steering capability.
In particular, the comparison is conducted based on the following principle performance metrics: (1) wave alignment: intrinsic feed-wave alignment and spatial phase uniformity, 
(2) parasitic mitigation: the reduction of edge diffraction, parasitic radiation, and ohmic losses, (3) enhancement in mutual-coupling suppression: mutual coupling suppression using highly directive FCC elements, and (4) application: capability to dynamically track moving targets. 

These characteristics arise primarily from the three dimensional hemispherical topology and the electromagnetic properties of the proposed FCC unit cells, which together enable more efficient aperture excitation, superior radiation pattern control, and consistent high-gain performance over a broad angular beam-steering range.
\subsection{Intrinsic Feed-Wave Alignment and Spatial Phase Uniformity}
\begin{figure}[htbp]
    \centering
    \subfloat[]{%
        \includegraphics[
            width=1\columnwidth,
            height=0.7\columnwidth]
            {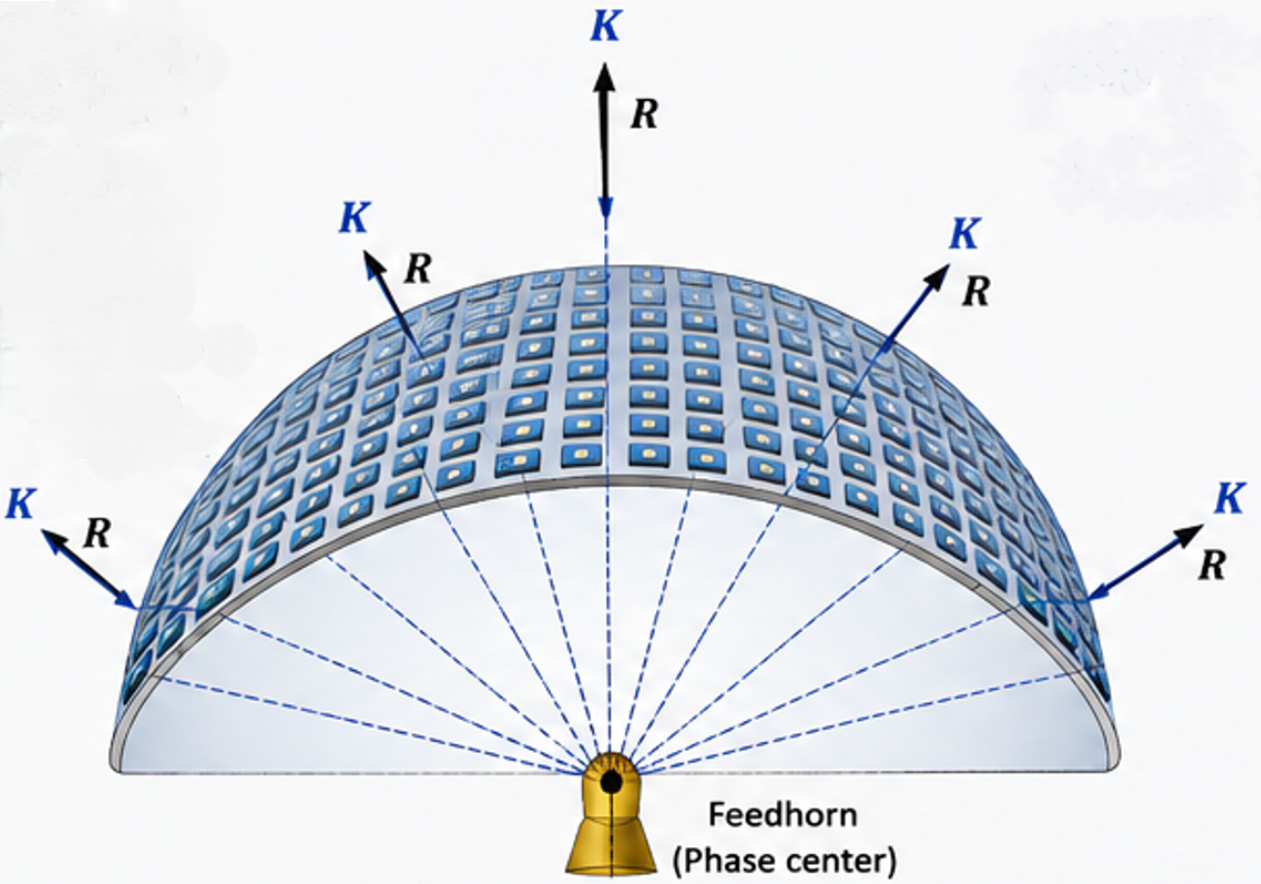}%
    }
    \hfill
    \subfloat[]{%
        \includegraphics[
            width=1\columnwidth,
            height=0.75\columnwidth]
            {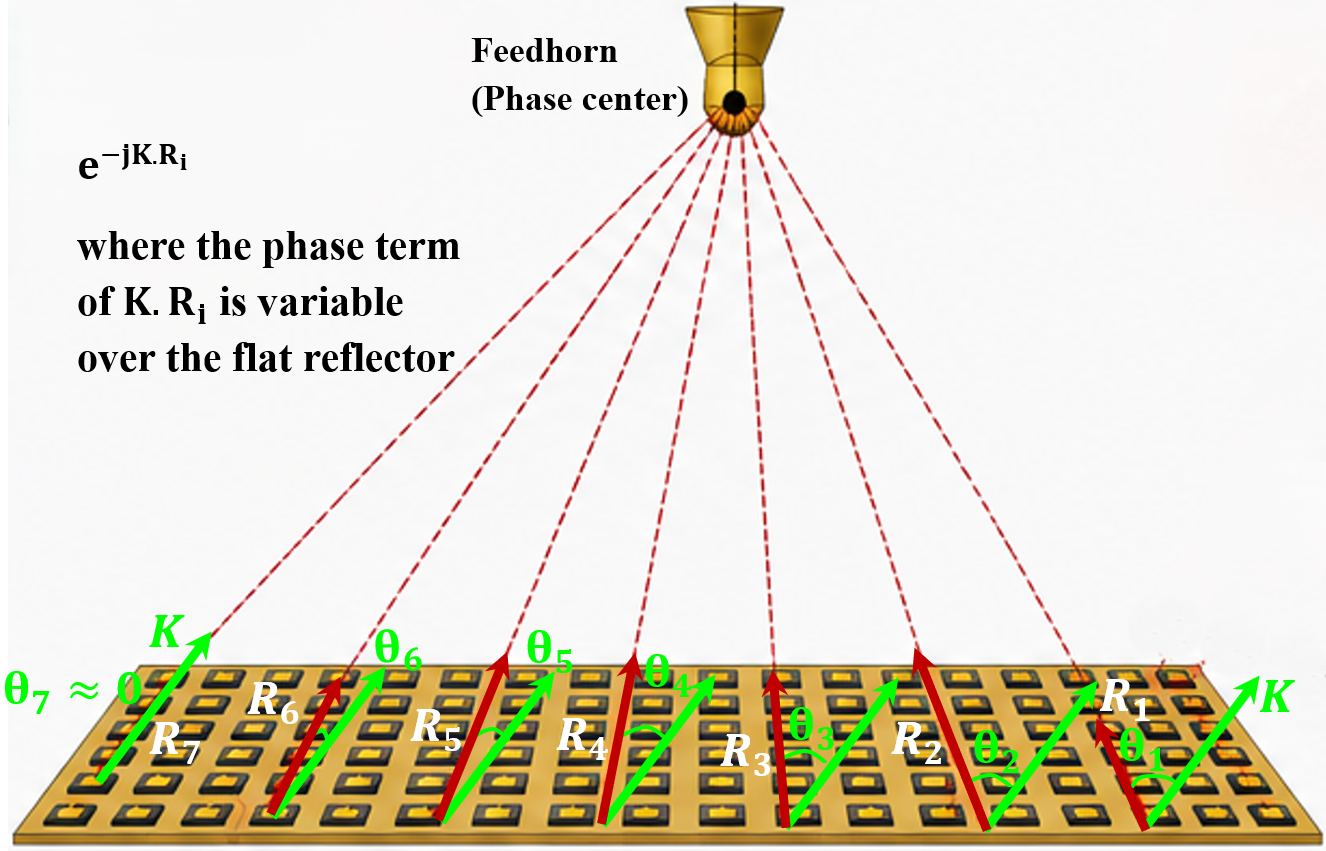}%
    }\\[2mm]
    \subfloat[]{%
        \includegraphics[width=0.5\columnwidth]{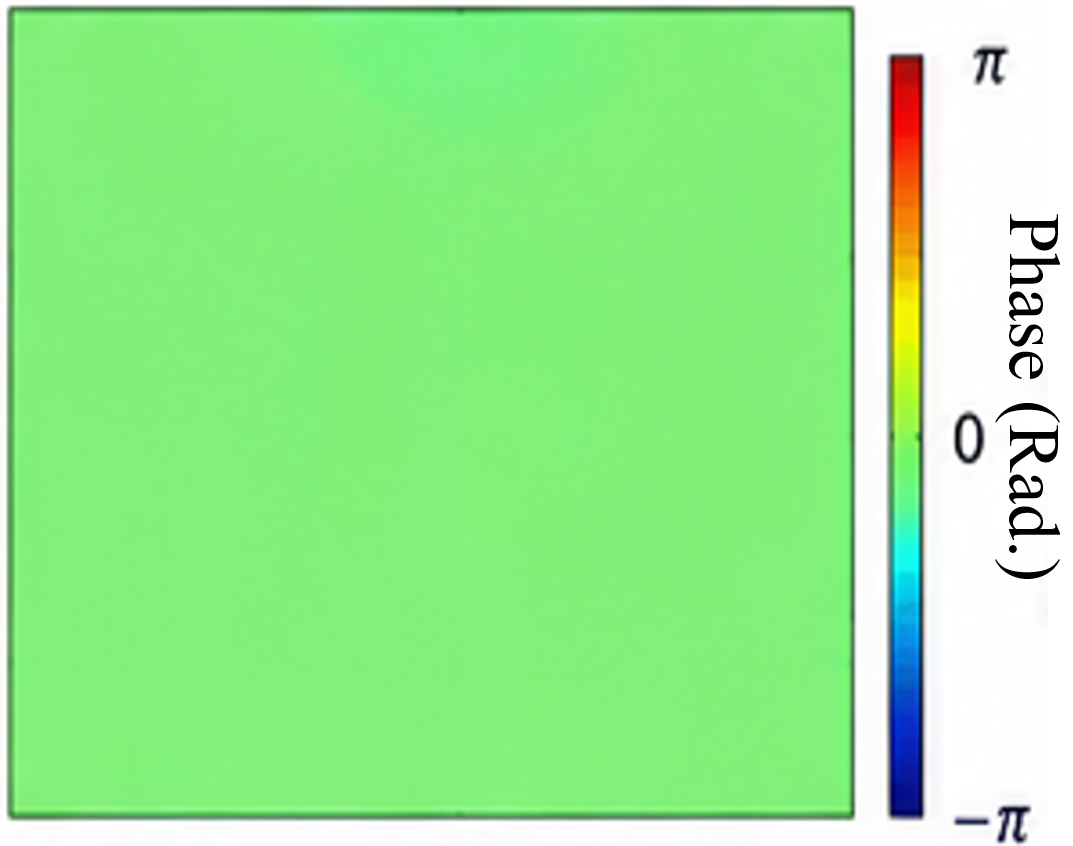}%
    }
    \hfill
    \subfloat[]{%
        \includegraphics[width=0.5\columnwidth]{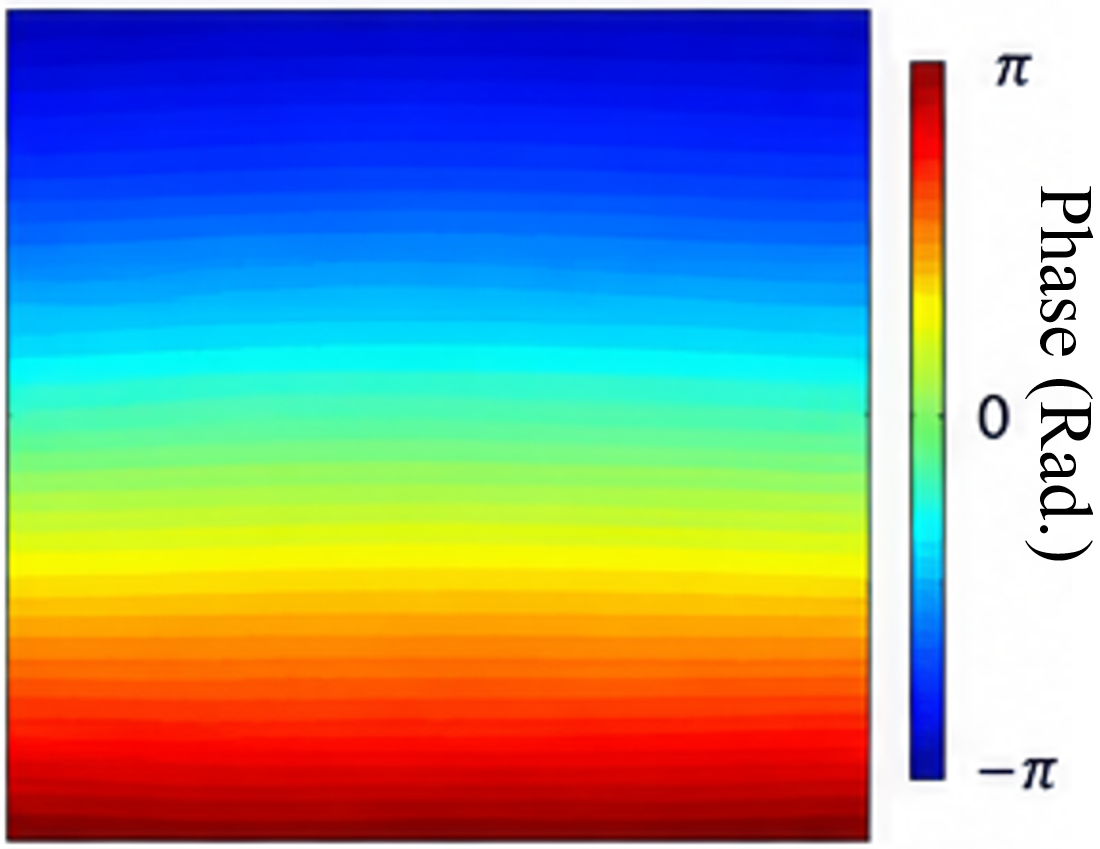}%
        \label{fig:vphase_reflector}%
    }
\caption{Performance comparison of planar and hemispherical apertures illuminated by the feedhorn: (a) an illustration of the intrinsic alignment between the incident wave vector $\mathbf{K}$ and the radial position vector $\mathbf{R}$ across the hemispherical transmitarray aperture. Since $\mathbf{K}$ is approximately aligned with $\mathbf{R}$ and normal to the local surface at each point, the feed-induced phase variation across the aperture is minimized, resulting in a nearly uniform incident phase distribution, (b) an illustration of the feed-induced spatial phase variation across the planar aperture. Different unit cells experience different propagation path lengths from the feed, resulting in nonuniform incident phases across the aperture, (c) an incident phase variation across the aperture of the hemispherical transmitarray lens shown in part $(a)$, with an RMS phase error of approximately $0^\circ$. The nearly uniform incident phase distribution minimizes feed-induced phase variations across the aperture, and (d) an incident phase variation across the aperture of the planar reflector shown in part $(b)$. The large phase variation results in a nonuniform incident phase distribution, thereby requiring additional phase compensation.}
\label{fig:transmitarray_reflector_K_R}
\end{figure}
\begin{table}[t]
\caption{Performance comparison of the effects of $\mathbf{K}$ and $\mathbf{R}$ alignment in the planar reflector of Fig.~\ref{fig:flat_reflector_configuration} and the hemispherical transmitarray antenna of Fig.~\ref{fig:hemispherical_transmitarray_2}.}
\label{tab:comparison_K_R}
\centering
\footnotesize
\setlength{\tabcolsep}{3pt}
\renewcommand{\arraystretch}{1.1}
\begin{tabular*}{\columnwidth}{@{\extracolsep{\fill}}
    p{0.26\columnwidth}
    p{0.31\columnwidth}
    p{0.35\columnwidth}
    @{}}
\hline
\textbf{Characteristic} &
\textbf{Planar Reflector} &
\textbf{Hemispherical Transmitarray Antenna} \\
\hline
Angles between $\mathbf{K}$ and $\mathbf{R}$ &
Varies across the aperture, $0^\circ \leq \theta \leq 90^\circ$ &
Approximately $0^\circ$ across the aperture due to $\mathbf{K} \parallel \mathbf{R}$ and thereby, $\mathbf{K}\cdot\mathbf{R}_n = K R_n = \text{constant}$ as a geometrical condition\\\\

Incident phase distribution across the aperture&
Nonuniform &
Nearly uniform \\

Feed-induced phase variation &
Significant and should be compensated&
Inherently minimized \\

Phase compensation and synthesis complexity &
High &
Low \\

Aperture efficiency &
Lower &
Higher \\

Beam accuracy &
Lower &
Higher \\

\hline
\end{tabular*}
\end{table}
As illustrated in Fig. 3.12 of Yamani~\cite{yamani2021reflectarray}, the phase distribution across a planar reflector is inherently nonuniform due to spatial variations in the feed-to-element path lengths, which introduce differential phase delays across the aperture~\cite{encinar2003broadband,carrasco2008bandwidth}. The nonuniform phase distribution across a planar reflector arises from spatial variations in the phase distribution across the individual array elements. This phase variation results from the unequal propagation paths from the horn antenna to the respective elements, which are determined by their relative positions with respect to the incident wavefront.
In contrast to planar reflectors, a hemispherical antenna maintains alignment between the incident-wave vector from the feedhorn and the position vectors of the array elements, resulting in a uniform phase distribution across the aperture lens.

Figure~\ref{fig:transmitarray_reflector_K_R} compares the relationship between the incident wave vector $\mathbf{K}$ from the feedhorn and the element position vectors $\mathbf{R}_{i}$ for both the planar reflector and the hemispherical antenna. 
Consequently, individual phase compensation is required to mitigate these delays and establish the desired aperture phase distribution.

For broadside radiation, this compensation yields an approximately uniform total aperture phase, whereas beam steering requires a corresponding spatial phase gradient.
However, employing individual phase adjustments increases the structural, biasing, and control complexity of planar reflectors, particularly for large arrays with independently controlled elements. As summarized in Table~\ref{tab:comparison_K_R}, the intrinsic alignment of the incident wave vector with the radial position vectors in the hemispherical transmitarray antenna suppresses feed-induced spatial phase errors, thereby enhancing aperture efficiency and beam-steering accuracy.
\subsection{Mitigation of Edge Diffraction and Parasitic Radiation}
In conventional planar reflectors and array antennas, the induced surface currents on the ground plane, together with mutual coupling among the array elements, can significantly influence the aperture fields and consequently degrade the radiation characteristics, particularly under wide-angle beam steering~\cite{stewart2022design}.
In this context, a major source of interference that can artificially increase the apparent spherical coverage arises from surface currents supported by the ground plane of planar array antennas and reflectors.
These currents propagate along the ground plane and diffract from its edges, producing spurious radiation that interferes with the desired radiation pattern of the antenna\cite{Camacho2020,Kuznetcov2024}. 
Consequently, the apparent spherical coverage becomes dominated by sidelobes and parasitic radiation rather than by the intended high-gain main beam, thereby reducing radiation efficiency, increasing interference, and degrading beam-steering performance.
Surface currents supported by the ground plane of planar arrays propagate toward the edges and diffract, generating spurious radiation that distorts the intended and useful main-beam pattern.
Thus, portions of the apparent spherical coverage may be dominated by sidelobes rather than by the intended main beam, resulting in ineffective angular coverage.
In contrast, the proposed hemispherical metasurface antenna effectively mitigates these limitations owing to its three-dimensional geometry.

The three-dimensional lens in Fig.~\ref{fig:hemispherical_transmitarray_2} inherently mitigates surface current propagation and edge diffraction, thereby ensuring that radiated power remains tightly confined to the intended and useful main beam over a broad angular scanning range. Consequently, the proposed transmitarray antenna achieves highly effective and robust spherical coverage in the THz regime. As detailed in this work and \cite{komeylian2026graphenebasedhemisphericaltransmitarrayantenna, komeylian2025high, komeylian2026activehemisphericalmetasurfacetransmitarray}, it outperforms planar arrays in both sidelobe suppression and beam-steering agility.
\subsection{Ohmic Loss Reduction Strategy}
As the operating frequency extends into the THz regime, conductor (ohmic) losses become a significant source of performance degradation; therefore, the relative contributions of conductor and dielectric losses should be meticulously considered and balanced according to the antenna configuration and constituent materials~\cite{freer2024loss}.

Indeed, at THz frequencies, conductor losses are substantially influenced by the distribution and concentration of electromagnetic fields and surface currents at conductive interfaces, where localized current crowding increases ohmic dissipation. 
Geometrical discontinuities, such as sharp corners and edges, alter the local field distribution and induce additional diffraction effects,~\cite{perlmutter1985electric}. 
Consequently, reducing the conductive area alone does not guarantee lower conductor loss; the resulting surface-current profile and electromagnetic response should be evaluated across the entire element geometry,~\cite{perlmutter1985electric}. 
To minimize conductive material while preserving necessary resonant paths and phase-control functionality, the proposed transmitarray antenna employs hollow FCC elements. 
Full-wave electromagnetic analysis accounts for the trade-offs between reduced metal volume, localized current concentration, edge diffraction, and mutual coupling. 
Furthermore, the aperture primarily utilizes dielectric materials, thereby avoiding the substantial ohmic dissipation typical of all-metallic implementations~\cite{perlmutter1985electric,freer2024loss}.
\subsection{High-Directivity FCC Elements for Mutual Coupling Suppression}
Severe atmospheric absorption and free-space path loss represent fundamental challenges to reliable wireless communication in the THz frequency regime\cite{dovelos2023superdirective}.
Although antenna arrays can compensate for these propagation losses by providing higher array gain through coherent beamforming, the radiation performance of individual array elements can degrade because of mutual electromagnetic coupling between neighboring elements\cite{chou2022radiation}. 
This mutual coupling can reduce the directivity of individual elements, distort the overall radiation pattern, and degrade the aperture efficiency of the array.

The FCC geometry improves the performance of the individual elements
while mitigating mutual coupling, thereby preserving the array radiation efficiency and beamforming capability over a wide scanning range. Consequently, the proposed hemispherical topology incorporating FCC elements enables high-gain, wide-angle beam steering while maintaining stable directivity, radiation efficiency, and sidelobe levels under large angular scanning conditions. These characteristics demonstrate the potential of the proposed transmitarray antenna for high-frequency wireless communication applications, including THz systems and emerging 6G and dense satellite communication networks.
\subsection{Application for tracking moving targets}
Another major advantage of our proposed hemispherical transmitarray antenna in Fig.~\ref{fig:hemispherical_transmitarray_2} consists of its suitability for dynamic moving-target tracking applications. For practical implementation of tracking moving targets, our hemispherical transmitarray antenna should provide not only the electromagnetic beam-steering performance but also sufficiently rapid and dynamic reconfiguration of its radiation pattern.
Its static electromagnetic beam-steering performance is essential to meet three primary functional requirements: (1) adaptive beam steering and target tracking: the proposed hemispherical transmitarray antenna dynamically steers its main beam to track the target's motion, thereby maintaining a stable communication link with high directional gain, (2) adaptive HPBW control: adaptive HPBW control enables the beamwidth to be dynamically adjusted, providing a broader beam for initial target acquisition and a narrower, higher-gain beam for subsequent lock-on tracking, thereby improving the received signal strength and SNR, and (3) dynamic sidelobe suppression and interference mitigation: dynamic radiation-pattern reconfiguration enables adaptive sidelobe suppression and null steering toward sources of multipath interference or intentional jamming. 
This spatial filtering capability enhances communication robustness and enhances the reliability of target tracking.

Furthermore, practical moving-target tracking necessitates evaluating several dynamic system parameters beyond electromagnetic beam-steering performance. 
These critical parameters include the bias-network switching speed,\cite{emara2024reconfigurable}, the RC charging time constant of the gated graphene devices, the update rate and latency of the electronic controller,\cite{rizza2021real}, and the execution speed of the real-time tracking algorithm used to estimate target trajectory and update beam-steering vectors. A comprehensive evaluation of these dynamic parameters is deferred to future work.
\section{Conclusion}
To conclude, this work presented a comprehensive comparison of the proposed hemispherical transmitarray antenna and its planar reflector counterpart for wide-angle beam steering in the THz regime. 

Our proposed hemispherical transmitarray antenna has been comprehensively investigated in\cite{komeylian2025high,komeylian2026graphenebasedhemisphericaltransmitarrayantenna,komeylian2026activehemisphericalmetasurfacetransmitarray}. 
These studies established the underlying analytical framework, phase-synthesis methodology for dynamic beam steering, achievable beam-steering range, RLC circuit model, and active array-factor formulation. Moreover, its radiation performance was systematically characterized in terms of gain, efficiency, and HPBW\cite{komeylian2025high,komeylian2026graphenebasedhemisphericaltransmitarrayantenna,komeylian2026activehemisphericalmetasurfacetransmitarray}.

To provide a systematic framework for evaluating the performance advantages of the proposed hemispherical transmitarray antenna over its planar reflector counterpart, we first establish the theoretical framework of the planar reflector.

The numerical results of the planar reflector yield a maximum elevation beam-steering range of $\pm60^\circ$ with full $360^\circ$ azimuthal coverage. 
Compared with the planar reflector, our proposed hemispherical transmitarray increases the elevation beam-steering range to $\pm78^\circ$ while preserving continuous $360^\circ$ azimuthal coverage.

These results demonstrate that the hemispherical geometry provides an additional degree of freedom for controlling the HPBW through the feedhorn and lens dimensions, while simultaneously maintaining more consistent radiation characteristics over a wide beam-steering range.
Quantitatively, the HPBW variation is substantially reduced from $26.48^\circ$ for the planar reflector to $13.1^\circ$ for the proposed hemispherical transmitarray antenna, corresponding to a reduction by more than a factor of two or a reduction of approximately $50.6\%$.
This significant reduction in the beamwidth variation enables a more stable and controllable radiation pattern over a wide beam-steering range, thereby preserving higher directional power concentration, directivity, and gain, particularly at large steering angles. 

The proposed transmitarray antenna supports a wide beam-steering frequency range of 200--300~GHz with $\lvert S_{11} \rvert < -10$~dB, corresponding to a 40\% fractional bandwidth, while maintaining wide-angle beam-steering capabilities \cite{komeylian2025high,komeylian2026graphenebasedhemisphericaltransmitarrayantenna,komeylian2026activehemisphericalmetasurfacetransmitarray}.
In this work, the beam-steering consistency and stability of the proposed hemispherical transmitarray antenna are verified across an extended frequency range of $240$-$360$ GHz with an operating frequency of $300$ GHz.

The superior beam-steering performance of the proposed transmitarray antenna stems primarily from the intrinsic $\mathbf{K}$ and $\mathbf{R}$ alignment in its hemispherical aperture, which promotes a more uniform phase distribution across the lens aperture and thereby facilitates accurate phase synthesis for dynamic beam steering.

Furthermore, the combination of the hemispherical geometry and FCC elements mitigates edge diffraction and parasitic radiation, substantially enhances the operating bandwidth, and reduces ohmic losses, thereby improving the overall beam-steering performance.
Comparisons with state-of-the-art planar reflectors,\cite{li2026wide}, and antennas,\cite{jiang2026breaking}, presented in the literature further highlight the advantages of the proposed architecture in terms of steering range, beamwidth stability, and directional performance.
Overall, the results establish the hemispherical transmitarray as a high-performance and viable architecture for electronically reconfigurable, wide-angle, three-dimensional beam steering in future THz and 6G wireless systems.
\bibliographystyle{IEEEtran}
\bibliography{References}
\end{document}